\documentclass[12pt]{article}

\usepackage{newtxtext,newtxmath}
\usepackage{graphicx}
\usepackage{multibib}
\usepackage{caption}
\usepackage[letterpaper,margin=1in]{geometry}
\usepackage{lineno}   

\renewenvironment{abstract}
	{\quotation}
	{\endquotation}

\date{}

\makeatletter
\renewcommand{\fnum@figure}{\textbf{Figure \thefigure}}
\renewcommand{\fnum@table}{\textbf{Table \thetable}}
\makeatother

\usepackage{scicite}
\usepackage{url}

\usepackage{amsmath}
\usepackage{multirow}
\usepackage{makecell}
\usepackage[figuresright]{rotating}
\usepackage{xcolor}
\usepackage{longtable}
\usepackage{utfsym}
\usepackage{placeins}

\def\scititle{
	Critical Thresholds in Non-Pharmaceutical Interventions for Epidemic Control
}
\title{\bfseries \boldmath \scititle}

\author{
	Jinghui~Wang$^{1}$,
	Yutian~Zeng$^{1}$,
	Cong~Xu$^{1}$,
	Zhanwei~Du$^{1}$,
	Xiyun~Zhang$^{2}$,\\
	Jiarong~Xie$^{3}$,
	Jiu~Zhang$^{1}$,
	Sen~Pei$^{4}$,
	Zijian~Feng$^{5}$,
	Yanqing~Hu$^{1,6\ast}$\and
	\small$^{1}$Department of Statistics and Data Science, Southern University of Science\\ \small and Technology, Shenzhen 518055, China.\and\
	\small$^{2}$Department of Physics, Jinan University, Guangzhou 510632, China.\and\
	\small$^{3}$Center for Computational Communication Research, Institute of Advanced Studies in Humanities\\ \small and Social Science, Beijing Normal University at Zhuhai, Zhuhai 519087, China.\and\
	\small$^{4}$Department of Environmental Health Sciences at Mailman School of Public Health, Columbia\\ \small University, New York 10032, USA.\and\
	\small$^{5}$School of Public Health and Emergency Management, Southern University of Science and\\ \small Technology, Shenzhen 518055, China.\and\
	\small$^{6}$Center for Complex Flows and Soft Matter Research, Southern University of Science and\\ \small Technology, Shenzhen 518055, China.\and
	\small$^\ast$Corresponding author. Email: yanqing.hu.sc@qq.com
}

\begin{document}

\maketitle



\begin{abstract} \bfseries \boldmath
Non-pharmaceutical interventions (NPIs), such as contact tracing and social distancing, are critical for controlling epidemic outbreaks, yet their dynamic interactions remain underexplored.
We introduce a probabilistic framework to analyze the synergy between contact tracing speed, quantified by the contact tracing period $\tau$, and the average number of close contacts, $\bar{k}_+$, reflecting social distancing measures.
We identify critical thresholds ($R = 1$) that separate pandemic and contained phases in the $\bar{k}_{+}$--$\tau$ plane, validated using high-resolution data from Shenzhen's 2022 Omicron outbreak (1{,}187 cases, 86{,}451 contacts).
Our findings indicate that contact tracing alone can contain outbreaks of pathogens with $R_0 < 3.10$ (95\% CI 3.01--3.17).
Combining contact tracing with strong social distancing extends controllability to pathogens with $R_0 < 12.88$ (95\% CI 12.33--13.37).
These thresholds are derived from empirical data in Shenzhen and may vary in other settings with different transmission modes.
These results, supported by empirical data, highlight the efficacy of rapid tracing and targeted social distancing as alternatives to mass Polymerase Chain Reaction (PCR) testing.
Our framework offers actionable insights for optimizing NPI strategies, though challenges in scaling to regions with higher tracing missing rates or weaker infrastructure underscore the need for adaptive, data-driven policies.
\end{abstract}

\noindent
The unpredictable transmission dynamics of emerging infectious diseases, combined with their unknown virulence, have necessitated broad non-pharmaceutical interventions (NPIs), such as case detection, contact tracing, contact isolation, and social distancing to safeguard public health~\cite{baker2022infectious,wu2020new,smith2015use,bell2006non}.
These measures have proven indispensable in delaying and containing the pandemic (e.g., severe acute respiratory syndrome--coronavirus 2), offering a real-time lesson in modern epidemiology~\cite{bell2006non,kraemer2020effect,hellewell2020feasibility,merler2015spatiotemporal,lai2020effect,polonsky2022feasibility,zhang2022shanghai,zou2025impact}.
However, the dynamic interactions and underlying mechanisms of NPIs remain poorly understood, particularly the critical factors governing their execution and effectiveness.
Elucidating how these interventions synergize could revolutionize proactive containment strategies, enabling more targeted and efficient responses to large-scale disease transmission.

Motivated by this objective, we introduce the contact tracing period $\tau$ to quantify the speed of close-contact tracing, and the average number of close contacts of a positive case $\bar{k}_+$ to capture the extent of social distancing measures.
The effective reproduction number $R$ (or $R_t$, where $t$ stands for time) of an infectious disease can be derived to be (see Materials and Methods for more details)
\begin{equation}
	R = \bar{k}_+ \, \beta \, H(\tau),
	\label{eq:R}
\end{equation}

\noindent where $\beta$ is the probability that a positive case would infect a healthy contact per day and $H(\tau)$ is a function of $\tau$ measuring the expected effective infectious period accounting for the implementation of NPIs.
Successful epidemic containment is equivalent to controlling $R$ to be less than 1, which can be achieved by speeding up the close-contact tracing process (making $H(\tau)$ smaller) or enhancing social distancing measures (making $\bar{k}_+$ smaller).
Consequently, for a given infectious disease, $R=1$ corresponds to a curve of $\bar{k}_+$ versus $\tau$, termed the ``critical line''.
The $\bar{k}_+$--$\tau$ plane is divided by the critical line into two areas: one is the pandemic phase with $R>1$, and the other is the contained phase with $R<1$.
Moreover, for any given value of $\bar{k}_+$, there is a critical contact tracing period $\tau_c$, such that the disease transmission can be successfully interrupted when $\tau < \tau_c$.
Similarly, there is also a critical value of $\bar{k}_+$, denoted as $\bar{k}_{+c}$, corresponding to any given value of $\tau$ (see section~\ref{sup:univeral_R_tau} for more details).


\subsection*{Results}
Integrating test-trace-quarantine dynamics, our framework is robustly supported by empirical data (figure~\ref{fig:real_control}, figure~\ref{fig:result}A and table~\ref{tab:R0_tauc}).
Our theory elucidates why NPIs effectively contained the Delta variant in 2021 but struggled against the Omicron variant in 2022 in Shenzhen.
We estimate the critical contact tracing period, $\tau_c$, at 26 hours (95\% CI 24.75--28.07) for Delta and 13 hours (95\% CI 11.63--14.58) for Omicron.
Given Shenzhen has the reaction times at 17 hours in 2021~\cite{control_manual_delta} and 11 hours in 2022~\cite{control_manual_omicron}, explaining the success against Delta and subsequent adaptation for Omicron.
These alignments validate our theory, highlighting the need for rapid policy adjustments to counter highly transmissible variants, as previous studies have shown that contact tracing diminishes in efficacy against more transmissible variants like Omicron~\cite{ge2023effects}.
Our framework, with predictions shown in figure~\ref{fig:result}A,~B and table~\ref{tab:R0_tauc}, provides actionable guidance for optimizing containment strategies.

\begin{center}
	\includegraphics{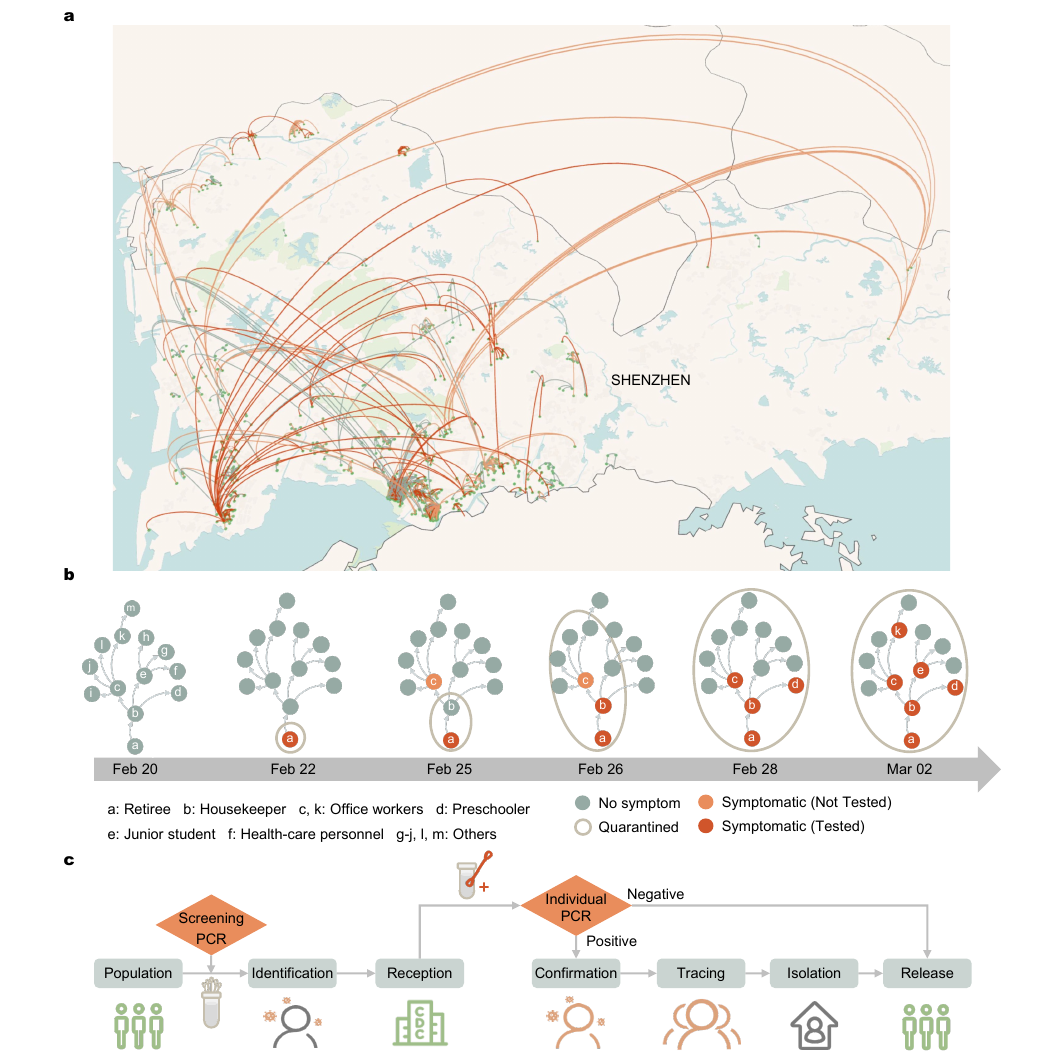}
    \captionof{figure}{
		\textbf{Epidemic control in Shenzhen, 2022.}
		(\textbf{A}) Contact trajectories of positive cases.
		Nodes are geolocated to cases' residential addresses.
		Edges represent epidemiological links indicating close contact history between them.
		The color of the edge (olive green, warm peach, burnt orange) indicates three time periods divided equally according to the chronological order of close contact.
		(\textbf{B}) A real transmission chain (see section~\ref{sup:dataset_spread_chain}).
		Muted teal nodes represent individuals who have no symptoms and have not been tested positive or have not been tested;
		burnt orange nodes represent individuals who have symptoms and have been tested positive;
		warm peach nodes represent individuals who have symptoms and have not been tested.
		Warm beige circles around nodes indicate that individuals within each circle have been under quarantine till the corresponding date, and the directed edges between nodes represent the disease transmission path.
		The transmission chain started from a retiree, confirmed through the community's PCR testing on February 22nd, who was immediately isolated.
		A housekeeper, who lived in the same building as the retiree, was identified as a close contact and was quarantined on February 25th.
		Subsequently, three family members served by the housekeeper (nodes c, k, and d), the housekeeper's daughter (node e), and three other close contacts (nodes i, j, and l) were also quarantined.
		Until February 28th, with all 13 people on the transmission chain (of whom six tested positive) in quarantine, this transmission chain was cut off successfully.
		(\textbf{C}) The process of isolating positive cases and quarantining close contacts according to Shenzhen's Prevention and Control Manual (see section~\ref{sup:real_control}).
	}
		\label{fig:real_control}
\end{center}

We analyzed transmission chains from Shenzhen's February--April 2022 Omicron outbreak, leveraging high-resolution contact tracing and PCR testing data (1{,}187 cases, 86{,}451 close contacts, figure~\ref{fig:real_control}), and tested epidemiological models via probabilistic methods (figure~\ref{fig:result}C, section~\ref{sup:dataset_spread_chain}).
The epidemic's trajectory reveals how NPIs transitioned the outbreak from a high-risk to a contained state, crossing the critical threshold into a non-outbreak phase.
However, lapses in enforcement, marked by increases in $\tau$ or $\bar{k}_+$, triggered rebounds, with multiple peaks in daily cases (figure~\ref{fig:result}C,~D).
Shenzhen's data align closely with our model, underscoring the utility of the $\bar{k}_+$--$\tau$ phase plot.
This tool not only evaluates NPI efficacy dynamically but also identifies operational gaps, enabling timely interventions like accelerated tracing or enhanced social distancing.

\begin{table}
	\centering
	\small
	\caption{\textbf{Estimated basic reproduction number $R_0$ and critical reaction time $\tau_c$.}
		The estimated $R_0$ of several variants of COVID-19 in different regions and the corresponding critical reaction time $\tau_c$ in hours (see section~\ref{sup:R0_beta_parameter}).
		Intervals in parentheses represent the 95\% confidence intervals.
		The asterisks represent the estimated $R_0$ (95\% confidence interval) of the Omicron transmission in Shenzhen (see section~\ref{sup:data_R0}).}
	\label{tab:R0_tauc}
	\begin{tabular}{@{}cccccc@{}}
		\\
		\hline
		Variants &
		Omicron (SZ) &
		Omicron BA.2 &
		Omicron BA.1 &
		Delta &
		Alpha \\
		\hline
		$R_0$ &
		\begin{tabular}[c]{@{}c@{}}20.17*\\(14.39*, 25.64*)\end{tabular} &
		\begin{tabular}[c]{@{}c@{}}18.21\\(16.25, 20.03)~\cite{wang2022reproduction}\end{tabular} &
		\begin{tabular}[c]{@{}c@{}}11.44\\(10.21, 12.58)~\cite{wang2022reproduction}\end{tabular} &
		\begin{tabular}[c]{@{}c@{}}5.50\\(4.91, 6.05)~\cite{liu2020reproductive,campbell2021increased}\end{tabular} &
		\begin{tabular}[c]{@{}c@{}}3.63\\(3.46, 3.71)~\cite{liu2020reproductive,campbell2021increased}\end{tabular} \\
		\hline
		$\tau_c$ &
		\begin{tabular}[c]{@{}c@{}}13.07\\(11.63, 14.58)\end{tabular} &
		\begin{tabular}[c]{@{}c@{}}13.78\\(13.15, 15.53)\end{tabular} &
		\begin{tabular}[c]{@{}c@{}}17.43\\(16.59, 18.50)\end{tabular} &
		\begin{tabular}[c]{@{}c@{}}26.19\\(24.75, 28.07)\end{tabular} &
		\begin{tabular}[c]{@{}c@{}}34.38\\(33.68, 35.35)\end{tabular} \\
		\hline \\
	\end{tabular}
\end{table}

Frequent large-scale PCR testing, aimed at detecting cases missed by contact tracing, imposes significant socioeconomic costs.
We explored whether contact tracing alone could suffice, assuming an 18.6\% missing rate, as observed in Shenzhen (see Materials and Methods), and a $\tau$ of 11 hours~\cite{control_manual_omicron}.
Our model indicates that diseases with an $R_0$ below 3.10 (95\% CI 3.01--3.17) can be controlled without PCR supplementation.
By reducing $\bar{k}_+$ from 177.64 (baseline, 95\% CI 155.40--199.88) to 51.14 (under Shenzhen's social distancing, 95\% CI 37.82--64.46), the control policy extends containment to diseases with $R_0$ up to 12.88 (95\% CI 12.33--13.37) (figure~\ref{fig:policy}A,~B).
Based on an extensive literature review (see Materials and Methods), most mean $R_0$ estimates for respiratory pathogens and their sub-lineages (e.g., variants, serotypes, and subtypes) remain below the social distancing threshold, with only a minority exceeding it (figure~\ref{fig:policy}D).

\begin{center}
	\includegraphics{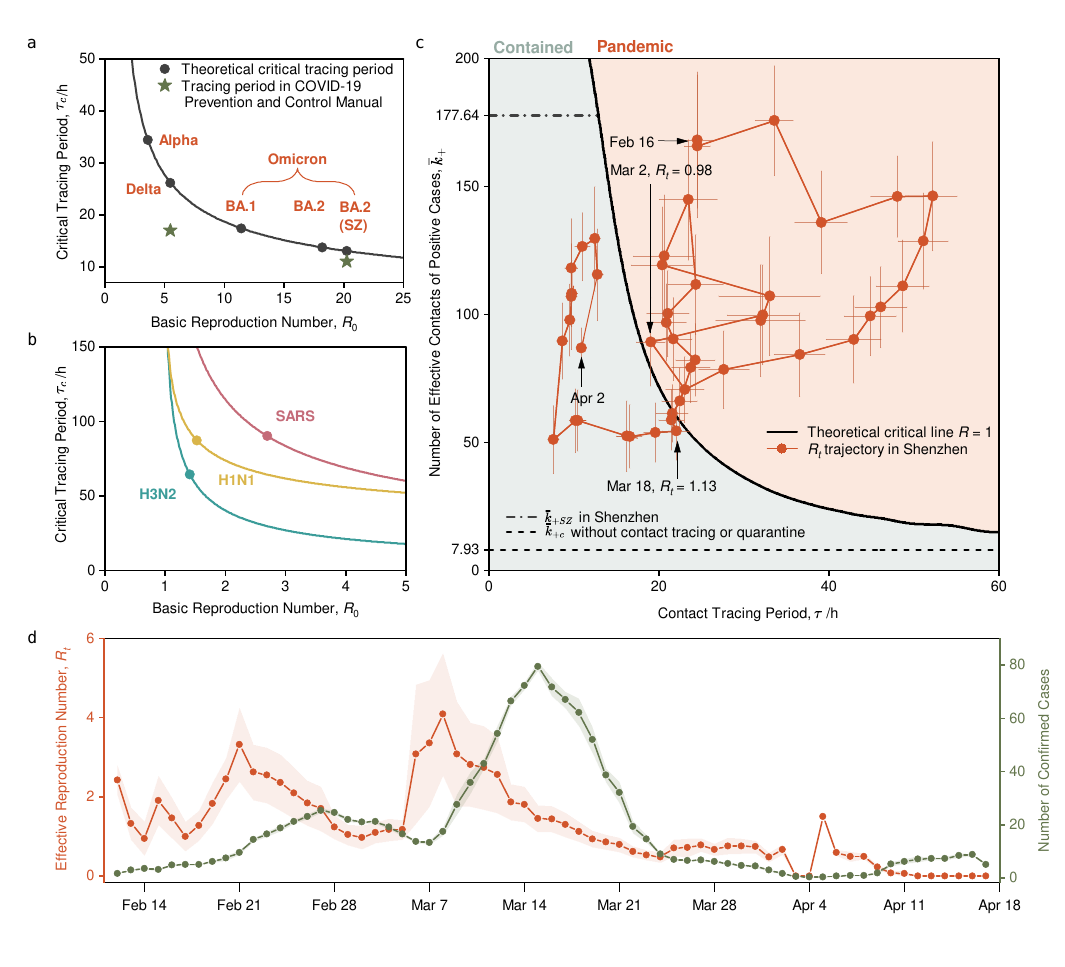}
    \captionof{figure}{\textbf{Theoretical results versus empirical data analysis results.}
		(\textbf{A}) Theoretical relationship between the critical reaction time $\tau_c$ (in hours) and the basic reproduction number $R_0$ for different variants of COVID-19 by eq.~\eqref{eq:criticalreactime}.
		The dark dots mark the $R_0$ and the corresponding $\tau_c$ of four COVID-19 variants and the one spread in Shenzhen as listed in table~\ref{tab:R0_tauc}.
		The olive green asterisks represent the tracing period set in the COVID-19 Prevention and Control Manual for Delta (17 hours) and Omicron (11 hours).
		(\textbf{B}) Theoretical relationship between $\tau_c$ and $R_0$ for three typical epidemics by eq.~\eqref{eq:criticalreactime}.
		The dots mark the $R_0$ and the corresponding $\tau_c$ of SARS, H1N1, H3N2 as listed in table~\ref{tab:R0_tauc}.
		Distributions of disease transmission state durations are provided in section~\ref{sup:dist_parameter}.
		(\textbf{C}) Phase plot of the containment policy on the $\bar{k}_+$--$\tau$ plane and $\bar{k}_{+t}$--$\tau_t$ trajectory of Shenzhen's epidemic control from February 16th to April 2nd, 2022, where each dot is calculated by averaging over the positive cases confirmed from day $t-3$ to day $t+3$.
		The dash-dotted line marks the actual value of the average number of effective contacts ($\bar{k}_{+SZ}$) in Shenzhen.
		The dashed line marks the critical value ($\bar{k}_{+c}$ by eq.~\eqref{eq:kc_infty}) with transmission rate $\beta_{SZ}$ in Shenzhen) in the absence of contact tracing or quarantine (i.e., $\tau \to \infty$).
		Solid black curve represents the theoretical critical line ($R = 1$);
		burnt orange points with error bars depict the mean values and their corresponding 95\% confidence intervals, aligned with the empirical trajectory based on real Shenzhen data.
		Please refer to section~\ref{sup:all_quantities} for the computation of the time-varying $\tau_t$ and $\bar{k}_{+t}$, as well as $\bar{k}_{+SZ}$ and $\beta_{SZ}$.
		(\textbf{D}) The daily effective reproduction number $R_t$ and number of newly confirmed cases (see section~\ref{sup:data_Rt} for more details).
	}
		\label{fig:result}
\end{center}

Contact tracing-based policies, previously adopted in Singapore, Japan, Australia, Belgium, and New Zealand~\cite{wang2021policy,stobart2022australia,luyten2022belgium,cumming2022going}, faltered due to imported cases.
Using Shenzhen's 11-hour $\tau$ benchmark, we assessed containment limits under higher missing rates: 39\% (UK~\cite{keeling2020efficacy}), 66\% (US~\cite{lash2021covid}), and 73\% (Hong Kong~\cite{yang2022universal}).
Contact tracing alone controls diseases with the basic reproduction number threshold ($R_0^c$) of 1.95 (95\% CI 1.87--1.97), 1.29 (95\% CI 1.24--1.32), and 1.18 (95\% CI 1.14--1.21) in these regions, respectively.
With Shenzhen-level social distancing, $R_0^c$ rises to 7.98 (95\% CI 7.50--8.15), 5.01 (95\% CI 4.81--5.17), and 4.53 (95\% CI 4.33--4.67) (figure~\ref{fig:policy}C).
These findings suggest that, despite regional variations, the control policy offers a viable framework for early-stage epidemic suppression globally.

\subsection*{Discussion}
This study advances epidemic control through a novel analysis of non-pharmaceutical interventions (NPIs).
We developed a dynamic model that captures the competition between disease transmission and NPIs, establishing an operational critical contact tracing period, $\tau_c$, functionally linked to the basic reproduction number, $R_0$.
The model quantifies the synergy between test-trace-quarantine and social distancing, which tracks the epidemic's trajectory from a high-risk to a non-outbreak state.
In the case study of Shenzhen, we demonstrate that contact tracing alone can contain diseases with 3.10 (95\% CI 3.01--3.17).
With strong social distancing reducing $\bar{k}_+$ from 177.64 to 51.14, containment extends to 12.88 (95\% CI 12.33--13.37), without extensive PCR testing (table~\ref{tab:assumptionC}).

\begin{center}
	\centering
	\includegraphics{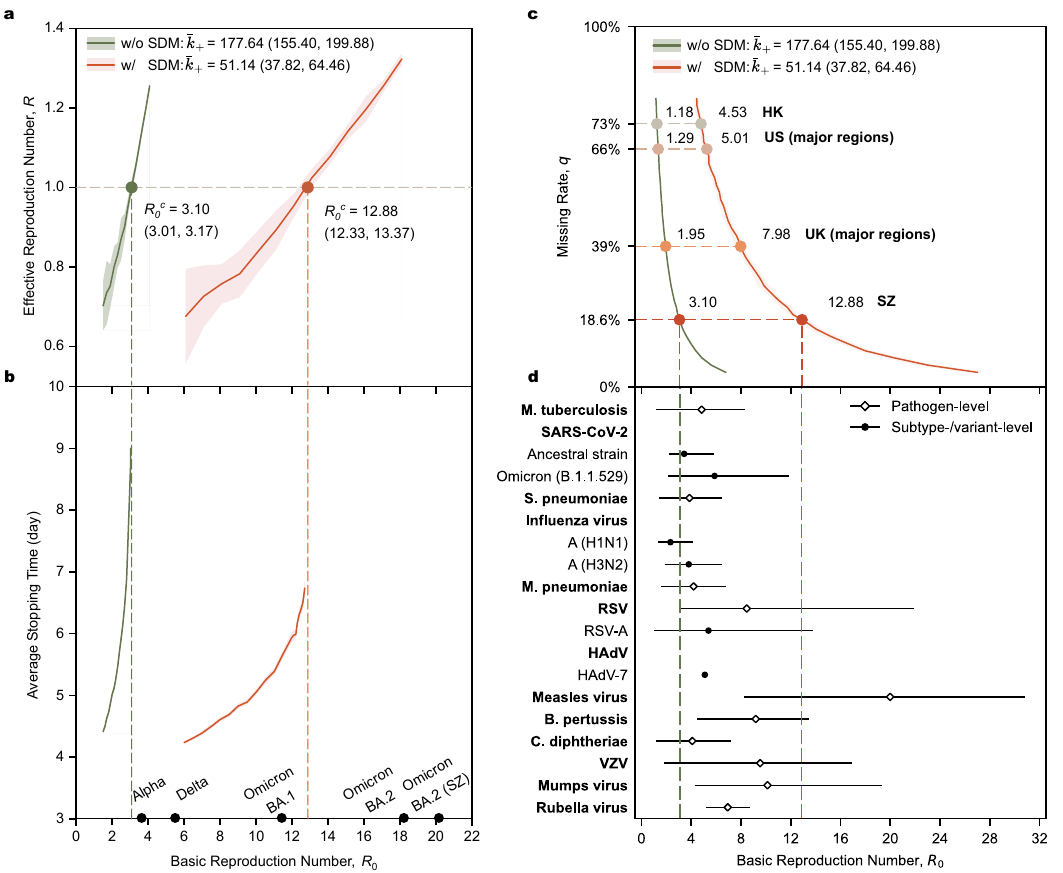}
    \captionof{figure}{
		\textbf{Policy with incomplete contact tracing.}
		The olive green and burnt orange lines in (\textbf{A})--(\textbf{D}) represent simulation results for the Shenzhen network under weak and strong social distancing measures (weak SDM and strong SDM), respectively.
		The intensity of social distancing is classified into two categories: ``Weak'', characterized by a contact rate of $\bar{k}_+ = 177.64$ and reflecting the least stringent measures implemented in Shenzhen, and ``Strong'', characterized by $\bar{k}_+ = 51.14$ and reflecting the most stringent measures observed during the period from March 21 to March 27, 2022.
		In each simulation, we miss a proportion $q$ of close contacts for each positive case during contact tracing.
		Based on Shenzhen's control data, we set $q = 18.6\%$ as the proportion of positive cases identified by large-scale PCR testing instead of contact tracing.
		The contact tracing periods in all simulations are set to $\tau = 11$ hours.
		(\textbf{A})--(\textbf{B}) Stabilized effective reproduction number (A) and average stopping time (B) for diseases with different $R_0$.
		The average stopping time is the average period from the start of each simulation until all positive cases in the simulation are in the R (removed) or Q (quarantined) state.
		The basic reproduction numbers of Alpha~\cite{liu2020reproductive,campbell2021increased}, Delta~\cite{liu2020reproductive,campbell2021increased}, Omicron BA.1~\cite{wang2022reproduction}, BA.2~\cite{wang2022reproduction} and BA.2 (SZ) variants are marked on the $x$-axis.
		(\textbf{C}) The relationship between the missing rate in contact tracing, versus the critical basic reproduction number $R_0^c$, i.e., the value of $R_0$ under which an infectious disease can be successfully controlled.
		Points with four different colors mark the $R_0^c$ corresponding to missing rates of Shenzhen, major regions in the UK~\cite{keeling2020efficacy}, major regions in the US~\cite{lash2021covid}, and Hong Kong~\cite{yang2022universal}.
		For details of the simulations, please refer to section~\ref{sup:simu} and section~\ref{sup:missing_rate}.
		Note that these results are theoretical results based on Shenzhen's $\bar{k}_+$, generalization to other regions may differ in practice.
		(\textbf{D}) Forest-style plot of reported $R_0$ estimates for pathogens (in bold text) and their lower-level categories (e.g., variants, serotypes, and subtypes).
		Pathogen-level denotes either an individual pathogen or, where conventionally used in the literature, a clinically recognized pathogen group.
		Open diamonds denote pathogen-level mean estimates and filled circles denote lower-level categories.
		Horizontal lines show the ranges (see table~\ref{SItable_pathogens_variants} for details).
		Vertical dashed lines indicate the two thresholds.
		Please refer to Materials and Methods for the procedure of selection of pathogens and their lower-level categories.
	}
		\label{fig:policy}
\end{center}

\begin{table}
	\centering
	\small
	\caption{\textbf{Critical basic reproduction number $R_0^c$.}
		Additional analyses were performed by varying the contact tracing period ($\tau$) (11, 24, 48, 72 hours), the missing rate of contacts ($q$) (18.6\%, 40\%, 50\%, 60\%, 70\%), and the intensity of social distancing (classified as ``Weak'' with contact rate $\bar{k}_+ = 177.64$, reflecting the least stringent measures in Shenzhen, and ``Strong'' with $\bar{k}_+ = 51.14$, reflecting the most stringent measures observed during the period from March 21 to March 27, 2022).
		For each scenario, 50{,}000 stochastic branching process simulations were conducted.
		The critical threshold $R_0^c$ was determined as the $R_0$ value at which $R = 1$ under the Assumption C (unlimited recursive forward tracing plus first-order backward tracing, further details in section~\ref{sup:simu} and section~\ref{sup:missing_rate}).}
	\label{tab:assumptionC}

	\begin{tabular}{@{}ccccccc@{}}
		\hline
		\begin{tabular}[c]{@{}c@{}}Contact tracing\\period, $\tau$/h\end{tabular} &
		\begin{tabular}[c]{@{}c@{}}Social distancing\\intensity\end{tabular} &
		\multicolumn{5}{c}{Missing rate, $q$} \\
		\cline{3-7}
		 &  & 18.6\% & 40\% & 50\% & 60\% & 70\% \\
		\hline
		11 & Weak &
		\begin{tabular}[c]{@{}c@{}}3.10\\(3.01, 3.17)\end{tabular} &
		\begin{tabular}[c]{@{}c@{}}1.92\\(1.84, 1.99)\end{tabular} &
		\begin{tabular}[c]{@{}c@{}}1.65\\(1.61, 1.69)\end{tabular} &
		\begin{tabular}[c]{@{}c@{}}1.43\\(1.39, 1.47)\end{tabular} &
		\begin{tabular}[c]{@{}c@{}}1.28\\(1.25, 1.31)\end{tabular} \\
		   & Strong &
		\begin{tabular}[c]{@{}c@{}}12.88\\(12.33, 13.37)\end{tabular} &
		\begin{tabular}[c]{@{}c@{}}7.88\\(7.48, 8.19)\end{tabular} &
		\begin{tabular}[c]{@{}c@{}}6.62\\(6.28, 6.84)\end{tabular} &
		\begin{tabular}[c]{@{}c@{}}5.62\\(5.38, 5.82)\end{tabular} &
		\begin{tabular}[c]{@{}c@{}}4.96\\(4.75, 5.13)\end{tabular} \\
		\hline
		24 & Weak &
		\begin{tabular}[c]{@{}c@{}}2.39\\(2.33, 2.44)\end{tabular} &
		\begin{tabular}[c]{@{}c@{}}1.68\\(1.62, 1.73)\end{tabular} &
		\begin{tabular}[c]{@{}c@{}}1.49\\(1.45, 1.52)\end{tabular} &
		\begin{tabular}[c]{@{}c@{}}1.34\\(1.30, 1.37)\end{tabular} &
		\begin{tabular}[c]{@{}c@{}}1.24\\(1.20, 1.26)\end{tabular} \\
		   & Strong &
		\begin{tabular}[c]{@{}c@{}}9.18\\(8.88, 9.45)\end{tabular} &
		\begin{tabular}[c]{@{}c@{}}6.49\\(6.21, 6.70)\end{tabular} &
		\begin{tabular}[c]{@{}c@{}}5.65\\(5.36, 5.87)\end{tabular} &
		\begin{tabular}[c]{@{}c@{}}5.20\\(4.94, 5.38)\end{tabular} &
		\begin{tabular}[c]{@{}c@{}}4.64\\(4.39, 4.82)\end{tabular} \\
		\hline
		48 & Weak &
		\begin{tabular}[c]{@{}c@{}}1.70\\(1.65, 1.73)\end{tabular} &
		\begin{tabular}[c]{@{}c@{}}1.39\\(1.34, 1.43)\end{tabular} &
		\begin{tabular}[c]{@{}c@{}}1.30\\(1.26, 1.33)\end{tabular} &
		\begin{tabular}[c]{@{}c@{}}1.23\\(1.19, 1.26)\end{tabular} &
		\begin{tabular}[c]{@{}c@{}}1.16\\(1.11, 1.18)\end{tabular} \\
		   & Strong &
		\begin{tabular}[c]{@{}c@{}}6.13\\(5.81, 6.48)\end{tabular} &
		\begin{tabular}[c]{@{}c@{}}5.10\\(4.84, 5.30)\end{tabular} &
		\begin{tabular}[c]{@{}c@{}}4.77\\(4.55, 4.92)\end{tabular} &
		\begin{tabular}[c]{@{}c@{}}4.51\\(4.24, 4.69)\end{tabular} &
		\begin{tabular}[c]{@{}c@{}}3.32\\(3.28, 3.40)\end{tabular} \\
		\hline
		72 & Weak &
		\begin{tabular}[c]{@{}c@{}}1.36\\(1.32, 1.40)\end{tabular} &
		\begin{tabular}[c]{@{}c@{}}1.23\\(1.20, 1.26)\end{tabular} &
		\begin{tabular}[c]{@{}c@{}}1.19\\(1.16, 1.21)\end{tabular} &
		\begin{tabular}[c]{@{}c@{}}1.14\\(1.11, 1.16)\end{tabular} &
		\begin{tabular}[c]{@{}c@{}}1.09\\(1.05, 1.12)\end{tabular} \\
		   & Strong &
		\begin{tabular}[c]{@{}c@{}}4.85\\(4.66, 4.97)\end{tabular} &
		\begin{tabular}[c]{@{}c@{}}4.41\\(4.11, 4.59)\end{tabular} &
		\begin{tabular}[c]{@{}c@{}}4.25\\(4.04, 4.41)\end{tabular} &
		\begin{tabular}[c]{@{}c@{}}4.08\\(3.83, 4.28)\end{tabular} &
		\begin{tabular}[c]{@{}c@{}}3.93\\(3.75, 4.06)\end{tabular} \\
		\hline
	\end{tabular}
\end{table}

Our findings extend prior NPI research~\cite{kraemer2020effect,hellewell2020feasibility,merler2015spatiotemporal,lai2020effect} by elucidating operational dynamics at unprecedented resolution.
Unlike models assuming uniform transmission~\cite{keeling2011modeling}, our probabilistic framework accounts for variable tracing efficiency, mirroring field observations in Singapore and New Zealand~\cite{wang2021policy,cumming2022going}.
The strategy diverges from PCR-reliant approaches, resonating with budget-friendly policies in Japan and Australia before import-driven disruptions~\cite{wang2021policy,stobart2022australia}.
For policymakers, the $\bar{k}_+$--$\tau$ plot offers a dynamic tool to monitor and adjust NPIs, critical in resource-constrained settings.
In regions with higher tracing missing rates, contact tracing alone controls diseases with $R_0^c$.
These results suggest that rapid tracing infrastructure, paired with adaptive distancing, can mitigate outbreaks in early phases, reducing reliance on costly measures like mass lockdowns.
Policymakers should invest in scalable tracing systems and public compliance campaigns, particularly for culturally diverse regions, to maximize NPI efficacy while preserving societal functions.

Our probabilistic framework seeks to address certain limitations observed in previous models of non-pharmaceutical interventions.
For instance, traditional compartmental models, such as that presented in prior work~\cite{lai2020effect}, typically employ aggregated and homogeneous population dynamics.
While models incorporating daily travel networks can effectively simulate diverse outbreak and intervention scenarios across China, they may overlook the heterogeneity in contacts that characterizes real-world transmission and consequently constrain their capacity to fully represent the subtleties of interventions like contact tracing and social distancing.
In comparison, our approach is grounded in empirical contact network data, with key properties derived directly from observations, thereby minimizing reliance on parametric assumptions.
Notably, the mean effective contact rate $\bar{k}_+$ serves as a dynamic indicator of contact intensity that adjusts according to intervention scenarios, allowing non-pharmaceutical interventions---such as tracing speed $\tau$ and social distancing---to be integrated at the core of the model.
This facilitates a more comprehensive assessment of their combined impact on epidemic trajectories.

Several limitations temper our conclusions.
The model assumes uniform epidemiological parameters (e.g., latent periods, transmission rates) across populations, potentially overestimating containment feasibility in regions with weaker health systems.
Shenzhen's 18.6\% tracing missing rate, supported by advanced surveillance, is lower than in settings like Hong Kong (73\%), limiting generalizability.
The exclusion of imported cases, which overwhelmed policies in Singapore and Australia~\cite{wang2021policy,stobart2022australia}, underestimates challenges in open economies.
Reliance on Shenzhen's data infrastructure may not translate to low-resource settings, where tracing delays or incomplete data could inflate $\tau$ beyond $\tau_c$.
These omissions imply that our $R_0$ thresholds are optimistic; regions with higher missing rates or logistical constraints may require supplementary interventions, such as targeted testing or partial lockdowns.
Practically, implementing the epidemic control demands robust coordination, real-time data systems, and public adherence, which are challenging in politically fragmented or low-trust environments.
Cultural and lifestyle variations further complicate uniform NPI execution, as seen in the UK and US~\cite{keeling2020efficacy,lash2021covid}.
Future models should incorporate import dynamics, heterogeneous compliance, and socioeconomic trade-offs to enhance global applicability.
Sensitivity analyses could quantify the impact of missing rates and delays, refining $R_0^c$ estimates for diverse contexts.

As highlighted by Fraser et al.~\cite{fraser2004factors}, the fraction of asymptomatic or presymptomatic infections is a key determinant of outbreak controllability, particularly for interventions like contact tracing, as these cases can propagate undetected transmission chains.
Our results rely on the efficacy of contact tracing measures, where unidentified events elevate the effective missing rate ($q$) and lower critical basic reproduction numbers ($R_0^c$).
Our sensitivity analyses (table~\ref{tab:assumptionC}) illustrate this: under weak social distancing and $\tau = 11$ hours, contact tracing alone controls $R_0$ below $\sim 1.92$ (95\% CI 1.84--1.99) at $q = 40\%$, $\sim 1.65$ (1.61--1.69) at 50\%, $\sim 1.43$ (1.39--1.47) at 60\%, and $\sim 1.28$ (1.25--1.31) at 70\%.
These results highlight the need for complementary measures (e.g., ventilation, masking) in indirect transmission settings.
Findings remain robust for close-contact respiratory outbreaks but warrant caution for the efficacy of social distancing measures.

We further clarify that, although the infectious period $H(\tau)$ and the generation interval are conceptually distinct (figure~\ref{SIfig_diagram_GI_Htau}), they are positively correlated in ways that meaningfully influence contact tracing effectiveness.
As demonstrated in prior studies~\cite{park2023inferring,harris2023time}, longer generation intervals denote the longer infectious period $H(\tau)$ and would increase the cumulative proportion of transmissions during $H(\tau)$.

We further verified the stability of our model to address the critical concern of backward bifurcation~\cite{feng2000model,siddiqui2008backward}---a phenomenon where disease persistence may occur even with $R_0 < 1$ in certain epidemiological frameworks.
Through detailed analysis, we confirm our unidirectional branching-process model does not exhibit this phenomenon, as it lacks nonlinear feedback mechanisms (e.g., reinfection, resource saturation) that induce bifurcation.

In summary, this study redefines epidemic management by decoding the interplay of NPIs and proposing the strategy for budget-friendly containment.
By linking $\tau_c$ to $R_0$ and mapping NPI dynamics, we offer a data-driven framework for proactive pandemic responses.
Despite challenges in scaling to diverse regions, our approach empowers policymakers to optimize interventions, balancing efficacy with societal needs.
The findings underscore the urgent need for global investment in tracing infrastructure and adaptive policies to prepare for future infectious threats.


\subsection*{Materials and Methods}

\subsubsection*{Dataset}
The dataset of the COVID-19 epidemic in Shenzhen between February and April 2022 includes 1{,}187 positive cases (81.4\% identified through contact tracing and 18.6\% via daily PCR testing) and 86{,}451 close contacts.
Close contacts were defined as individuals without effective protection who had close interactions with positive cases from four days before symptom onset or PCR confirmation until the quarantine of the positive cases.
The dataset records the time of symptom onset, confirmation through PCR testing, and initiation of quarantine for both positive cases and their close contacts, if applicable.
The Shenzhen CDC employed various methods, such as on-site and telephone investigations, to accurately identify close contacts of positive cases.
All data used in this study were de-identified and did not contain any personally identifiable information such as names or identification numbers.

\subsubsection*{The effective reproduction number $R$}
In a probabilistic framework, $R$ is expressed as the expected number of individuals that can further spread the disease in the contact network of a positive case, i.e., $R = \bar{k}_+ \times \Pr(\text{a contact can further} \allowbreak \text{spread the disease})$.
The average number of close contacts of a positive case is derived in section~\ref{sup:num_contact}.
The probability is derived by considering an arbitrary transmission chain from the disease transmission trajectory.
Specifically, we focus on the dynamics of two adjacent nodes, denoted as $l$ and $l+1$ (see figure~\ref{SIfig_diagram}).
Given that $l$ is infected, two conditions need to be satisfied so that the transmission chain continues:
(i) $l+1$ must be infected by $l$ before $l$ is quarantined;
(ii) $l+1$ must become infectious before it is traced and quarantined.
Let $T^l$ be the infectious period of $l$, $T_s^{l+1}$ be the duration from when $l$ becomes infectious until $l+1$ is infected, and $T_e^{l+1}$ be the latent period of $l+1$, then the two conditions can be expressed mathematically as
\begin{align}
	& T^{l} > T^{l+1}_{s}, \label{eq:condition1} \\
	(T^l + \tau) - & (T_s^{l+1} + T_e^{l+1}) > 0. \label{eq:condition2}
\end{align}

\noindent Only when the first condition is satisfied would it be meaningful to discuss the infectious period of $l+1$, which is expressed as
\begin{equation}
	T^{l+1} = \max\left\{0,\ (T^l + \tau) - (T_s^{l+1} + T_e^{l+1})\right\}.
	\label{eq:transperiod}
\end{equation}

\noindent Two scenarios when the transmission chain is interrupted are provided in section~\ref{sup:scenarios_interrupted}.
Define $f^{l+1}(t;\tau)$ as the conditional probability density function (PDF) of $T^{l+1}$ given the two conditions, i.e.,
\begin{equation}
	f^{l+1}(t;\tau) = \Pr\left(T^{l+1} = t \,\bigg|\, T^l > T_s^{l+1},\ T^{l+1} > 0\right).
	\label{eq:ffunc_prob}
\end{equation}

\noindent Assuming that $T_s^{l+1}$, $T_e^{l+1}$, and $T^l$ are independent, we express $f^{l+1}(t;\tau)$ as a recursive function of $f^l(t;\tau)$.
Then, solving by iteration where $T^l$ and $T^{l+1}$ follow the same distribution, we obtain $f(t;\tau)$ as the limit of $f^l(t;\tau)$, which typically converges in a few steps.
Furthermore, assuming an exponential distribution $f_s(t_s) = \beta e^{-\beta t_s}$ for $T_s$ and a linear approximation of $f_s(t_s) \approx \beta$ since $\beta$ is typically small ($\beta_{SZ} = 0.038$ from Shenzhen's Omicron control, see section~\ref{sup:data_beta}), we have $\Pr(\text{a contact can further spread the disease}) \approx \beta H(\tau)$ with
\begin{equation}
	\begin{aligned}
		H(\tau) & = \int_0^{\infty} \int_0^{\tau} t \, f(t;\tau) \, f_e(t_e) \, dt_e \, dt \\
		        & \quad + \int_0^{\infty} \int_{\tau}^{t + \tau} (t + \tau - t_e) \, f(t;\tau) \, f_e(t_e) \, dt_e \, dt,
	\end{aligned}
	\label{eq:eff_infect_period}
\end{equation}

\noindent where $f_e(t_e)$ is the PDF of the latent period $T_e$.
More details on the derivation of $H(\tau)$ are provided in section~\ref{sup:detail_gtau}.

\subsubsection*{Parameter uncertainty}
Parameter uncertainty is quantified through Monte Carlo sampling, enabling calculation of generational reproduction numbers $R^l$ for each chain.
The reported critical thresholds are estimated as the mean values averaged over 50{,}000 independent chains, with 95\% confidence intervals provided in section~\ref{sup:simu} and section~\ref{sup:missing_rate}.

\subsubsection*{Literature search and study selection}
Respiratory pathogens were identified by sequentially filtering the Global Burden of Disease Level 3 causes~\cite{GBD2023Results}, linked pathogen entities~\cite{sirota2025global,calderaro2022respiratory,cdcTuberculosisTB,WHO_Coronavirus,CDC_Measles_Clinical_2026,CDC_Pertussis_About,CDC_aboutDiphtheria_2026,CDC_Shingles_2026}, and, where relevant, lower-level classifications (such as variants, serotypes, and subtypes)~\cite{rojas2025mycobacterium,markov2023evolution,WHO_SARS_CoV2_Variants_2023,CDC_Flu_Viruses_2025,velusamy2020expanded,fatima2025macrolide,CDC_Hib_Clinicians_2025,nuttens2024differences,CDC_StrepA_emmTyping_2024,gern2010abcs,gaunt2010epidemiology,CDC_Adenovirus_Outbreaks_2025,CDC_HPIV_Clinical_2026,sloots2006human,verduin2002moraxella,schildgen2013human,CDC_measles_2026,guiso2001fimbrial,CDC_diphtheria_2026,breuer2010proposal,cdcMumps,world2013rubella}.
For each retained pathogen, we screened PubMed records retrieved via predefined keywords by title, abstract, and full text to identify reports containing a point estimate of $R_0$.
If multiple eligible reports existed for the same pathogen or lower-level category, we selected up to five with the highest citation counts.
Detailed procedures are provided in section~\ref{sup:pathogen_lit_search}.





\bibliography{manuscript}
\bibliographystyle{sciencemag}


\section*{Acknowledgments}
\paragraph*{Funding:}
Y.H.~was supported by the National Natural Science Foundation of China (NSFC).
\paragraph*{Author contributions:}
Y.H. designed research;
J.W., Y.Z., Z.D., J.X., Z.F., and Y.H. performed research;
C.X., X.Z., Z.D., S.P., and Z.F. contributed new reagents/analytic tools;
J.W., Y.Z., and J.Z. analyzed data;
C.X., X.Z., Z.D., and Y.H. wrote the paper.
\paragraph*{Competing interests:}
There are no competing interests to declare.
\paragraph*{Data, code and materials availability:}
We disclose the relevant data and codes included in the figures.
They are available at \url{https://github.com/SpaceInCode/critical_thresholds_NPI}.


\subsection*{Supplementary materials}
Supplementary Text (Sections~S1 to S11)\\
Figures~S1 to S37\\
Tables~S1 to S11


\newpage


\renewcommand{\thefigure}{S\arabic{figure}}
\renewcommand{\thetable}{S\arabic{table}}
\renewcommand{\theequation}{S\arabic{equation}}
\renewcommand{\thepage}{S\arabic{page}}
\renewcommand{\thesubsection}{S\arabic{subsection}}
\renewcommand{\thesubsubsection}{\thesubsection\Alph{subsubsection}}
\setcounter{secnumdepth}{3}
\setcounter{figure}{0}
\setcounter{table}{0}
\setcounter{equation}{0}
\setcounter{subsection}{0}
\setcounter{page}{1}


\begin{center}
	\section*{Supplementary Materials for\\ \scititle}

	Jinghui~Wang,
	Yutian~Zeng,
	Cong~Xu,
	Zhanwei~Du,
	Xiyun~Zhang,
	Jiarong~Xie,
	Jiu~Zhang,
	Sen~Pei,
	Zijian~Feng,
	Yanqing~Hu$^{\ast}$\\
	\small$^\ast$Corresponding author. Email: yanqing.hu.sc@qq.com
\end{center}

\subsubsection*{This PDF file includes:}
Supplementary Text (Sections~S1 to S11)\\
Figures~S1 to S37\\
Tables~S1 to S11


\newpage


\subsection{Operational critical tracing time $\tau_c$ and critical average number of effective contact $\bar{k}_{+c}$}\label{sup:univeral_R_tau}

Following $
R = \bar{k}_+ \times \Pr(\text{a contact of a positive case can spread the disease})$ 
in Materials and Methods in the main text, we have
\begin{equation}\label{eq:repronum3}
R \approx \bar{k}_+ \beta H(\tau),
\end{equation}
where $\beta$ is the probability that a positive case would infect a healthy contact per day.
Considering the case when $\tau\to\infty$, then $R_0\approx\bar{k}_+ \beta \bar{T}_{\infty}$ represents the expected number of individuals a positive case can infect when there is no contact tracing or quarantine, with $\bar{T}_{\infty}=H(\infty)$ being the expected transmissible period of an infected individual in this case. 

\begin{figure}[htbp]
    \centering
    \includegraphics{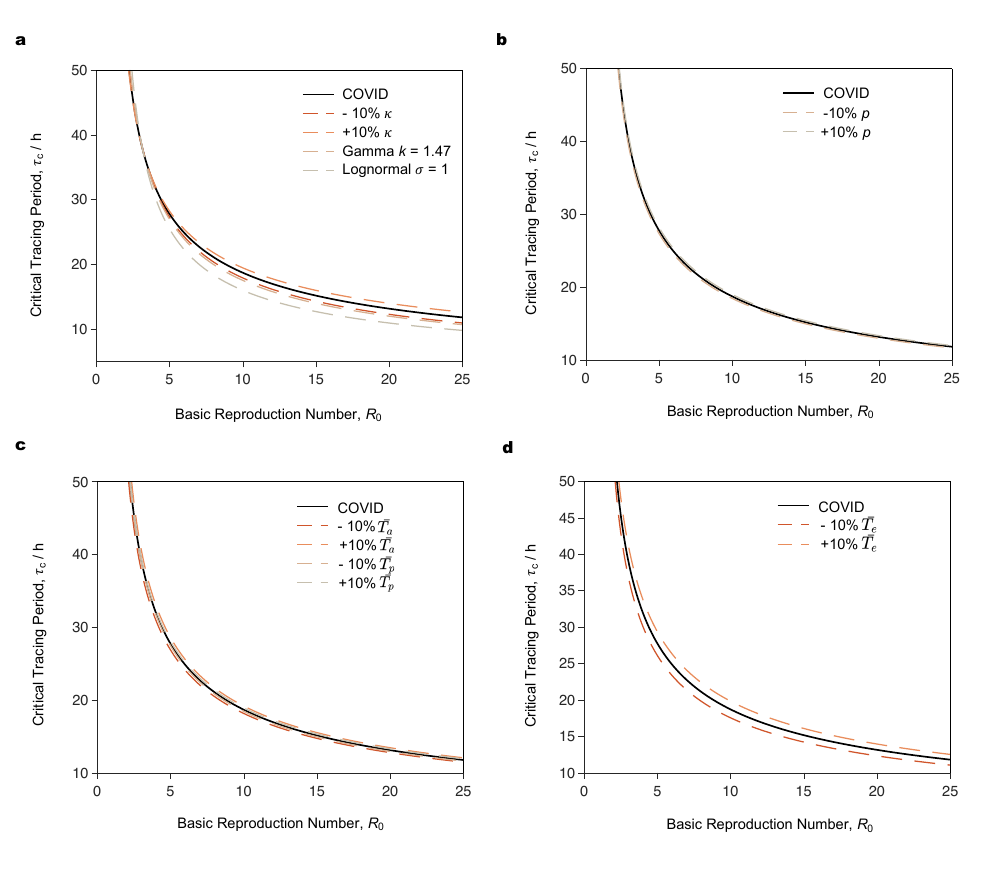}
	\caption{\textbf{The impact of model parameters on $\tau_c$.} 
 The curves are obtained based on Eq.~(8) in the main text. 
 The solid curve in each panel is exactly the olive green line in figure~2a in the main text, derived based on the parameter settings of COVID-19 (see section~\ref{sup:dist_parameter}).
 \textbf{(a)} 10\% increase or decrease in the parameter $\kappa$ of the Weibull distribution for the duration of each state, along with the cases when assuming a Gamma distribution or a lognormal distribution instead of a Weibull distribution for the duration of each state. 
 The mean duration of each state remains the same when parameter values or form of distributions are changed. 
 \textbf{(b)} 10\% increase or decrease in $p$, the probability of transitioning to the AS state, with other parameters remaining the same. 
 \textbf{(c)} 10\% increase or decrease in the mean duration of the AS/PS state with other parameters remaining the same. 
 \textbf{(d)} 10\% increase or decrease in the mean duration of the E state with other parameters remaining the same.}
	\label{SIfig_diff_dist}
\end{figure}

Plugging $R_0$ into \eqref{eq:repronum3}, $R$ can be expressed in terms of $R_0$ and $\tau$:
\begin{equation}\label{eq:repronum5}
R \approx \frac{R_0}{\bar{T}_{\infty}}H(\tau).
\end{equation}
Finally, the critical contact tracing period (also termed the critical reaction time) $\tau_c$ is obtained by solving $R=1$:
\begin{equation}\label{eq:criticalreactime}
\tau_c \approx H^{-1}\left(\frac{\bar{T}_{\infty}}{R_0}\right).
\end{equation}
We then discuss the impact of model parameters on the critical contact tracing period $\tau_c$.
figure~\ref{SIfig_diff_dist} illustrates the impact of various parameters (i.e., $p,\bar{T}_{a},\bar{T}_{p},\bar{T}_e$) and the distributions of the duration of disease transmission states on the theoretical $\tau_c$ \eqref{eq:criticalreactime}. 
As shown in figure~\ref{SIfig_diff_dist}a, when the mean duration of each state is held constant, the robustness of $\tau_c$ is evident for small $R_0$. 
However, as $R_0$ increases, the influence of the form of distribution on $\tau_c$ becomes more pronounced. 
On the other hand, $\tau_c$ is very robust to changes in the probability of transitioning to the AS state $p$, as displayed in figure~\ref{SIfig_diff_dist}b. 
An increase in $p$ slightly extends $\tau_c$, which is consistent with our intuition as the expected transmissible period is longer for positive cases of the AS state ($\bar{T}_{a}=5.64$ and $\bar{T}_{p}=2.22$ as shown in table~\ref{SItable_dist_parameter}). 
figure~\ref{SIfig_diff_dist}c-d demonstrate that $\tau_c$ is very robust to changes in the expected transmissible period but more sensitive to changes in the expected latent period.

On the other hand, for a fixed reaction time $\tau$, the critical value of $\bar{k}_+$ for a successful epidemic control, denoted by $\bar{k}_{+c}$, can also be determined by solving $R = 1$:
\begin{equation}\label{eq:kc_infty}
    \bar{k}_{+c} \approx \frac{1}{\beta H(\tau)}.
\end{equation}
This result can be used to guide the implementation of social distancing measures.

\begin{figure}[ht!]
    \centering
    \includegraphics[width=0.9\linewidth]{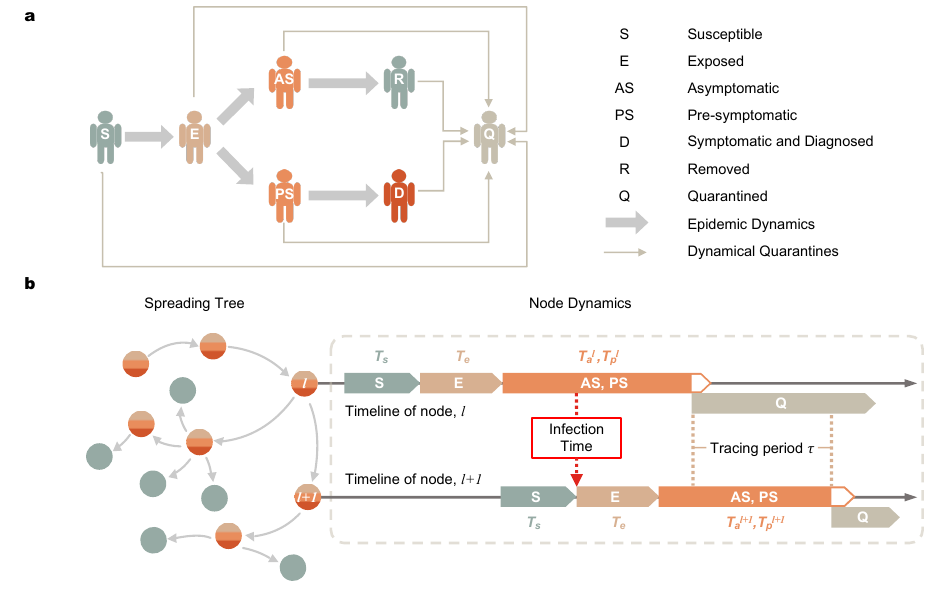}
    \caption{\textbf{Schematic diagram of disease transmission under the control policy}. 
    \textbf{(a)} The state transition process of each individual. 
    Grey arrows represent the natural state transitions, and sage green arrows show the process of quarantine.
    Susceptible individuals (S state) become infected after exposure to infectious contacts but are not yet infectious themselves (E state).
    After a latent period, these infected individuals transition to an infectious state (AS or PS state).
    Individuals in the PS state may be identified or quarantined (Q state) upon diagnosis due to symptom onset (D state) or through contact tracing.
     In contrast, individuals in the AS state never develop symptoms and are always identified and quarantined through contact tracing, either before or after recovery (R state).
    Individuals in all states may be quarantined through contact tracing.
    \textbf{(b)} The state transition dynamics on a transmission chain. 
    Muted teal and warm peach nodes in the spreading tree represent individuals who are not infected and those who are infected, respectively.
    The grey arrows represent the connections between individuals and their close contacts.
    The burnt orange dotted arrow stands for the transmission process, and the sage green dotted line indicates the process of quarantining close contacts. 
    For the transmission chain to continue, two conditions need to be met: (i) $T^l>T_s^{l+1}$ (node $l+1$ must be infected by node $l$ before node $l$ is quarantined) and (ii) $T^l+\tau>T_s^{l+1}+T_e^{l+1}$ (node $l+1$ must become infectious before it is traced and quarantined).
    } \label{SIfig_diagram}
\end{figure}

\begin{figure}[ht!]
    \centering
    \includegraphics[width=0.9\linewidth]{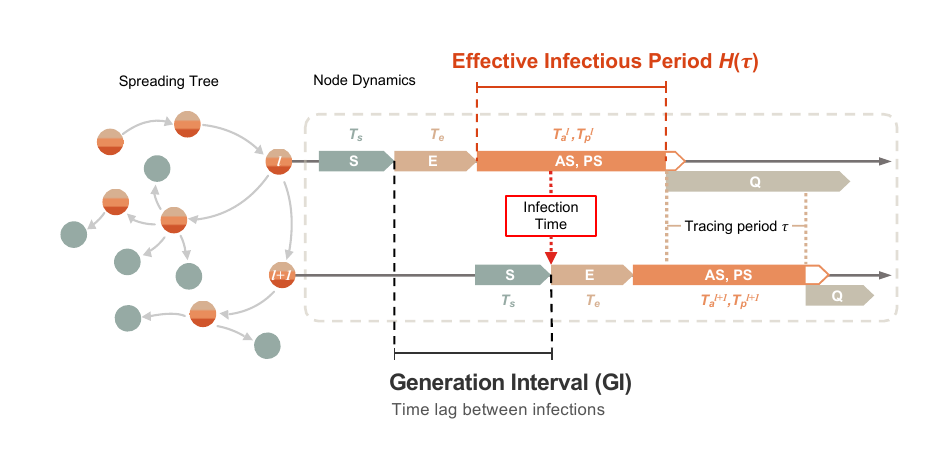}
    \caption{
\textbf{Schematic distinguishing the effective infectious period $H(\tau)$ from the generation interval (GI) in our model.} Muted teal and warm peach nodes in the spreading tree represent individuals who are not infected and those who are infected, respectively. The grey arrows represent the connections between individuals and their close contacts. The burnt orange dotted arrow stands for the transmission process, and the sage green dotted line indicates the process of quarantining close contacts.
}
    \label{SIfig_diagram_GI_Htau}
\end{figure}

The relationship between the critical contact tracing period $\tau_c$ and the basic reproduction number $R_0$ \eqref{eq:criticalreactime} is not explicit due to the involvement of the $H(\cdot)$ function.
To intuitively understand how $\tau_c$ depends on $R_0$, here we derive a simplified version of \eqref{eq:criticalreactime}. 
We focus on two arbitrary adjacent nodes on the entire tree-structured disease transmission trajectory, i.e., node $l$ and node $l+1$ (see figure~\ref{SIfig_diagram} and figure~\ref{SIfig_diagram_GI_Htau}).
Specifically, we consider a simplified situation when $T_s$, $T_e$, and the transmissible periods of all individuals in the transmission chain are constants.
Assume that the latent and transmissible periods are constants, i.e., $T_e=\bar{T}_e$ (the expected latent period) and $T^l=T^{l+1}=t_0>0$.
First, an appropriate constant value assumed for $T_s$ needs to be determined. 
In the main text, $T_s$ is assumed to follow an exponential distribution with PDF $f_s(t)=\beta e^{-\beta t}$ and $T^l>T_s^{l+1}$ is a condition required so that the transmission chain continues (i.e., Eq.~(1) in the main text).
Hence, we consider the PDF of $T_s$ conditional on $T_s<t_0$:
\begin{eqnarray}
    f_s(t|T_s<t_0) & = & \lim\limits_{dt\to 0}\frac{\Pr\left(t\leq T_s\leq t+dt|T_s<t_0\right)}{dt} = \lim\limits_{dt\to 0}\frac{\Pr\left(t\leq T_s\leq t+dt,T_s<t_0\right)}{\Pr(T_s<t_0)dt} \nonumber \\
    & = & \frac{\beta e^{-\beta t}}{1-e^{-\beta t_0}}, \forall t\in[0, t_0].
\end{eqnarray}
When $\beta\to 0$, the conditional distribution of $T_s$ given $T_s < t_0$ tends towards a uniform distribution on $[0, t_0]$ and the conditional expectation of $T_s$ given $T_s < t_0$ is then $t_0/2$.
Therefore, the constant value assumed for $T_s$ is $T_s=t_0/2$.
Since $T^{l+1} = (T^l+\tau)-(T_s^{l+1}+T_e^{l+1})$, we have $t_0 = t_0+\tau-t_0/2-\bar{T}_e$ and consequently, $t_0=2(\tau-\bar{T}_e)$.


\begin{table}[ht!]
\centering
\caption{Parameter values used for the duration distribution of different states in the diesase transmission. The duration of different states are modeled with the Weibull distributions (with PDF $f(t;\kappa, \lambda) =\frac{\kappa}{\lambda}\left( \frac{t}{\lambda} \right)^{\kappa- 1}e^{- \left( \frac{t}{\lambda} \right)^{\kappa}}, t>0$). The value of $\lambda$ is determined based on the mean duration and the value of $\kappa$.}
\begin{tabular}{@{}lccc@{}}
\hline
State  & Mean duration (days)                        &  $\kappa$                              &  $\lambda$   \\
\hline
E  & 1.2 \cite{cai2022modeling}                     & 1.47 \cite{meidan2021alternating}  &  1.33   \\
AS & 5.64 \cite{cai2022modeling}                    & 1.47 \cite{meidan2021alternating}  &  6.23   \\
PS & 2.22 \cite{cai2022modeling, wu2022incubation}  & 1.47 \cite{meidan2021alternating}  &  2.45    \\
\hline
\end{tabular}
\label{SItable_dist_parameter}
\end{table}

Then, we can derive the effective reproduction number $R$ under this simplified case. 
Let $\bar{T}_{\infty}$ be the expected transmissible period of an infected individual when there is no contact tracing or quarantine, it follows naturally that 
\begin{equation}
    R = R_0 \times \frac{t_0}{\bar{T}_{\infty}} = \frac{2R_0(\tau - \bar{T}_e)}{\bar{T}_{\infty}}.
\end{equation}
The critical contact tracing period $\tau_c$ is obtained by solving $R=1$:
\begin{equation}\label{eq:simp_tauc}
    \tau_c = \frac{\bar{T}_{\infty}}{2R_0}+\bar{T}_e.
\end{equation}
Finally, we would like to add a comment on the validity of this equation. 
The equation is actually only valid when $\bar{T}_e < \tau < \bar{T}_e + \bar{T}_{\infty} / 2$, where $\tau > \bar{T}_e$ is required for $t_0 > 0$ and $\tau < \bar{T}_e + \bar{T}_{\infty} / 2$ is required for $t_0 < \bar{T}_{\infty}$ (since the transmissible period is expected to be shorter with contact tracing compared to scenarios without contact tracing). Note here that this simplified version of the $\tau_c$, which is still usually quite different from the $\tau_c$ calculated using empirical distributions \eqref{eq:criticalreactime}, should be used with caution.

\subsection{Determination of model parameters}

\subsubsection{Distribution of disease transmission state duration}\label{sup:dist_parameter}

Parameter values used for simulations and model construction are obtained from analyses of historical data in existing literature. 
The pooled percentage of asymptomatic infections is 32.40\% \cite{shang2022percentage}. 
Therefore, individuals in the E state would transition to the AS state with probability $p = 0.324$, or to the PS state with probability $1-p=0.676$. 
We assume the Weibull distribution, with probability density function (PDF) $f(t;\kappa, \lambda) =\frac{\kappa}{\lambda}\left( \frac{t}{\lambda} \right)^{\kappa- 1}e^{- {\left( \frac{t}{\lambda} \right)}^{\kappa}} (t>0)$, for the latent period (duration of the E state) and the transmissible period (duration of the AS or PS state) \cite{meidan2021alternating}.
Parameter values used for $\kappa$ and $\lambda$ are summarized in table~\ref{SItable_dist_parameter}.
The mean duration of the E state and AS state, i.e., $\bar{T}_e = 1.2$ days and $\bar{T}_{a} = 5.64$ days, are obtained from a study of the Omicron variant in Shanghai \cite{cai2022modeling}.
Based on a study of 57 cases of the Omicron BA.1 variant, the mean incubation period—the duration from infection to symptom onset—for symptomatic individuals is 3.42 days \cite{wu2022incubation}.
Therefore, we derive the mean transmissible period of symptomatic individuals, i.e., the mean duration of the PS state $\bar{T}_{p}$, as the difference between the incubation period and the latent period, which gives $3.42 - 1.2 = 2.22$ days.
We use the same set of parameters for different variants.

For three other typical epidemics, i.e., SARS (severe acute respiratory syndrome), H1N1 (influenza A virus subtype H1N1 in 2009), and H3N2 (influenza A subtype H3N2), the probability of transitioning from the E state to the AS state is assumed to be $p=0$ for SARS \cite{wu2021unique}, $p=0.398$  for H1N1 \cite{tan2013modeling}, and $p=0$ for H3N2 \cite{loeb2012longitudinal}
, suggesting that individuals infected with SARS and H3N2 would always exhibit symptoms. Moreover, we follow existing literature to assume the following distributions for the duration of the E and AS/PS states: Gamma distribution (with PDF $f(t;a,b)=\frac{b^a}{\Gamma(a)}t^{a-1}\exp(-b t),t>0$) for SARS\cite{riley2003transmission}, the log-normal distribution (with PDF $f(t;\mu,\sigma) =\frac{1}{t\sigma \sqrt{2\pi}}\exp\left[-\frac{(\ln t-\mu)^2}{2\sigma^2}\right], t>0$) for H1N1\cite{tuite2010estimated}, and the exponential distribution (with PDF $f(t;\lambda)=\frac{1}{\lambda} \exp(-t/\lambda), t>0$) for H3N2 \cite{iwanaga2018analysis}. 

The parameter values for the duration distribution of different states are provided in table~\ref{SItable_dist_parameter_SARS} for SARS, table~\ref{SItable_dist_parameter_H1N1} for H1N1, and table~\ref{SItable_dist_parameter_flu} for H3N2, respectively. Note that for H1N1, the mean duration of the E state and AS state, i.e., $\bar{T}_e  = 2.62$ days and $\bar{T}_a  = 3.38$ days, are obtained from a study of the H1N1 epidemic in Hong Kong \cite{tuite2010estimated}.
The study also reports a mean incubation period of 4.3 days, leading to a mean transmissible period of $4.3-2.62=1.68$ days for symptomatic individuals calculated as 
the difference between the incubation period and the latent period.

\begin{table}[ht!]
\centering
\caption{ Parameter values used for the duration distribution of different states in the transmission of SARS. The duration of different states are modeled with the Gamma distributions (with PDF $f(t;a,b)=\frac{b^a}{\Gamma(a)}t^{a-1}\exp(-b t),t>0$). The value of $b$ is determined based on the mean duration and the value of $a$.}
\begin{tabular}{@{}lccc@{}}
\hline
State  & Mean duration (days)               &  $a$                                  &  $b$   \\
\hline
E  & 5.36 \cite{riley2003transmission}  &   2 \cite{riley2003transmission}  &  0.373   \\
PS & 1 \cite{riley2003transmission}     &   1 \cite{riley2003transmission}  &   1    \\
\hline
\end{tabular}
\label{SItable_dist_parameter_SARS}
\end{table}

\begin{table}[ht!]
\centering
\caption{ Parameter values used for the duration distribution of different states in the transmission of H1N1. The duration of different states are modeled with the log-normal distributions (with PDF $f(t;\mu,\sigma) =\frac{1}{t\sigma \sqrt{2\pi}}\exp\left[-\frac{(\ln t-\mu)^2}{2\sigma^2}\right], t>0$). The value of $\mu$ is determined based on the mean duration and the value of $\sigma^2$.
}
\begin{tabular}{@{}lccc@{}}
\hline
State  & Mean duration (days)             &  $\mu$    &  $\sigma^2$  \\
\hline
E  & 2.62 \cite{tuite2010estimated}   &   0.891   &    0.145 \cite{tuite2010estimated}   \\
AS & 3.38 \cite{tuite2010estimated}   &   1.15    &    0.145 \cite{tuite2010estimated}   \\
PS & 1.68 \cite{tuite2010estimated}   &   0.447   &    0.145 \cite{tuite2010estimated}   \\
\hline
\end{tabular}
\label{SItable_dist_parameter_H1N1}
\end{table}

\begin{table}[ht!]
\centering
\caption{ Parameter values used for the duration distribution of different states in the transmission of H3N2. The duration of different states are modeled with the exponential distributions (with PDF $f(t;\lambda)=\frac{1}{\lambda} \exp(-t/\lambda), t>0$). The value of $\lambda$ is determined based on the mean duration.}
\begin{tabular}{@{}lccc@{}}
\hline
State  & Mean duration (days)           &  $\lambda$  \\
\hline
E  & 1 \cite{iwanaga2018analysis}   &   1     \\
PS & 2 \cite{iwanaga2018analysis}   &   2     \\
\hline
\end{tabular}
\label{SItable_dist_parameter_flu}
\end{table}

\subsubsection{Basic reproduction number $R_0$}\label{sup:R0_beta_parameter}
Estimating the reproduction numbers (including both the basic and effective reproduction numbers) 
is not a straightforward task. 
Different studies have estimated the reproduction numbers for different variants based on data from various regions, resulting in no consistent results. 
Therefore, this section provides a detailed explanation of how we derived the basic reproduction numbers 
used in our study (as presented in table~1 and figure~2a of the main text) based on multiple research outcomes.
Nevertheless, the precision in estimating the basic reproduction number does not affect our major theoretical conclusion.

Our derivation involves finding the ratios of the effective reproduction number between different 
variants from existing literature, and then multiplying these ratios by the estimated basic reproduction number of the original strain to obtain the basic reproduction number estimates for different 
variants.
Based on data from 64 countries, the effective reproduction numbers of the Alpha and Delta variants, specifically the Alpha B.1.1.7 and Delta B.1.617.2 variants, are 1.29 (95\% CI: 1.24--1.33) and 1.97 (95\% CI: 1.76--2.17) times that of the original 
strain, respectively \cite{campbell2021increased}.
For the Omicron BA.1 and BA.2 variants, the effective reproduction numbers, averaged over multiple studies, are 2.08 (95\% CI: 1.57–2.58) and 3.31 (95\% CI: 2.83–3.79) times that of the Delta variant, respectively \cite{wang2022reproduction}. 
Since the estimated basic reproduction number for the original 
strain ranges between 2 and 3, with a median of 2.79 \cite{liu2020reproductive}, we estimate the $R_0$ of the Alpha variant with $1.29 * 2.79 = 3.63$ (95\% CI: 3.46--3.71) and the Delta variant with $1.97 * 2.79=5.50$ (95\% CI: 4.91--6.05).
Subsequently, the estimated $R_0$ for the Omicron BA.1 and BA.2 variants are $2.08 * 5.50 = 11.44$ (95\% CI: 8.49--14.39) and $3.31 * 5.50 = 18.21$ (95\% CI: 15.06--21.35), respectively.

The basic reproduction numbers for three other typical epidemics (as presented in table~1 and figure~2b of the main text) are obtained directly from existing literature.
Based on data from the SARS epidemic in Hong Kong in 2003, the $R_0$ for SARS is estimated to be 2.7 (95\% CI: 2.2--3.7) \cite{riley2003transmission}.
The estimated $R_0$ of H1N1 is obtained as 1.525 (95\% CI: 1.448--1.602) by fitting to data from the H1N1 pandemic in Guangdong province in 2009 \cite{tan2013modeling}.
For H3N2, a model-fitting result suggests a median $R_0$ of 1.41 with a 95\% confidence interval of (0.92, 2.19).

\subsection{Dataset on Shenzhen's Omicron epidemic control}\label{sup:dataset_spread_chain}


The dataset utilized in this study comprises detailed epidemiological investigations of positive cases and their close contacts during the COVID-19 epidemic in Shenzhen from February to April 2022. The relevant survey data on quarantining positive cases, tracing and quarantining their close contacts support the dynamic analysis of the prevention and control policy. The dataset includes a total of 1,187 positive cases and 86,451 close contacts with negative PCR (polymerase chain reaction) testing results. 
Positive cases in the dataset were found in two ways: 966 cases (81.4\%) were discovered through contact tracing, and 221 cases (18.6\%) were detected through large-scale screening with PCR testing.

In the dataset, 68.3\% of the positive cases have clearly documented sources of infection.
The first reported positive case in this epidemic developed a sore throat on February 11, 2022, and was confirmed positive the next day through PCR testing. 
This infected individual is a driver handling cross-border cargo, with 665 close contacts, while the average number of close contacts for all positive cases in the dataset is 83.27.
The method of tracing the source of infection includes comprehensively investigating the historical travel trajectory, determining the transmission relationships, and inferring the transmission chains based on the epidemiological connections between cases, exposure and onset times, virus genetic sequencing,  PCR testing for viral load, etc. \cite{control_manual_omicron}
Specifically, the Center for Disease Control and Prevention (CDC) of Shenzhen employed questionnaires, phone interviews, and big data tracing techniques to precisely investigate the contacts of positive cases.
Close contacts are individuals who had close contact (including sharing meals, living together, working together, traveling together, having medical appointments together, or in close proximity at the same time and place \cite{control_manual_omicron}) with positive cases without effective protection, starting from four days before the onset of symptoms or confirmation until the quarantine of positive cases. 

The investigations of positive cases and their contacts include the time of symptom onset (i.e., the first appearance of 
symptoms), the time of 
confirmation (by PCR testing), and starting time of quarantine for the positive cases, as well as the close contacts of positive cases, the time of last contact between positive cases and their close contacts, and the starting time of quarantine for the close contacts. The temporal resolution for the confirmation of cases and the onset of symptoms is measured in days, whereas that for the initiation of quarantine is measured in seconds. 
figure~1a of the main text illustrates real contact trajectories of positive cases. From our dataset, these trajectories span a period from February 14 to March 30 (45 days in total). The olive green, warm peach, and burnt orange edges represent the first, second, and third 15-day intervals, respectively.
figure~1b of the main text illustrates a real transmission chain from our dataset, spanning from February 20 to March 2. The figure shows the process of virus transmission and the subsequent interruption of the transmission chain as a result of the test-trace-quarantine intervention. 


\subsection{Control policy implementation in Shenzhen}\label{sup:real_control}

We now provide details on how Shenzhen implemented the control policy (figure~1c in the main text). The implementation of the policy is divided into three steps: (1) when a suspected case is identified, it is immediately isolated and reported to the CDC; epidemiological investigators are required to promptly arrive at the location of the suspected case upon receiving the epidemic report, including the place of discovery, residence, workplace, etc.; (2) the investigators must quickly carry out epidemiological investigations to identify close contacts and secondary close contacts of the suspected cases; meanwhile, the CDC would conduct a PCR testing and re-testing on the suspected case to confirm the infection status; (3) after completion of the epidemiological investigations and positive case confirmation, the close contacts are transported to the quarantine centers. 

To effectively control the spread of the epidemic, it is imperative that the three steps be completed as swiftly as possible. Prior to Shenzhen's Omicron outbreak in 2022, the COVID-19 Prevention and Control Manual of China (8th edition) for managing the Delta variant mandated that the third step—transporting and quarantining close contacts—be completed within 12 hours \cite{control_manual_delta}. However, the manual did not specify the time frame for completing the first two steps. After successfully containing the Omicron epidemic, Shenzhen issued the COVID-19 Prevention and Control Handbook of Shenzhen, stipulating that the investigators must arrive at the location within 1 hour and the epidemiological investigations and positive case confirmation should finish within 4 hours \cite{control_manual_omicron}. Moreover, the time requirement for transporting and quarantining close contacts was reduced to 6 hours in the updated handbook issued by Shenzhen \cite{control_manual_omicron}.

Therefore, based on the details of the control measures above, the required reaction time from identifying an infected individual to transporting and quarantining his or her close contacts is $1+4+12=17$ hours for the Delta variant. For the Omicron variant, this time is reduced to $1+4+6=11$ hours.

\subsection{Average number of effective contacts $\bar{k}_+$} \label{sup:num_contact}

Here we provide the details on how $\bar{k}_+$, the average number of effective contacts of positive cases in the contact network, is calculated.
Denote the degree distribution (i.e., the probability of having $k$ contacts) of the positive cases and all individuals in a population by $q_k$ and $p_k$, respectively. 
It is natural to assume that the probability of a node being infected in a contact network is proportional to the degree of that node, i.e., $P(\text{A node is infected} | \text{Node degree} = k) = ck$ ($c$ is a constant). Then, by the Bayes' rule and the law of total probability:
\begin{equation}
\begin{split}
q_k & = P(\text{Node degree} = k | \text{The node is infected}) \\
& = \frac{P(\text{A node is infected} | \text{Node degree} = k)\times P( \text{Node degree} = k)}{\sum_k P(\text{A node is infected} | \text{Node degree} = k)\times P( \text{Node degree} = k)} \\
& = \frac{ck\times p_k}{\sum_k ck\times p_k} = \frac{ kp_{k}}{\sum_k kp_k}
\end{split}
\end{equation}
Since a positive case with degree $k$ can potentially infect $k-1$ people (excluding the individual who infected him/her), the average number of effective contacts defined in the main text is given by  
\begin{equation}\label{eq:num_eff_contact}
 \bar{k}_{+}=\sum_k (k-1) q_k =\frac{\sum_k k (k - 1) p_k }{\sum_k k p_k}.
\end{equation}

\subsection{Quantities from Shenzhen's Omicron epidemic control}\label{sup:all_quantities}
In this section, we provide the details on how some quantities are obtained from Shenzhen's Omicron epidemic control data. These quantities are reported in table~1 and figure~2 of the main text.

\subsubsection{Average transmission rate $\beta_{SZ}$}\label{sup:data_beta}
The average transmission rate $\beta_{SZ}$ is used as an estimate of the probability that a positive case would infect a healthy contact per day.
It is obtained by taking the average of the transmission rate of 221 positive cases detected through large-scale screening via PCR testing rather than contact tracing, better reflecting the situation in a natural state.
For a positive case $i$, four quantities are required to compute its transmission rate $\beta_i$: $M_i$ - the total number of close contacts of case $i$; $m_i$ - the number of positive cases that were infected by case $i$; $T^{\text{first}}_i$ and $T^{\text{last}}_i$ - the first and last date of infection among the people infected by case $i$, respectively.
The estimated transmissible period for case $i$ is naturally $T_{i} = T^{\text{last}}_i - T^{\text{first}}_i$ and the corresponding transmission rate is computed as $\beta_{i} = \frac{m_{i}/T_{i}}{M_{i}}$. 
Leveraging on the real data of Shenzhen's Omicron control in 2022, we get $\beta_{SZ}=0.0379$ with a margin of error (corresponding to the 95\% confidence interval) of $0.015$ obtained by applying the bootstrapping method (let $\hat{\sigma}$ denote the standard error in estimating $\beta_{SZ}$, then the margin of error is $1.96\times\hat{\sigma}$).

\subsubsection{Average number of effective contacts $\bar{k}_{+SZ}$} \label{sup:data_k}

How the average number of effective contacts of positive cases in a contact network is calculated has been provided in \eqref{eq:num_eff_contact} of section~\ref{sup:num_contact}, which requires knowing the degree distribution of the population in the contact network.
Here we provide the resulting average number of effective contacts of positive cases obtained directly from the real data of Shenzhen's Omicron control in 2022, $\bar{k}_{+SZ} = 177.64$ (95\% CI 155.40–199.88), which is shown as the horizontal solid line in figure~3d of the main text. 
It is obtained by averaging the number of close contacts of the 219 positive cases detected through large-scale PCR testing (2 of the 221 positive cases, which reported zero close contacts, are excluded due to data acquisition errors), and then subtracting one (to exclude the contact who infected each positive case).
While the average number of close contacts for all positive cases in the data is 83.27, here we consider only the cases detected by PCR testing because they better reflect the behavior of infected individuals in the absence of contact tracing.

\subsubsection{Basic reproduction number $R^{SZ}_0$}\label{sup:data_R0}
In section~\ref{sup:univeral_R_tau}, we have a formula $R_0\approx \bar{k}_+\beta \bar{T}_{\infty}$, which is an approximation when $\beta$ is small. Here, we would provide a more accurate formula for calculating $R_0$ without assuming $\beta$ is small. 
By definition, the basic reproduction number $R_0$ is the expected number of individuals who are infected by a positive node in the contact network during his/her entire transmissible period when there is no contact tracing or quarantine.
Therefore, $R_0=\bar{k}_+\Pr(\text{a contact of a positive case is infected})$.
By referring to figure~\ref{SIfig_diagram} and considering the case when there is no contact tracing or quarantine, node $l+1$ would be infected by node $l$ if $T_s^{l+1}\leq T^{l}$ (recall that $T^l$ is the transmissible period of node $l$, which is a generic notation of $T_a^l$ for the AS state and $T_p^l$ for the PS state). Let $f_{ap}(t)=pf_a(t)+(1-p)f_p(t)$ be the probability density function (PDF) of $T^l$, where $p$ is the probability that an individual in the E state would transition to the AS state and $f_a(t), f_p(t)$ are the PDFs of $T_a^l$ and $T_p^l$, respectively. Assuming that $T_s^{l+1}$ and $T^l$ are independent, and the PDF of $T_s^{l+1}$ is $f_s(t)=\beta e^{-\beta t}$, then:
\begin{eqnarray}\label{eq:R0}
   R_0 &=&\bar{k}_+  \Pr(T_s^{l+1}\leq T^l) = \bar{k}_+ \int_0^\infty \Pr(T_s^{l+1}\leq t|T^l=t)f_{ap}(t) dt \nonumber\\
   & = &\bar{k}_+\int_0^{\infty} (1 - e^{- \beta t}) f_{ap} (t) d t.   
\end{eqnarray}
Given that the transmission rate $\beta_{SZ}=0.0379$ (see section~\ref{sup:data_beta}) and the average number of effective contacts $\bar{k}_{+SZ}=177.64$ (see section~\ref{sup:data_k}), we only need to determine $f_{ap}(t)$ to calculate $R_0^{SZ}$.
However, our real data on Shenzhen's epidemic control is insufficient for the determination of $f_{ap}(t)$, so we seek help from the existing literature.
In the absence of contact tracing or quarantine, $T_a^l$ and $T_p^l$ are assumed to follow the Weibull distribution with different parameters \cite{meidan2021alternating} and $p=0.324$ \cite{shang2022percentage} (see section~\ref{sup:dist_parameter}) .
Plugging these into \eqref{eq:R0}, we obtain $R_0^{SZ}=20.17$ as shown in table~1 in the main text.
The 95\% confidence interval of $R_0^{SZ}$ is calculated accordingly by considering the margin of error of 0.015 when estimating $\beta_{SZ}$ (see section~\ref{sup:data_beta}).

\subsubsection{Time-varying contact tracing period $\tau_t$ and number of effective contacts of the positive cases $\bar{k}_{+t}$}\label{sup:data_tau}

In figure~2c of the main text, the contact tracing period and the number of effective contacts of the positive cases on different dates are displayed. 
Here we provide the details on how these time-varying values are calculated based on all 1,187 positive cases confirmed between February 12 and April 17, 2022.

The contact tracing period $\tau_t$ on a specific day $t$ is determined based on all positive cases confirmed on that day and their close contacts.
For a positive case $i$ confirmed on day $t$ and one of his/her close contacts $j$, two quantities are used in the calculation: (1) $t_i^I$, the time when individual $i$ was identified, and (2) $t_j^Q$, the time when close contact $j$ is quarantined.
Then the contact tracing period based on the tracing path $(i,j)$ is naturally $\tilde{\tau}_{ij}= t_j^Q-t_i^I$.
However, not all $t_i^I$ and $t_j^Q$ are accurately recorded in our data.
Only 954 such $(i,j)$ tracing paths are available to calculate the corresponding tracing periods $\tilde{\tau}_{ij}$. 
To obtain stable estimates of $\tau_t$, we perform the calculation by taking the average of each tracing period through a 7-day sliding window method, i.e., $\tau_t$ is the average of all $\tilde{\tau}_{ij}$'s where case $i$ is confirmed within 3 days before and after day $t$ (from day $t-3$ to day $t+3)$. 
The 95\% confidence interval of $\tau_t$ is again obtained using the bootstrap method.

\begin{figure}[ht!]
	\centering
	\includegraphics[width=0.7\linewidth]{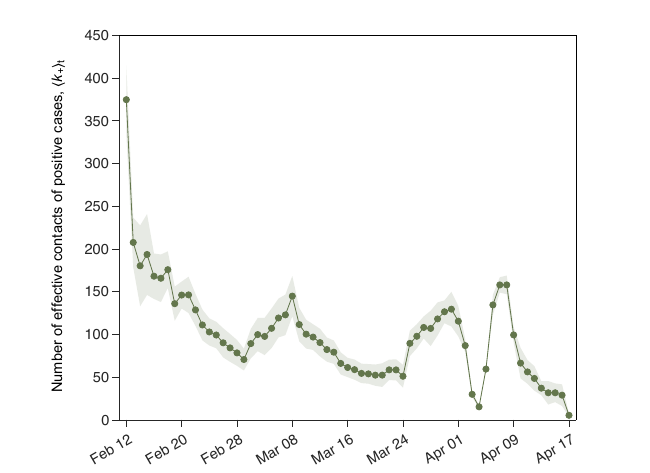}
	\caption{The daily average number of effective contacts of all positive cases.}
	\label{SIfig_Ntau_vs_date}
\end{figure}

Likewise, the number of effective contacts of the positive cases $\bar{k}_{+t}$ on day $t$ is calculated by averaging over the number of close contacts of all positive cases confirmed from day $t-3$ to day $t+3$ and then subtracting one to exclude the individual who infected them.
Figure~\ref{SIfig_Ntau_vs_date} shows the resulting $\bar{k}_{+t}$ and the corresponding 95\% confidence intervals obtained by the bootstrap method.
Notice that almost all $\bar{k}_{+t}$ are less than the average number of effective contacts of positive cases detected by large-scale screening $\bar{k}_{+SZ}=177.64$ (see section~\ref{sup:data_k}).
This is consistent with the fact that social distancing measures would reduce the average number of effective contacts \cite{kraemer2020effect,anderson2020will}.

In figure~2c of the main text, February 16th to April 2rd are selected to depict the trajectory of Shenzhen's epidemic control, with $\tau_t$ on the $x$-axis and $\bar{k}_+$ on the $y$-axis.

\subsubsection{Daily effective reproduction number $R_t$ and number of confirmed cases}\label{sup:data_Rt}

In figure~2d of the main text, the effective reproduction number and the number of newly confirmed cases on different dates are displayed. These values are calculated based on all 1,187 positive cases.

The effective reproduction number is the average number of secondary cases infected by a positive case.
Therefore, again using the 7-day sliding window method as in section~\ref{sup:data_tau}, the effective reproduction number $R_t$ on day $t$ is calculated by averaging over the number of secondary cases infected by a positive case who is confirmed from day $t-3$ to day $t+3$.
The 95\% confidence intervals are again obtained by using the bootstrap method.
Similarly, the number of confirmed cases on day $t$ is calculated by averaging over the 7 daily numbers of confirmed cases from day $t-3$ to day $t+3$.

\subsection{Scenarios when the transmission chain is interrupted}\label{sup:scenarios_interrupted}
\begin{figure}[ht!]
    \centering
    \includegraphics{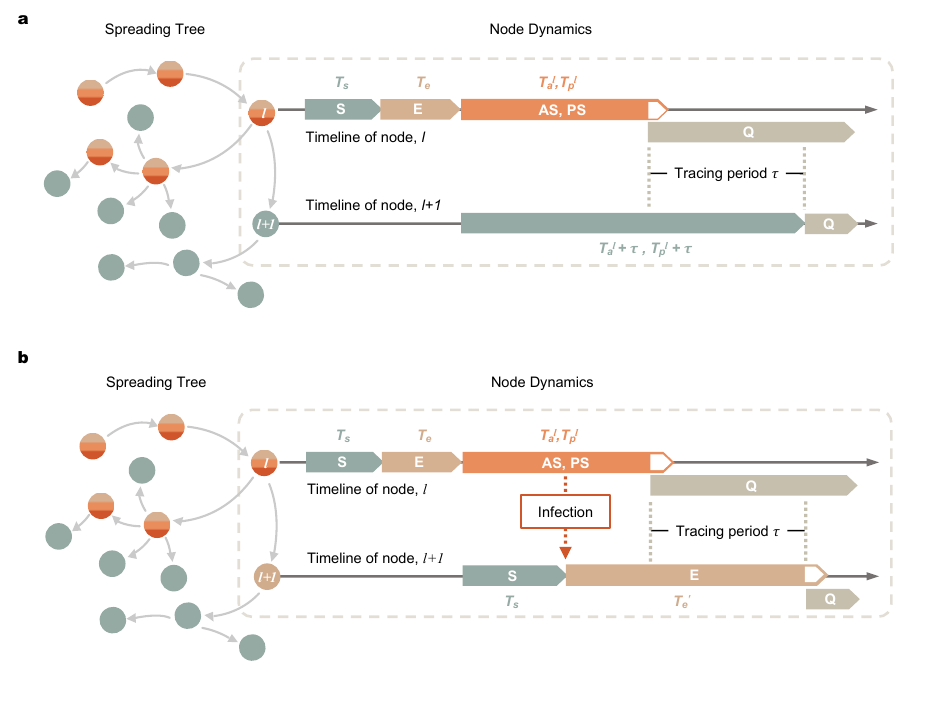}
    \caption{
    \textbf{Two scenarios when the transmission chain is successfully interrupted.} The warm peach and muted teal nodes on the spreading trees represent infected individuals and healthy individuals, respectively. \textbf{(a)} Node $l+1$ is not infected during the entire transmissible period ($T^l_a$ or $T^l_p$) of node $l$, but is quarantined as the close contact of node $l$. \textbf{(b)} Node $l+1$ is infected by node $l$ but would be quarantined as the close contact of node $l$ before becoming infectious. $T'_e$ is the duration from being infected to being quarantined.}
	\label{SIfig_succuss_control}
\end{figure}

In this section, we explain why the transmission chain would be interrupted if the two conditions by Eq.~(1) and (2) in the main text are not met, as illustrated by figure~\ref{SIfig_succuss_control}.
In figure~\ref{SIfig_succuss_control}a, Eq.~(2) in the main text is not met, i.e., $T^l \leq T_s^{l+1}$, suggesting that node $(l+1)$ remains in the S state throughout the entire transmissible period of node $l$. 
This means that node $(l+1)$ is not infected by node $l$ but would be quarantined as the closed contact of node $l$, thereby terminating the transmission chain.
In figure~\ref{SIfig_succuss_control}b, Eq.~(1) in the main text is satisfied but Eq.~(2) is not, i.e., $T^l+\tau \leq T_s^{l+1} + T_e^{l+1}$. 
This indicates that node $(l+1)$ transits to the E state due to infection by node $l$, however, it is quarantined as the close contact of node $l$ before becoming infectious, thereby eliminating the possibility of it further spreading the disease.
 
\subsection{Expected effective transmissible period $H(\tau)$}\label{sup:detail_gtau}


From the Materials and Methods in the main text, we have 
\begin{align}\label{methodeq:ffunc}
    f^{l+1}(t;\tau) &= \Pr\left(T^{l+1}=t \bigg| T^l>T_s^{l+1},T^{l+1}>0\right) \nonumber  \\
    &= \frac{g^{l+1}(t;\tau)}{\int_0^\infty g^{l+1} (t;\tau)dt}, 
\end{align}
where
\begin{align}\label{methodeq:gfunc}
&g^{l+1}(t;\tau) = \Pr(T^{l+1}=t, T^l>T_s^{l+1},T^{l+1}>0) \nonumber\\
    &= \int_0^\infty\left(\int_{t_s}^\infty f^l(t^\prime;\tau)f_e(t^\prime+\tau-t_s-t)dt^\prime\right)f_s(t_s)dt_s
\end{align}
is derived by assuming that $T_s^{l+1}$, $T_e^{l+1}$, and $T^l$ are independent.
Through iterations of \eqref{methodeq:ffunc} and \eqref{methodeq:gfunc}, $f(t;\tau)=\lim\limits_{l\to\infty}f^l(t;\tau)$ and $g(t;\tau)=\lim\limits_{l\to\infty}g^l(t;\tau)$ can be obtained, and by definition, $\int_0^\infty g(t;\tau)dt$ is the probability that a contact of an infected individual can spread the disease.
$T_s$ is typically modeled by an exponential distribution, i.e., $f_s(t_s)=\beta e^{-\beta t_s}$\cite{riley2003transmission}.
Moreover, $\beta_{SZ}=0.038$ (95\% CI 0.023-0.053) is obtained from the real data of Shenzhen's Omicron control (see section~\ref{sup:data_beta}), which is small enough to lead to a linear approximation of $f_s(t_s)\approx \beta$ and thus
\begin{align}\label{methodeq:gfunclim}
\int_0^\infty g(t;\tau)dt = \beta\int_0^\infty\int_0^\infty\int_{t_s}^\infty f(t^\prime;\tau)
f_e(t^\prime+\tau-t_s-t)dt^\prime dt_s dt.
\end{align}
Define $H(\tau)$ as the integration part of the right hand side of \eqref{methodeq:gfunclim}, i.e., 
\begin{equation}\label{methodeq:G}
H(\tau) = \int_0^\infty \int_0^\infty\int_{t_s}^\infty f(t^\prime;\tau)f_e(t^\prime+\tau-t_s-t) dt^\prime dt_s dt.
\end{equation}
Since $\beta$ is the probability that a positive case would infect a healthy contact per day, then $H(\tau)$ has a clear physical meaning which is the effective transmissible period of an infected individual.


Detailed derivation on the expected effective transmissible period $H(\tau)$, i.e., Eq.~(5) in Materials and methods of the main text, 
is provided here. 
From figure~\ref{SIfig_diagram}b, we have $t' + \tau = t_s + t_e + t$. So, substituting $t'+\tau-t_s-t$ by $t_e$ in \eqref{methodeq:gfunc} and letting $l\rightarrow \infty$, we have
\begin{equation}
  g(t;\tau) = \beta \int_0^{\infty} \int_0^{\infty} \int_0^{t'} f (t';\tau) f_e (t_e) \delta (t' + \tau - t_s - t_e - t) dt_s dt' dt_e,
\end{equation}
where $\delta(x)$ is the Dirac delta function, i.e., $\delta(x)=+\infty$ at $x=0$, $\delta(x)=0$ when $x\neq 0$, and $\int_{-\infty}^{+\infty} \delta(x)dx=1$.
Since $0 \leq  t_s \leq t'$ indicates $\tau - t_e \leq t \leq t' + \tau - t_e$, then
\begin{equation} \label{eq:gt}
  g(t;\tau)  = \beta \int_0^{\infty} \int_0^{\infty} f(t';\tau) f_e(t_e) \mathbf{1}_{[\tau - t_e, t' + \tau - t_e]}(t)  dt' dt_e,
\end{equation} 
where $\mathbf{1}_A(x)$ is the indicator function, i.e., $\mathbf{1}_A(x)=1$ if $x\in A$ and $\mathbf{1}_A(x)=0$ if $x\notin A$. 
By definition, $H(\tau)$ is defined as the integration of $g(t;\tau)$ omitting $\beta$, then following \eqref{eq:gt} we have 
\begin{eqnarray}\label{eq:expectation}
  H(\tau) & = & \int_0^{\infty} \int_0^{\infty} \left[ \int_0^{\infty} 
    \mathbf{1}_{[\tau - t_e , t' + \tau - t_e]}(t) dt \right]
    f(t';\tau) f_e (t_e) 
dt' d t_e \nonumber\\
  & = & \int_0^{\infty} \int_0^{\infty} \left( \int_{\max \{0, \tau - t_e \}}^{t' + \tau - t_e} 1 d t \right) f(t';\tau) f_e (t_e) dt' dt_e \nonumber\\
  & = & \int_0^{\infty} \int_0^{\infty} t_{\text{eff}} f(t';\tau) f_e (t_e)
  dt' dt_e.
\end{eqnarray}
Then it remains to show that the $H(\tau)$ expressed in \eqref{eq:expectation} can be decomposed to the two integrals as listed in Eq.~(5) of the main text.
The first integral is obtained under the case when $0\leq t_e \leq \tau$ where $t_{\textrm{eff}}=(t'+\tau-t_e)-(\tau-t_e)=t'$. 
In this case, as long as node $l+1$ is infected by node $l$ (i.e., $t_s\leq t'$), it is capable of spreading the disease (i.e., $t_s+t_e\leq t'+\tau$).
The second integral is obtained under the case when $t_e > \tau$, where $t_{\textrm{eff}}=(t' + \tau - t_e) - 0 = t' + \tau - t_e$. 
However, we also need $t' + \tau - t_e > 0$ in this case, so $\tau < t_e < t' + \tau$, this is why the integration with respect to $t_e$ in the second integral in Eq.~(5) of the main text goes from $\tau$ to $t+\tau$.

Note that in the derivation above, we do not distinguish between the AS and PS states, as a generic notation $T^l$ is used to represent both $T_a^l$ and $T_p^l$. 
In our model setup, the primary difference between positive cases of the AS and PS states lies in how they would be quarantined: AS cases are always identified and quarantined through contact tracing, whereas PS cases may also be quarantined upon diagnosis due to symptom onset.
However, assuming the contact tracing process is efficient enough, i.e., $\tau$ is relatively short, positive cases of both the AS and PS states are identified and quarantined solely through contact tracing.
This assumption is reasonable in reality. 
Our data shows that it typically takes 2 days or more for PS cases to develop symptoms after infection (see section~\ref{sup:dist_parameter}), which is significantly longer compared to the $\tau$ observed during the implementation of the control policy.

For completeness, a thorough discussion of the situation when $\tau$ is relatively long is provided.
Here, we provide the corresponding derivations without making this assumption.
Again, we focus on the dynamics of two adjacent nodes on the entire tree-structured disease transmission trajectory, i.e., node $l$ and node $l+1$, as we do in the main text (see figure~\ref{SIfig_diagram}).
The two conditions required for node $l+1$ to be capable of spreading the disease, as stated in Eq.~(1) and Eq.~(2) in the main text, also apply here. 
First of all, the conditional probability density function (PDF) of the transmissible period of node $l+1$, given that the two conditions are satisfied, needs to be determined. 
For the general situation, we should consider the AS and PS states separately.

For the case in which node $l+1$ will transition to the PS state, let $T^{l+1}_{ep}$ be the latent period (from being infected to becoming infectious) and $T^{l+1}_{dp}$ be the duration from becoming infectious to being diagnosed and quarantined due to symptom onset when there is no contact tracing. 
Assuming that an infected individual always becomes infectious before symptom onset, we have $T^{l+1}_{dp}>0$ and denote its PDF by $f_{dp}(t)$.
Following Eq.~(3) in the main text, the transmissible period of node $l+1$ is
\begin{equation}
T^{l+1}_p = \min\left\{\max\left\{0, (T^l+\tau)-(T^{l+1}_s+T^{l+1}_{ep})\right\}, T^{l+1}_{dp}\right\} = \min\left\{\tilde{T}^{l+1}_p, T^{l+1}_{dp}\right\},
\end{equation}
considering the fact that he/she may be quarantined upon diagnosis due to symptom onset before being identified through contact tracing (note that $\tilde{T}^{l+1}_p = \max\{0, (T^l+\tau)-(T^{l+1}_s+T^{l+1}_{ep})\}$ is the transmissible period used in the main text, assuming a short contact tracing period $\tau$).
The PDF of $T^{l+1}_p$ jointly with the two conditions in Eq.~(1) and Eq.~(2) in the main text is ($t>0$)
\begin{eqnarray}
g^{l+1}_p(t;\tau) & = & -\frac{d\left\{\Pr\left(T^{l+1}_p > t, T^{l}>T^{l+1}_s, T^l+\tau>T^{l+1}_s+T^{l+1}_{ep}\right)\right\}}{dt} \nonumber \\
& = & - \frac{d\left\{\Pr\left(\tilde{T}^{l+1}_p>t, T^l>T^{l+1}_s\right)\Pr\left(T^{l+1}_{dp}>t\right)\right\}}{dt} \nonumber \\
& = & f_{dp}(t)\Pr\left(\tilde{T}^{l+1}_p>t, T^l>T^{l+1}_s\right) - \Pr\left(T^{l+1}_{dp}>t\right)\frac{d\left\{\Pr\left(\tilde{T}^{l+1}_p>t, T^l>T^{l+1}_s\right)\right\}}{dt} \nonumber \\
& = & f_{dp}(t)\int_t^{\infty}\tilde{g}^{l+1}_p(u;\tau)du + \tilde{g}^{l+1}_p(t;\tau) \int_t^{\infty}f_{dp}(u)du,
\end{eqnarray}
where $\tilde{g}^{l+1}_p(t;\tau)$ is the PDF of $\tilde{T}^{l+1}_p$ jointly with the two conditions. Following \eqref{methodeq:gfunc}, we have
\begin{equation}
\tilde{g}^{l+1}_p(t;\tau) = \int_0^{\infty}\int_{t_s}^{\infty}f^l(t';\tau)f_s(t_s)f_{ep}(t'+\tau-t_s-t)dt'dt_s.
\end{equation}

Similarly, for the case in which node $l+1$ will transition to the AS state, let $T^{l+1}_{ea}$ be the latent period and $T^{l+1}_{ra}$ with PDF $f_{ra}(t)$ be the duration from becoming infectious to being recovered and no longer infectious when there is no contact tracing. Then, the transmissible period of node $l+1$ is
\begin{equation}
T^{l+1}_a = \min\left\{\max\left\{0, (T^l+\tau)-(T^{l+1}_s+T^{l+1}_{ea})\right\}, T^{l+1}_{ra}\right\} = \min\left\{\tilde{T}^{l+1}_a, T^{l+1}_{ra}\right\},
\end{equation}
considering the fact that he/she may already recover and is no longer infectious when being identified through contact tracing ($\tilde{T}^{l+1}_a = \max\{0, (T^l+\tau)-(T^{l+1}_s+T^{l+1}_{ea})\}$ is again the transmissible period assuming a short contact tracing period $\tau$).
Following the same derivation as for the PS state, the PDF of $T^{l+1}_a$ jointly with the two conditions in Eq.~(1) and Eq.~(2) in the main text is ($t>0$)
\begin{equation}
g^{l+1}_a(t;\tau)=f_{ra}(t)\int_t^{\infty}\tilde{g}^{l+1}_a(u;\tau)du + \tilde{g}^{l+1}_a(t;\tau) \int_t^{\infty} f_{ra}(u)du,
\end{equation}
where $\tilde{g}^{l+1}_a(t;\tau)$ is the PDF of $\tilde{T}^{l+1}_a$ jointly with the two conditions:
\begin{equation}
\tilde{g}^{l+1}_a(t;\tau) = \int_0^{\infty}\int_{t_s}^{\infty}f^l(t';\tau)f_s(t_s)f_{ea}(t'+\tau-t_s-t)dt'dt_s.
\end{equation}

Putting the AS and PS states together, the PDF of $T^{l+1}$ (the transmissible period of node $l+1$) jointly with the two conditions is $g^{l+1}(t;\tau)=pg^{l+1}_a(t;\tau) + (1-p)g^{l+1}_p(t;\tau)$, where $p$ is the probability of transitioning to the AS state.
The conditional PDF of $T^{l+1}$ given the two conditions is then
\begin{equation}
    f^{l+1}(t;\tau) = \frac{g^{l+1}(t;\tau)}{\int_0^\infty g^{l+1}(t;\tau)dt}.
\end{equation}
By iteration, $g(t;\tau)=\lim\limits_{l\to\infty}g^l(t)$ can be obtained, and assuming $f_s(t_s)\approx \beta$, the expected effective transmissible period $H(\tau)$ is again defined as the integration of $g(t;\tau)$ with respect to $t$ divided by $\beta$.

\subsection{Simulation studies}\label{sup:simu}

\subsubsection{Building tree-structured disease transmission trajectories} 
\begin{figure}[ht!]
	\centering
	\includegraphics[width=0.64\linewidth]{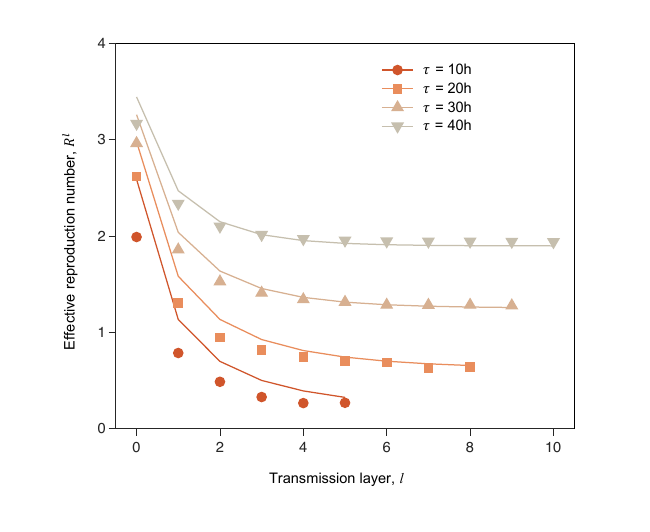}
	\caption{
 \textbf{The theoretical effective reproduction numbers and those obtained through simulation.}
 The solid curves represent the theoretical effective reproduction numbers $R^l \approx \bar{k}_+ \beta H^{l+1} (\tau)$ ($H^{l+1}(\tau)$ is derived from $g^{l+1}(t;\tau)$, see section~\ref{sup:detail_gtau}), while the points are obtained through simulation (see \eqref{eq:Rtau_sim}, averaged over 500 simulation rounds). 
 Parameters used in both the simulation study and the theoretical derivation are: $\bar{k}_+ = 177.64$ (the number of effective contacts), $\beta=0.0095$ (the transmission rate per day), $p=0.324$ (the probability of transitioning to the AS state), the Weibull distributions for the duration of the E/AS/PS states (see table~\ref{SItable_dist_parameter}).}
	\label{SIfig_Rlayer}
\end{figure}

\begin{figure}[ht!]
    \centering
    \includegraphics[width=0.5\linewidth]{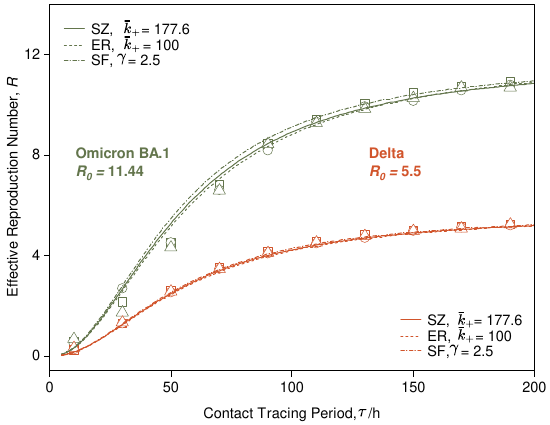}
    \caption{
  \textbf{Stabilized reproduction number after multi-layer transmission.} 
    The symbols (circles, squares and triangles) are the simulation results given three different network structures, i.e., the Shenzhen network (SZ), the ER network, and the scale-free network (SF, with degree distribution $p_k\propto k^{-\gamma}$) respectively.
  The lines are the theoretical results from Eq.~(\ref{eq:repronum3}). 
  }    \label{SIfig_R_tau}
\end{figure}

Other than all the theoretical results, computer simulation studies are also performed.
Our simulation implements a branching process model to build tree-structured disease transmission trajectories (see figure~\ref{SIfig_Rlayer} and \ref{SIfig_R_tau}), which consist of only positive cases (nodes) and transmission paths between them (edges) \cite{hellewell2020feasibility}.
We construct tree-structured transmission trajectories for the following reason. In the early transmission stage, the structure of the transmission trajectory appears in a tree-like form.
We initiate the transmission trajectory in one round of simulation with a node in the E state, which serves as the source of transmission, and the layer number of it is set to 0.  
Each time when there is a newly infected node, the transmission trajectory is extended, and the layer number of this newly infected node is set to 1 plus the layer number of the node that infected it.
The number of effective contacts of a node $k_+$ is sampled from a degree distribution, and the continuous timeline is discretized into time steps of length $\Delta t = 0.25$ hour, i.e., 15 minutes.

At each time step $t$, we find the nodes that are currently infectious (i.e., in the AS/PS state), and for each of them, we add a newly infected node to it with probability derived in the following. 
Let node $i$ be one of the nodes that are infectious at $t$ and $t_i < t$ be the time step when it became infectious (i.e., transitioned from the E state to the AS/PS state). 
Let $\beta$ be the transmission rate as defined in the main text (i.e., the probability that a positive node would infect a healthy contact per day), then the probability that node $i$ would infect a healthy contact per time step is $\tilde{\beta}=\beta\Delta t / 24$.
Consequently, the probability of adding a node infected by node $i$ at $t$ to the transmission trajectory is approximately
\begin{equation}\label{eq:infection_rate}
    k_+\tilde{\beta} (1-\tilde{\beta})^{t-t_i-1} \left[1 - \tilde{\beta} (1-\tilde{\beta})^{t-t_i-1}\right]^{\bar{k}_+ - 1} \approx k_+ \tilde{\beta} (1-\tilde{\beta})^{t-t_i-1}, 
\end{equation}
where $k_+$ is the number of ways of selecting one from all effective contacts of node $i$, $\tilde{\beta} (1-\tilde{\beta})^{t-t_i-1}$ is the probability that the selected one is infected by node $i$ at $t$ but not at any time step between $t_i$ to $t-1$, and $\left[1 - \tilde{\beta} (1-\tilde{\beta})^{t-t_i-1}\right]^{\bar{k}_+ - 1}\approx 1$ (since $\tilde{\beta}$ is very small) is the probability that the other contacts of node $i$ are not infected at $t$.
Note that we only consider adding at most one infected node to an infectious node at each time step because the probability of infecting two or more contacts within such a short time interval (0.25 hours) is negligible.

If a newly infected node, denoted as node $j$, is added at time step $t$, its time of infection is recorded as $t$, and we determine whether it would be asymptomatic by sampling from $\text{Bernoulli}(p)$ ($p$ is the probability of transitioning to the AS state).
Then, the latent period $t_{je}$ and the transmissible period without contact tracing of node $j$ are obtained by sampling from the PDFs $f_{ea}(t)$ and $f_a(t)$ if it is asymptomatic (otherwise, sample from PDFs $f_{ep}(t)$ and $f_p(t)$).
The time of quarantine of node $j$ is naturally the time of quarantine of the node that infected it plus the contact tracing period $\tau$, which is compared to $t+t_{je}$ to determine whether node $j$ can further spread the disease and if so, how long is the effective transmissible period.

The simulation would be terminated if any one of the two stopping criteria is met: (1) all nodes on the transmission trajectory are in the Q state (i.e., the quarantined state) so that the disease cannot spread further, and (2) the number of transmission layers reaches a prespecified cut-off value $l^*$.
Based on the simulated transmission trajectories, an estimation of the effective reproduction number in layer $l< l^*$, denoted by $\hat{R}^l$, can be calculated as
$\hat{R}^l$ is defined by:
\begin{align}\label{eq:Rtau_sim}
    \hat{R}^l=\frac{\text{Number of nodes in layer $l+1$ with non-zero transmissible period} }{\text{Number of nodes in layer $l$ with non-zero transmissible period} }.
\end{align}

\subsubsection{Simulation results}
First, we report the effective reproduction numbers obtained through simulation in comparison with those from our theoretical results (see figure~\ref{SIfig_Rlayer}).

The theoretical analysis in section~\ref{sup:univeral_R_tau} leads to a surprising result: $R$ and $\tau_c$ in \eqref{eq:repronum5} and \eqref{eq:criticalreactime} only depend on the epidemiological characteristics of the infectious disease (e.g., the latent period and the proportion of asymptomatic infections, etc., as reflected in $H(\cdot)$), and are mathematically irrelevant to the contact network structure of the population. 
It is also shown in figure~\ref{SIfig_R_tau} that while keeping $R_0$ fixed, $R$ increases as $\tau$ increases and the $R-\tau$ curves under networks of different structures (e.g., different degree distributions or different average degrees) almost overlap. 
It should be emphasized that this does not imply network structure lacks influence on disease transmission dynamics or containment efficacy; rather, its entire epidemiological impact is already encapsulated within $R_0$.

Specifically, the input parameters used in both the simulation study and the theoretical derivation are related to COVID-19: the PDFs of the duration of the E/AS/PS states are assumed to be Weibull distributions with parameters given in table~\ref{SItable_dist_parameter}, $\bar{k}_+ = \bar{k}_{+SZ} =177.64$, $p=0.324$, $\beta=0.0095$ (obtained based on \eqref{eq:R0} with $R_0=5.50$, which is the basic reproduction number of the Delta variant).
The results suggest that the simulation outcomes align with our theoretical results, and both the simulated and theoretical effective reproduction numbers converge when $l$ increases.

The alignment between the theoretical results and simulation outcomes is also demonstrated in figure~\ref{SIfig_R_tau}.
Two values of $R_0$ ($R_0=5.50$ for the Delta variant and $R_0=11.44$ for the Omicron BA.1 variant) and three distributions of $k_+$ are considered.
Specifically, for the ER network, $\bar{k}_+=100$ (by definition, for an ER network with average degree $\bar{k}_+$ and total number of nodes $N$, any two nodes are connected with probability $\bar{k}_+/(N-1)$, resulting in a Poisson degree distribution with mean $\bar{k}_+$ for large $N$); for the scale-free network, we assume a degree distribution $p_k\propto k^{-\gamma}$ with $\gamma=2.5$; for the Shenzhen network, $k_+$ is randomly sampled from a set of values consists of the observed number of effective close contacts of the 219 positive cases identified through large-scale PCR testing (see Sec \ref {sup:data_k}, with an average of $\bar{k}_+=177.64$).
Note that our simulation only builds tree-structured transmission trajectories instead of the entire contact network, so the three networks marked as SZ/ER/SF only differ in the distribution of $k_+$ in the simulation, and other differences in terms of the network structure are not considered.
With the value of $R_0$ and $\bar{k}_+$, the corresponding transmission rate $\beta$ can be determined using \eqref{eq:R0} (see table~\ref{SItable_r0_beta_parameters} for the case when $\bar{k}_+=177.64$) and is used as an input parameter for the simulation.

\begin{table}[ht!]
\centering
\caption{ The basic reproduction number $R_0$ of the Delta and Omicron BA.1 variant as well as the corresponding transmission rate \(\beta\) with 95 \% CI ($\bar{k}_+=\bar{k}_{+SZ}=177.64$).}
\begin{tabular}{@{}ccc@{}} \hline
Variants  & Delta   & Omicron BA.1 \\ \hline $R_0$    & 
\begin{tabular}[c]{@{}c@{}}5.50\\ (4.91,6.05)\end{tabular} & 
\begin{tabular}[c]{@{}c@{}}11.44\\ (10.21,12.58)\end{tabular}           
\\ $\beta$/day & 
\begin{tabular}[c]{@{}c@{}}0.0095\\ (0.0085,0.0105)\end{tabular} &    
\begin{tabular}[c]{@{}c@{}}0.0204\\ (0.0181, 0.0226)\end{tabular}      
\\ \hline
\end{tabular}
\label{SItable_r0_beta_parameters}
\end{table}

\begin{figure}[ht!]
	\centering
	\includegraphics[width=0.7\textwidth]{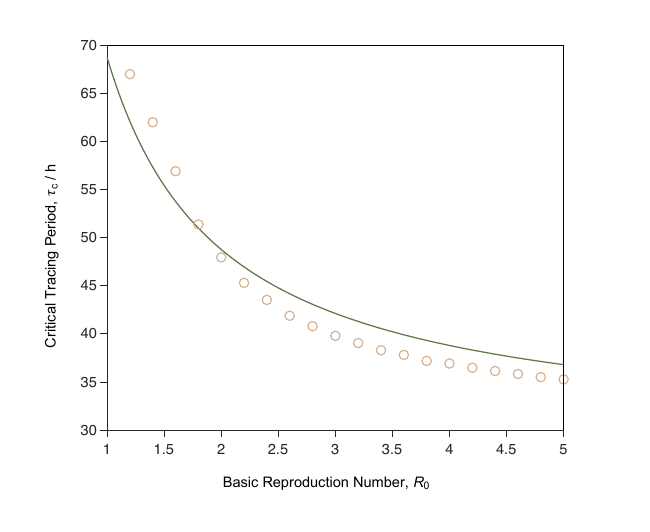}
	\caption{
 \textbf{The theoretical simplified critical contact tracing period and those obtained through simulation for the Delta variant.}
 The solid curve represents the simplified $\tau_c$ obtained from \eqref{eq:simp_tauc}, where the constant latent period is $\bar{T}_{e}=1.2$ days and the constant transmissible period is $\bar{T}_\infty=0.324*5.64+0.676*2.22=3.328$ days (see section~\ref{sup:dist_parameter}).
 The warm beige circles show the results of the simulation.
 For each value of $\beta$ (corresponding to an $R_0$), we perform a grid search on different values of $\tau$ to find the one such that the effective reproduction number $\hat{R}^l$ (averaged over 500 simulation rounds) is closest to 1.
 The transmission layer $l$ we are looking at varies with $R_0$, which is chosen when the resulting $\hat{R}^l$ at different layers are stabilized.
 The constant latent period used in the simulation is again $\bar{T}_{e}=1.2$ days, and the constant transmissible period is $\bar{T}_a = 5.64$ days for the AS state and $\bar{T}_{p} = 2.22$ days for the PS state (see section~\ref{sup:dist_parameter}).
 }
	\label{SIfig_verify}
\end{figure}

Then, we also consider the simplified case as stated in section~\ref{sup:univeral_R_tau}, where the duration of the S/E/AS/PS states are assumed to be constants.
The results are shown in figure~\ref{SIfig_verify}, indicating that the analytical simplified $\tau_c$ given in \eqref{eq:simp_tauc} is also consistent with the simulation outcomes.

\subsubsection{Monte Carlo stochasticity in the simulation}\label{sup:monte_carlo}
In the following, we summarize the procedures which incorporate probabilistic random sampling in our Monte Carlo simulations. These procedures account for stochastic variability and parameter uncertainty in our study.

\begin{enumerate}
  \item We make a probabilistic decision based on \eqref{eq:infection_rate} on whether the transmission chain continues from an infectious individual to one of her effective close contact.
  \item Each newly infected individual undergoes a random Bernoulli sampling with probability $p$ to determine whether the resulting case is asymptomatic.
  \item The number of effective close contacts $k_+$ for each new infection is generated through random sampling from either a predefined probability distribution (such as Poisson or power-law) or an empirical discrete sequence derived from observed positive cases.
  \item For each new infection, key durations—including the latent period, the transmissible period, and the delay to symptom onset—are independently generated by random sampling from assigned continuous distributions (e.g., Lognormal, Gamma, or Weibull), thereby determining the timings of state transitions.
\end{enumerate}

Together, these procedures serve as a standard Monte Carlo simulation framework, allowing the model to effectively reflect individual-level heterogeneity and randomness.

The layer-specific effective reproduction number $\hat{R}^l$ and its 95\% confidence interval is based on multiple independent simulations as described below. Let $X_i^l$ denote the number of nodes in layer $l$ with non-zero transmissible period of the $i$th simulated transmission chain, following \eqref{eq:Rl}, we have
\begin{equation}\label{eq:Rl}
  \hat{R}^l =\frac{\sum_{i=1}^M X_i^{l+1}}{\sum_{i=1}^M X_i^l}= \frac{\bar X^{l+1}}{\bar X^l}
\end{equation}
where $M$ is the total number of simulated transmission chains, $\bar X^l$ and $\bar X^{l+1}$ are the average number of nodes in layers $l$ and $l+1$. To obtain the confidence interval of $\hat{R}^l$, the delta method is applied to calculate the approximated variance of $\hat{R}^l$:
\begin{equation}
    \widehat{\text{Var}}(\hat{R}^{l}) \approx \frac{1}{M (\bar{X}^{l})^2} \left[ \hat{\sigma}_{l+1}^2- 2 \hat{R}^{l} \hat{\sigma}_{l, l+1} + (\hat{R}^{l})^2 \hat{\sigma}_{l}^2  \right]
\end{equation}
where $\hat{\sigma}_{l}^2$ and $\hat{\sigma}_{l+1}^2$ are the sample variance of the number of nodes in layers $l$ and $l+1$, $\hat{\sigma}_{l,l+1}$ is the sample covariance between the number of nodes in layers $l$ and $l+1$. Based on the asymptotic normality assumption, the 95\% confidence interval of $\hat{R}^l$, $(\hat{R}^l_{L}, \hat{R}^l_{U})$, is calculated as
\begin{equation}\label{sup:eq_ci_Rl}
  \left( \hat{R}^{l} - 1.96 \sqrt{\widehat{\text{Var}}(\hat{R}^{l})}, \hat{R}^{l} + 1.96 \sqrt{\widehat{\text{Var}}(\hat{R}^{l})} \right)
\end{equation}

\subsection{Contact tracing with missing cases}\label{sup:missing_rate}
\subsubsection{Updated simulation studies} \label{sup: updatedsimu}
Finally, we discuss the relationship between contact tracing and large-scale PCR testing. 
During the implementation of contact tracing in practice, it is inevitable that a certain proportion of infected individuals will be missed.
In such cases, large-scale PCR testing can assist in identifying the missed infections. However, due to the extremely high socio-economic costs of large-scale PCR testing, we further discuss the necessary scenarios for its implementation.

Specifically, we perform simulation studies by introducing a new input parameter $q$, the proportion of missing close contacts when searching for the close contacts of positive cases.
When building a tree-structured disease transmission trajectory as in section~\ref{sup:simu}, a newly infected node would now have a probability $q$ of not being traced or quarantined.
With Shenzhen's control data, we obtain $q = 18.6\%$.
Since some nodes in the simulated transmission trajectory may fail to be quarantined, the stopping criteria (1) mentioned in section~\ref{sup:simu} is revised to be ``all nodes on the transmission trajectory are in the Q (quarantined) state or the R (removed) state."

In order to correspond to the two scenarios, i.e., weak social distancing measures (Weak SDM) and strong social distancing measures (Strong SDM), we consider two sets of values from which the number of effective close contacts $k_+$ is sampled. 
For the case ``Weak SDM'', the set of values consists of the observed number of effective close contacts of the 219 positive cases identified through large-scale PCR testing (see Sec \ref {sup:data_k}, with an average of $\bar{k}_+=177.64$).
For the case ``Strong SDM'', the set of values consists of the observed number of effective close contacts of the 225 positive cases identified within 3 days before and after March 24 (with an average of $\bar{k}_+=51.14$ under Shenzhen's social distancing, and 95\% CI 37.82–64.46), representing the case under the strongest SDM in Shenzhen.
This average value of 51.14 corresponds to the point with the smallest value on the y-axis in figure~2c of the main text.
In figure~3ab of the main text, the olive green and burnt orange lines are for the case Weak SDM and the case Strong SDM, respectively, each based on $50,000$ simulated transmission trajectories.

\subsubsection{Critical basic reproduction number $R^c_0$}\label{sup:critical_R0c}
The critical basic reproduction number $R^c_0$ is determined by the $R_{0}$ that makes the stabilized effective reproduction number $\hat{R}$ closest to 1. 
Here, we briefly introduce how $\hat{R}$, and subsequently $R^c_0$ with 95\% confidence intervals, are determined.

Given $R_0$ and $\bar{k}_+$ = 177.64, the infection rate $\beta$ can be obtained by \eqref{eq:R0}. 
Considering the infection rate of different infectious diseases and the two effective close contact numbers mentioned in section~\ref{sup: updatedsimu}, 
50,000 stochastic branching process simulations (section~\ref{sup: updatedsimu}) are conducted.
The effective reproduction number in transmission layer $l$, denoted by $\hat{R}^{l}$, with 95\% CI $(\hat{R}^l_{L}, \hat{R}^l_{U})$ can be calculated (see section~\ref{sup:monte_carlo}).
As the transmission layer $l$ increases, the stabilized effective reproduction number $\hat{R}$ can be obtained  as well as its 95\% CI $(\hat{R}_{L}, \hat{R}_{U})$. 
At different values of $R_0$, three fourth-degree polynomials ($f_{\text{lower}}$, $f_{\text{mean}}$, $f_{\text{upper}}$) are fitted to the scatter-plot of $(R_0, \hat{R})$, $(R_0, \hat{R}_{L})$ and $(R_0, \hat{R}_{U})$ (as shown in figure~3a of the main text) to get smooth functional relationships, i.e., $\hat{R}_L\approx f_{\text{lower}}(R_0)$, $\hat{R}\approx f_{\text{mean}}(R_0)$, and $\hat{R}_U\approx f_{\text{upper}}(R_0)$.
The critical basic reproduction number $R^c_0$ is obtained by solving $f_{\text{mean}}(R_0) = 1$.
As for its 95\% CI, the lower and upper bound are obtained by solving $f_{\text{upper}}(R_0) = 1$ and $f_{\text{lower}}(R_0) = 1$, respectively.

On the other hand, given different values of the missing proportion $q$, the corresponding $R^c_0$ can be obtained. 
Simulations with $q$ ranging from 5\% to 80\% are performed to obtain the $R^c_0$ (see figure~3c in the main text). 

\subsubsection{Missing proportion of the four regions}\label{sup:missing_prop}

For Shenzhen, the missing proportion $q$ being 18.6\% is calculated as the ratio of the number of positive cases detected through large-scale PCR testing to the total number of positive cases (See section~\ref{sup:dataset_spread_chain}).

For the UK, based on the social network data in the UK, researchers predicted that an average of 61\% of the close contacts of infected cases could be identified \cite{keeling2020efficacy}. So, 39\% is used as the missing proportion for the UK.

For the US, based on a cross-sectional study of the US local COVID-19 surveillance data, researchers estimated that 66\% of the close contacts might have been missed \cite{lash2021covid}. Therefore, 66\% is used as the missing proportion for the US.

For Hong Kong, positive cases are routinely detected through active clinical and public health surveillance (CDPHS) from July to September 2020, including tracing close contacts of confirmed cases. 
Based on the data from the Universal Community Nucleic Acid (UCTP) conducted from September 1 to 14, 2020, the estimated proportion of infections that were detected by the CDPHS is 27\% (95\% CI: 22\%, 34\%)\cite{yang2022universal}. Therefore, 73\% is used as the missing proportion for Hong Kong.

\subsubsection{Four assumptions on contact tracing}\label{subsec:four_tracing_machism}

For comprehensive evaluation, we introduce table~\ref{SItable_four_tracing}, clarifying four assumptions on the implementation of contact tracing (A, B, C, D, from weakest to strongest). While Assumption D (Full recursive bidirectional tracing) represents the strongest intervention, it is prohibitively resource-intensive and thus impractical. The remaining three are all feasible and can be observed in real-world practice. Given its balance of feasibility and close alignment with actual implementation standards, Assumption C (Forward + 1st-order Backward) is selected as our primary assumption for analysis. This prioritizes exhaustive downstream tracing, mitigating impacts from undetected upstream chains and resulting in higher critical basic reproduction number $R^c_0$.

\begin{sidewaystable}[htbp]
\centering
\footnotesize
\caption{Four assumptions on the implementation of contact tracing}
\begin{tabular}{@{}llll@{}}
\hline
ID & Assumption                                                       & Description        & Real-World Example   \\
\hline
A  & Forward Only                                                     & \makecell[l]{Upon detection of a confirmed case, tracing proceeds\\ only downward to secondary infections (child generations),\\ with no backtracing of upstream sources. Weakest capability.}    & \makecell[l]{During the 2022 COVID-19 outbreak in Beijing,\\ a confirmed case was identified as a food delivery rider.\\ The local CDC only tracked his secondary infections.\\ No backtracing was conducted to identify\\ his upstream infection source \cite{chinanews2022fangshan}. } \\
\hline
B  & \makecell[l]{Recursive Bidirectional \\(permanent untraceability\\ upon omission)} & \makecell[l]{Bidirectional recursive tracing to parent and child generations,\\ but any omitted node is permanently marked untraceable.}        & \makecell[l]{During the 2021 Beijing COVID-19 outbreak, an infected\\ case concealed his operation of a childcare catering \\service for over 20 children, and trouble the local CDC implemented\\ contact tracing and identifying his infection source \cite{yiyang2022epidemiccases}. } \\
\hline
C  & Forward + 1st-order Backward                                     & \makecell[l]{Unlimited recursive forward tracing \\ (child generations) combined with backward tracing \\ restricted to the direct infector\\(first-order only). Adopted in the main analysis.} & \makecell[l]{On January 15 2022, Beijing CDC reported one locally \\ transmitted Omicron case. Following the case discovery, nucleic \\ acid testing of close contacts and relevant \\ risk individuals identified secondary infections.\\ Simultaneously, both the inner and outer packaging, along with\\ documents, of an international \\ mail item received by the case on January 11\\ tested positive, confirming the mail as the source of infection \cite{report_beijing}.} \\
\hline
D  & Full recursive bidirectional                                     & \makecell[l]{Bidirectional recursive tracing with temporary omissions only,\\ allowing rediscovery via alternative paths. Strongest capability.} & Only in theory.\\                                                 
\hline
\end{tabular}
\label{SItable_four_tracing}
\end{sidewaystable}

\clearpage

\subsubsection{Sensitivity analysis under the four assumptions on contact tracing}\label{subsec:four_tracing_sensitivity}



\begin{figure}[ht!]
	\centering
	\includegraphics[width=0.9\textwidth]{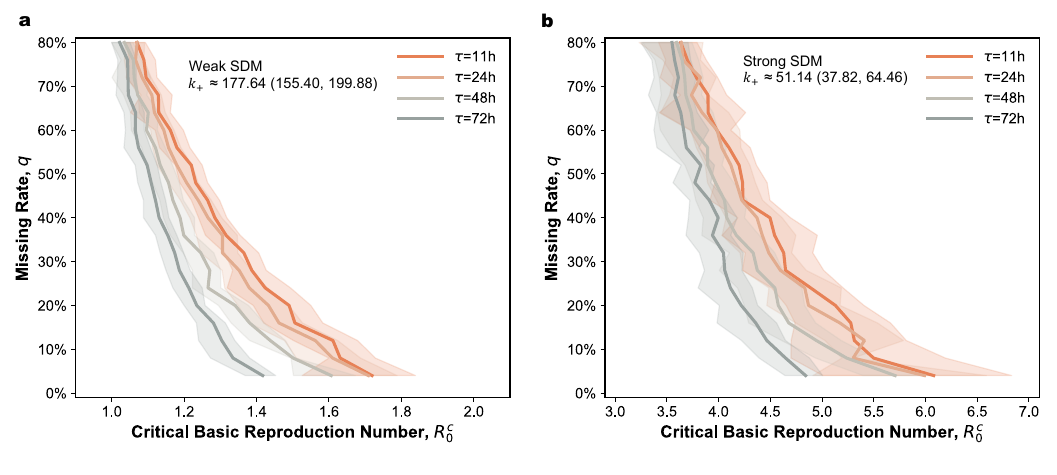}
	\caption{
    \textbf{Critical basic reproduction number $R^c_0$ in Shenzhen under Assumption A on contact tracing.}
  Analyses were performed by varying the contact tracing period (11, 24, 48, 72 hours), missing rates from 5\% to 80\%. Similarly, 50,000 stochastic branching process simulations were conducted under the two social distancing conditions ("Weak" and "Strong") in Shenzhen. 
 }
	\label{SIfig_A}
\end{figure}

\begin{figure}[ht!]
	\centering
	\includegraphics[width=0.9\textwidth]{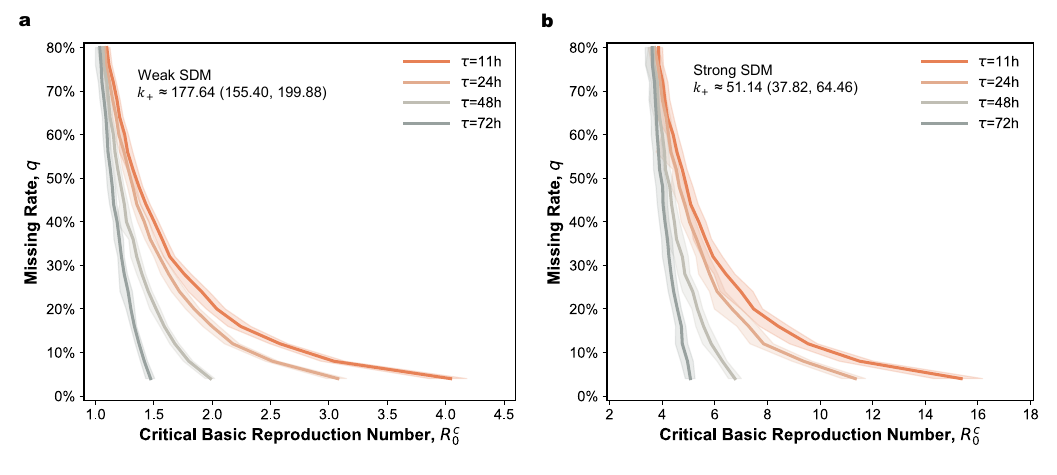}
	\caption{
    \textbf{Critical basic reproduction number $R^c_0$ in Shenzhen under Assumption B on contact tracing.}
  Analyses were performed by varying the contact tracing period (11, 24, 48, 72 hours), missing rates from 5\% to 80\%. Similarly, 50,000 stochastic branching process simulations were conducted under the two social distancing conditions ("Weak" and "Strong") in Shenzhen. 
 }
	\label{SIfig_B}
\end{figure}

\begin{figure}[ht!]
	\centering
	\includegraphics[width=0.9\textwidth]{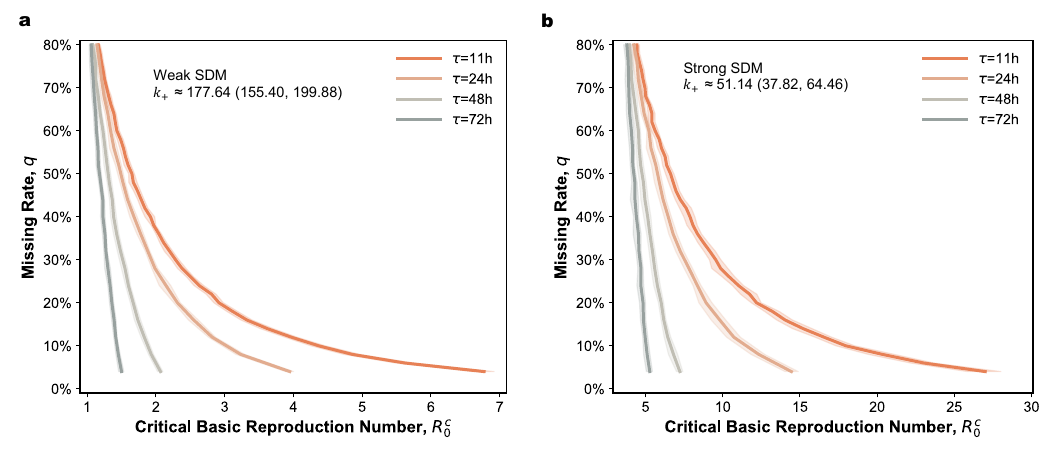}
	\caption{
    \textbf{Critical basic reproduction number $R^c_0$ in Shenzhen under Assumption C on contact tracing.}
  Analyses were performed by varying the contact tracing period (11, 24, 48, 72 hours), missing rates from 5\% to 80\%. Similarly, 50,000 stochastic branching process simulations were conducted under the two social distancing conditions ("Weak" and "Strong") in Shenzhen.  
 }
	\label{SIfig_C}
\end{figure}

\begin{figure}[ht!]
	\centering
	\includegraphics[width=0.9\textwidth]{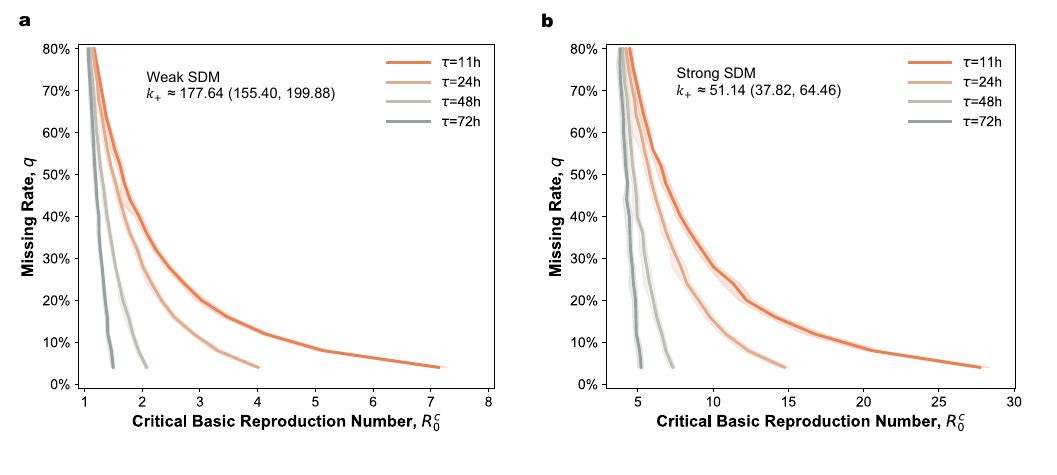}
	\caption{
    \textbf{Critical basic reproduction number $R^c_0$ in Shenzhen under Assumption D on contact tracing.}
  Analyses were performed by varying the contact tracing period (11, 24, 48, 72 hours), missing rates from 5\% to 80\%. Similarly, 50,000 stochastic branching process simulations were conducted under the two social distancing conditions ("Weak" and "Strong") in Shenzhen. 
  }
	\label{SIfig_D}
\end{figure}

We first conduct analyses on how the critical basic reproduction number $R_0^c$ changes with the contact tracing period, the missing rate in contact tracing, and the intensity of social distancing measures under the four assumptions on contact tracing. The results are shown in Figures~\ref{SIfig_A}, ~\ref{SIfig_B}, ~\ref{SIfig_C}, ~\ref{SIfig_D} and tables~\ref{SItable_A}, ~\ref{SItable_B}, ~2, ~\ref{SItable_D}.

\clearpage

\clearpage

\begin{table}[ht!]
\centering
\small
\caption{
\textbf{Critical basic reproduction number $R_0^c$.} Additional analyses were performed by varying the contact tracing period ($\tau$) (11, 24, 48, 72 hours), the missing rate of contacts ($q$) (18.6\%, 40\%, 50\%, 60\%, 70\%), and the intensity of social distancing (classified as "Weak" with contact rate $\bar{k}_+$ = 177.64, reflecting the least stringent measures in Shenzhen, and "Strong" with $\bar{k}_+$ = 51.14, reflecting the most stringent measures observed during the period from March 21 to March 27, 2022). 
  For each scenario, 50,000 stochastic branching process simulations were conducted.
  The critical threshold $R_0^c$ was determined as the $R_0$ value at which $R = 1$ under the Assumption A (further details in 
  section~\ref{sup:simu} and ~\ref{sup:missing_rate}). 
  }
  \begin{tabular}{@{}ccccccc@{}}
\hline
\multirow{2}{*}{\makecell[c]{Contact tracing\\period, $\tau$/h}} & \multirow{2}{*}{\makecell[c]{Social distancing\\intensity}} & \multicolumn{5}{c}{Missing rate, $q$} \\ \cline{3-7}
 &  & 18.6\% & 40\% & 50\% & 60\% & 70\% \\ \hline
\multirow{2}{*}{11} & Weak & \makecell{1.50 \\ (1.41, 1.56)} & \makecell{1.28 \\ (1.24, 1.32)} & \makecell{1.21 \\ (1.17, 1.25)} & \makecell{1.16 \\ (1.12, 1.20)} & \makecell{1.11 \\ (1.06, 1.13)} \\ \cline{2-7}
 & Strong & \makecell{5.19 \\ (4.73, 5.58)} & \makecell{4.47 \\ (4.15, 4.68)} & \makecell{4.14 \\ (3.75, 4.42)} & \makecell{3.98 \\ (3.69, 4.16)} & \makecell{3.81 \\ (3.56, 4.02)} \\ \hline
\multirow{2}{*}{24} & Weak & \makecell{1.45 \\ (1.38, 1.51)} & \makecell{1.27 \\ (1.21, 1.30)} & \makecell{1.18 \\ (1.14, 1.22)} & \makecell{1.14 \\ (1.10, 1.17)} & \makecell{1.10 \\ (1.06, 1.14)} \\ \cline{2-7}
 & Strong & \makecell{4.91 \\ (4.55, 5.14)} & \makecell{4.40 \\ (4.00, 4.63)} & \makecell{4.09 \\ (3.80, 4.29)} & \makecell{4.01 \\ (3.76, 4.21)} & \makecell{3.80 \\ (3.48, 3.97)} \\ \hline
\multirow{2}{*}{48} & Weak & \makecell{1.35 \\ (1.30, 1.40)} & \makecell{1.19 \\ (1.13, 1.22)} & \makecell{1.15 \\ (1.11, 1.18)} & \makecell{1.10 \\ (1.06, 1.13)} & \makecell{1.06 \\ (1.04, 1.09)} \\ \cline{2-7}
 & Strong & \makecell{4.67 \\ (4.25, 4.95)} & \makecell{4.06 \\ (3.81, 4.24)} & \makecell{3.95 \\ (3.72, 4.10)} & \makecell{3.82 \\ (3.47, 4.02)} & \makecell{3.70 \\ (3.42, 3.87)} \\ \hline
\multirow{2}{*}{72} & Weak & \makecell{1.24 \\ (1.20, 1.28)} & \makecell{1.13 \\ (1.09, 1.16)} & \makecell{1.11 \\ (1.07, 1.13)} & \makecell{1.07 \\ (1.04, 1.10)} & \makecell{1.04 \\ (1.00, 1.07)} \\ \cline{2-7}
 & Strong & \makecell{4.20 \\ (4.01, 4.31)} & \makecell{3.98 \\ (3.67, 4.18)} & \makecell{3.79 \\ (3.51, 3.97)} & \makecell{3.67 \\ (3.38, 3.92)} & \makecell{3.61 \\ (3.40, 3.76)} \\ 
 \hline
\end{tabular}
\label{SItable_A}
\end{table}

\clearpage

\begin{table}[ht!]
\centering
\small
\caption{
\textbf{Critical basic reproduction number $R_0^c$.} Additional analyses were performed by varying the contact tracing period ($\tau$) (11, 24, 48, 72 hours), the missing rate of contacts ($q$) (18.6\%, 40\%, 50\%, 60\%, 70\%), and the intensity of social distancing (classified as "Weak" with contact rate $\bar{k}_+$ = 177.64, reflecting the least stringent measures in Shenzhen, and "Strong" with $\bar{k}_+$ = 51.14, reflecting the most stringent measures observed during the period from March 21 to March 27, 2022). 
  For each scenario, 50,000 stochastic branching process simulations were conducted.
  The critical threshold $R_0^c$ was determined as the $R_0$ value at which $R = 1$ under the Assumption B (further details in section~\ref{sup:simu} and ~\ref{sup:missing_rate}). 
}
\begin{tabular}{@{}ccccccc@{}}
\hline
\multirow{2}{*}{\makecell[c]{Contact tracing \\ period, $\tau$/h}} & 
\multirow{2}{*}{\makecell[c]{Social distancing \\ intensity}} & 
\multicolumn{5}{c}{Missing rate, $q$} \\ \cline{3-7} 
 &  & 18.6\% & 40\% & 50\% & 60\% & 70\% \\ \hline
\multirow{2}{*}{11} & Weak & \makecell{2.08 \\ (2.01, 2.14)} & \makecell{1.50 \\ (1.46, 1.54)} & \makecell{1.36 \\ (1.31, 1.39)} & \makecell{1.25 \\ (1.21, 1.28)} & \makecell{1.17 \\ (1.12, 1.20)} \\ \cline{2-7} 
 & Strong & \makecell{7.79 \\ (7.28, 8.18)} & \makecell{5.39 \\ (5.07, 5.67)} & \makecell{4.83 \\ (4.49, 5.07)} & \makecell{4.42 \\ (4.18, 4.60)} & \makecell{4.14 \\ (3.89, 4.33)} \\ \hline
\multirow{2}{*}{24} & Weak & \makecell{1.89 \\ (1.82, 1.94)} & \makecell{1.42 \\ (1.36, 1.46)} & \makecell{1.28 \\ (1.25, 1.31)} & \makecell{1.20 \\ (1.16, 1.23)} & \makecell{1.13 \\ (1.10, 1.16)} \\ \cline{2-7} 
 & Strong & \makecell{6.86 \\ (6.44, 7.14)} & \makecell{5.04 \\ (4.72, 5.28)} & \makecell{4.61 \\ (4.32, 4.81)} & \makecell{4.24 \\ (4.00, 4.42)} & \makecell{4.00 \\ (3.71, 4.19)} \\ \hline
\multirow{2}{*}{48} & Weak & \makecell{1.54 \\ (1.50, 1.57)} & \makecell{1.26 \\ (1.17, 1.31)} & \makecell{1.20 \\ (1.16, 1.24)} & \makecell{1.15 \\ (1.11, 1.18)} & \makecell{1.09 \\ (1.05, 1.12)} \\ \cline{2-7} 
 & Strong & \makecell{5.44 \\ (5.08, 5.66)} & \makecell{4.46 \\ (4.18, 4.67)} & \makecell{4.20 \\ (3.95, 4.39)} & \makecell{4.01 \\ (3.76, 4.18)} & \makecell{3.80 \\ (3.52, 3.99)} \\ \hline
\multirow{2}{*}{72} & Weak & \makecell{1.33 \\ (1.29, 1.36)} & \makecell{1.18 \\ (1.15, 1.21)} & \makecell{1.13 \\ (1.10, 1.16)} & \makecell{1.10 \\ (1.06, 1.13)} & \makecell{1.07 \\ (1.03, 1.09)} \\ \cline{2-7} 
 & Strong & \makecell{4.61 \\ (4.32, 4.81)} & \makecell{4.11 \\ (3.90, 4.26)} & \makecell{3.97 \\ (3.74, 4.14)} & \makecell{3.81 \\ (3.59, 3.97)} & \makecell{3.70 \\ (3.49, 3.84)} \\ \hline
\end{tabular}
\label{SItable_B}
\end{table}

\clearpage

\begin{table}[ht!]
\centering
\small
\caption{
\textbf{Critical basic reproduction number $R_0^c$.} Additional analyses were performed by varying the contact tracing period ($\tau$) (11, 24, 48, 72 hours), the missing rate of contacts ($q$) (18.6\%, 40\%, 50\%, 60\%, 70\%), and the intensity of social distancing (classified as "Weak" with contact rate $\bar{k}_+$ = 177.64, reflecting the least stringent measures in Shenzhen, and "Strong" with $\bar{k}_+$ = 51.14, reflecting the most stringent measures observed during the period from March 21 to March 27, 2022). 
  For each scenario, 50,000 stochastic branching process simulations were conducted.
  The critical threshold $R_0^c$ was determined as the $R_0$ value at which $R = 1$ under the Assumption D (further details in section~\ref{sup:simu} and ~\ref{sup:missing_rate}). 
}
\begin{tabular}{@{}ccccccc@{}}
\hline
\multirow{2}{*}{\makecell[c]{Contact tracing\\period, $\tau$/h}} & \multirow{2}{*}{\makecell[c]{Social distancing\\intensity}} & \multicolumn{5}{c}{Missing rate, $q$} \\ \cline{3-7}
 &  & 18.6\% & 40\% & 50\% & 60\% & 70\% \\ \hline
\multirow{2}{*}{11} & Weak & \makecell{3.21 \\ (3.09, 3.31)} & \makecell{1.94 \\ (1.89, 1.99)} & \makecell{1.64 \\ (1.59, 1.67)} & \makecell{1.44 \\ (1.40, 1.48)} & \makecell{1.30 \\ (1.25, 1.33)} \\ \cline{2-7}
 & Strong & \makecell{13.07 \\ (12.60, 13.48)} & \makecell{7.84 \\ (7.53, 8.07)} & \makecell{6.60 \\ (6.37, 6.80)} & \makecell{5.74 \\ (5.42, 5.99)} & \makecell{5.01 \\ (4.74, 5.22)} \\ \hline
\multirow{2}{*}{24} & Weak & \makecell{2.40 \\ (2.35, 2.45)} & \makecell{1.69 \\ (1.64, 1.73)} & \makecell{1.50 \\ (1.46, 1.54)} & \makecell{1.35 \\ (1.30, 1.39)} & \makecell{1.24 \\ (1.21, 1.27)} \\ \cline{2-7}
 & Strong & \makecell{9.33 \\ (8.91, 9.67)} & \makecell{6.53 \\ (6.28, 6.71)} & \makecell{5.77 \\ (5.45, 5.98)} & \makecell{5.17 \\ (4.90, 5.40)} & \makecell{4.69 \\ (4.47, 4.85)} \\ \hline
\multirow{2}{*}{48} & Weak & \makecell{1.70 \\ (1.65, 1.73)} & \makecell{1.39 \\ (1.35, 1.42)} & \makecell{1.30 \\ (1.25, 1.34)} & \makecell{1.21 \\ (1.17, 1.25)} & \makecell{1.16 \\ (1.13, 1.18)} \\ \cline{2-7}
 & Strong & \makecell{6.06 \\ (5.83, 6.24)} & \makecell{5.00 \\ (4.77, 5.18)} & \makecell{4.77 \\ (4.53, 4.94)} & \makecell{4.47 \\ (4.26, 4.64)} & \makecell{4.22 \\ (3.96, 4.45)} \\ \hline
\multirow{2}{*}{72} & Weak & \makecell{1.36 \\ (1.32, 1.40)} & \makecell{1.25 \\ (1.21, 1.27)} & \makecell{1.18 \\ (1.15, 1.21)} & \makecell{1.15 \\ (1.11, 1.18)} & \makecell{1.11 \\ (1.08, 1.13)} \\ \cline{2-7}
 & Strong & \makecell{4.77 \\ (4.57, 4.94)} & \makecell{4.45 \\ (4.14, 4.67)} & \makecell{4.29 \\ (4.14, 4.43)} & \makecell{4.02 \\ (3.85, 4.14)} & \makecell{3.91 \\ (3.72, 4.04)} \\ 
 \hline
\end{tabular}
\label{SItable_D}
\end{table}


Furthermore, to quantify the impact of asymptomatic cases on the effectiveness of contact tracing, we construct a branching process model that distinguishes between cases with and without symptoms. Let $p$ be the proportion of asymptomatic infections (AS) in the population, and $q$ be the overall probability that a case is missed by contact tracing. Moreover, let $p_a$ denote the proportion of asymptomatic infections among all missed cases. Consequently, an asymptomatic case is missed by contact tracing with probability $p_a q/p$, whereas a symptomatic case is missed with probability $(1-p_a )q/(1-p)$. This formulation allows us to reflect the heterogeneity in contact tracing efficiency due to symptom presentation by adjusting the parameter $p_a$.

\begin{figure}[htbp!]
	\centering
	\includegraphics[width=0.9\textwidth]{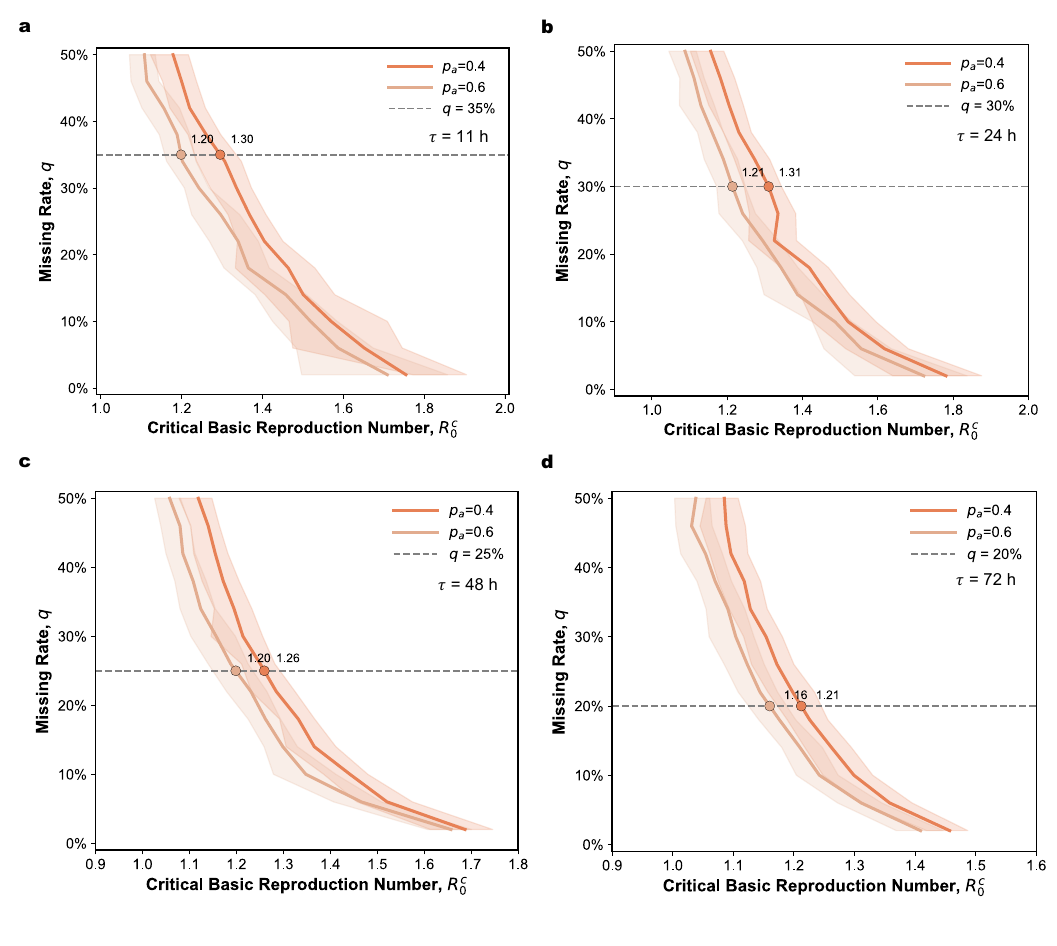}
	\caption{
        \textbf{Critical basic reproduction number $R^c_0$ in Shenzhen under different proportions of asymptomatic infections among all missed cases $p_a$.}
  Panels a–d correspond to $\tau=$ 11 h, 24 h, 48 h, and 72 h, respectively. Solid lines represent simulated mean $R^c_0$ for varying proportions of asymptomatic missing $p_a$ (0.4, 0.6); shaded areas indicate 95\% confidence intervals. 
  Horizontal dashed lines mark specific missing rates $q$. 
  Annotated circles on each curve show the corresponding $R_0^c$. 
  All simulations are under the Assumption A on contact tracing (see table~\ref{SItable_four_tracing}) and "Weak" social distancing with contact rate $\bar{k}_+=177.64$.
 }
	\label{SIfig_A_pasym}
\end{figure}

\begin{figure}[ht!]
	\centering
	\includegraphics[width=0.9\textwidth]{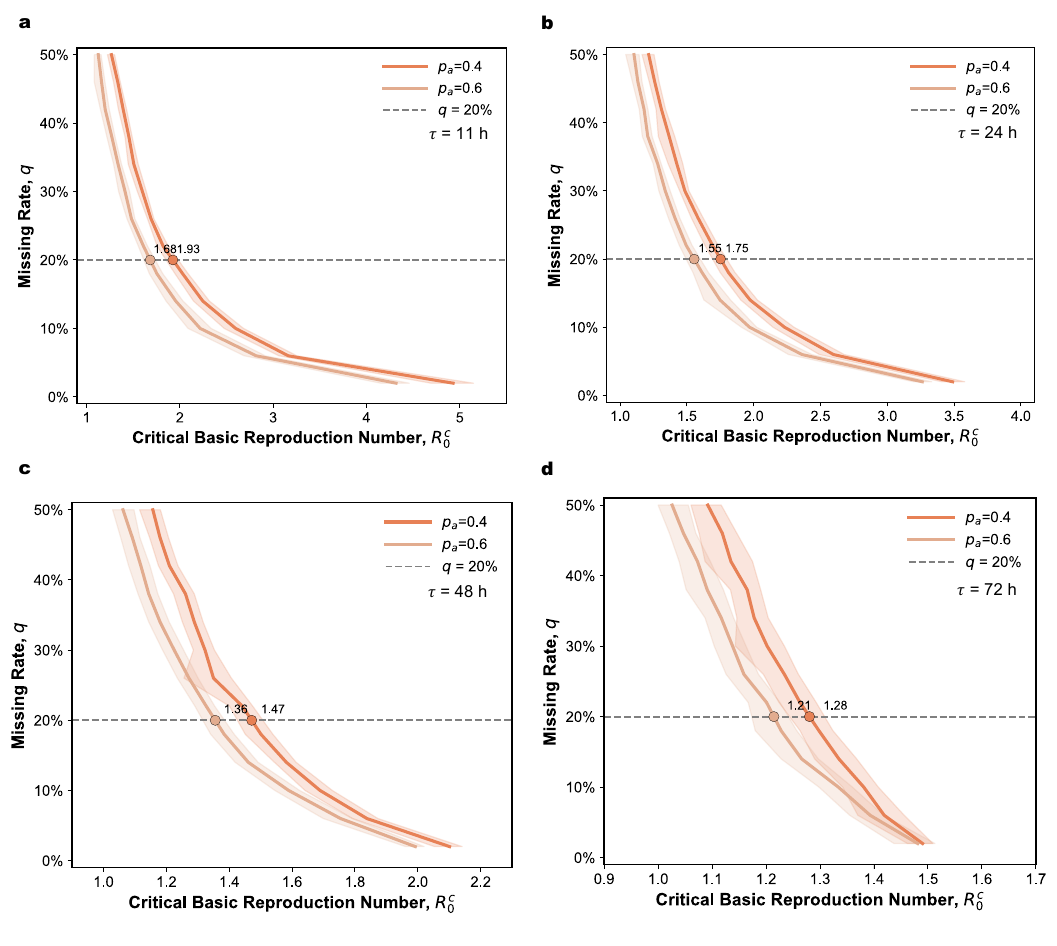}
	\caption{
        \textbf{Critical basic reproduction number $R^c_0$ in Shenzhen under different proportions of asymptomatic infections among all missed cases $p_a$.}
  Panels a–d correspond to $\tau=$ 11 h, 24 h, 48 h, and 72 h, respectively. Solid lines represent simulated mean $R^c_0$ for varying proportions of asymptomatic missing $p_a$ (0.4, 0.6); shaded areas indicate 95\% confidence intervals. 
  Horizontal dashed lines mark specific missing rates $q$. 
  Annotated circles on each curve show the corresponding $R_0^c$. 
  All simulations are under the Assumption B on contact tracing (see table~\ref{SItable_four_tracing}) and "Weak" social distancing with contact rate $\bar{k}_+=177.64$.
 }
	\label{SIfig_B_pasym}
\end{figure}

\begin{figure}[ht!]
	\centering
	\includegraphics[width=0.9\textwidth]{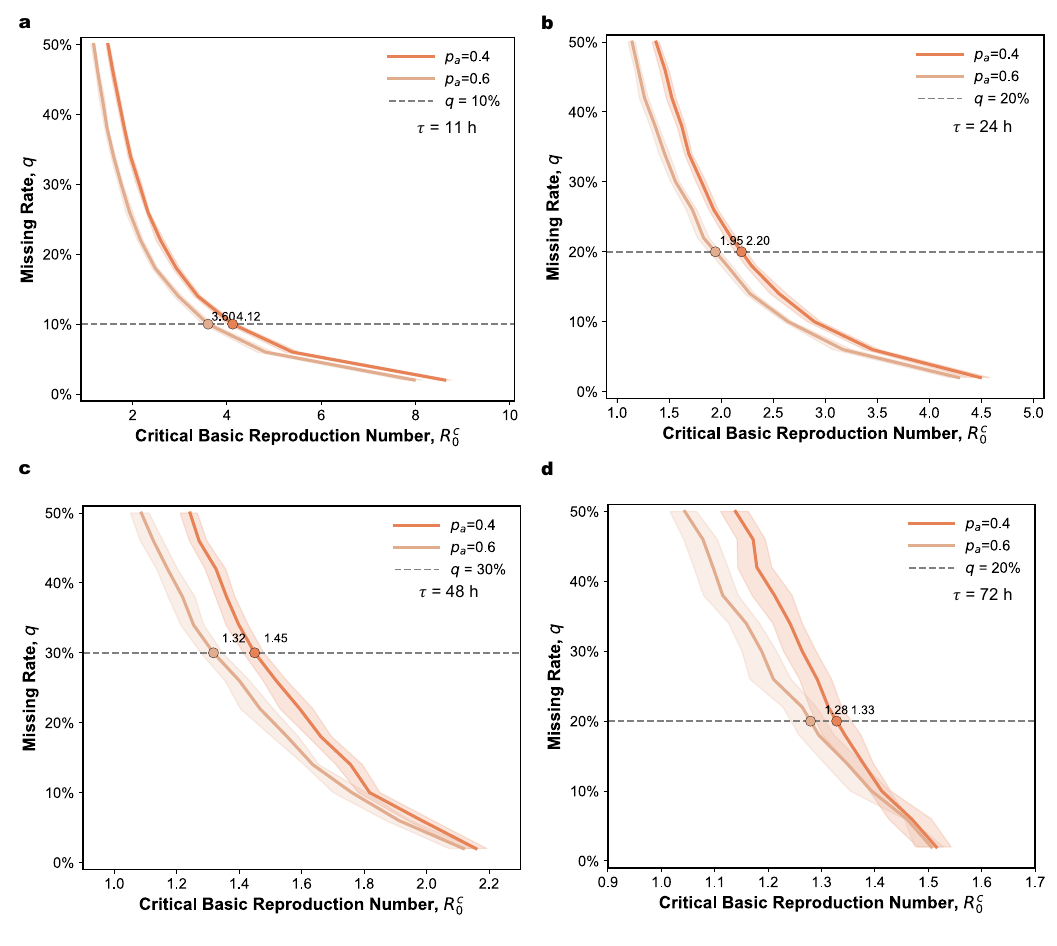}
	\caption{
    \textbf{Critical basic reproduction number $R^c_0$ in Shenzhen under different proportions of asymptomatic infections among all missed cases $p_a$.}
  Panels a–d correspond to $\tau=$ 11 h, 24 h, 48 h, and 72 h, respectively. Solid lines represent simulated mean $R^c_0$ for varying proportions of asymptomatic missing $p_a$ (0.4, 0.6); shaded areas indicate 95\% confidence intervals. 
  Horizontal dashed lines mark specific missing rates $q$. 
  Annotated circles on each curve show the corresponding $R_0^c$. 
  All simulations are under the Assumption C on contact tracing (unlimited recursive forward tracing plus first-order backward tracing; see table~\ref{SItable_four_tracing}) and "Weak" social distancing with contact rate $\bar{k}_+=177.64$.
 }
	\label{SIfig_C_pasym}
\end{figure}

\begin{figure}[ht!]
	\centering
	\includegraphics[width=0.9\textwidth]{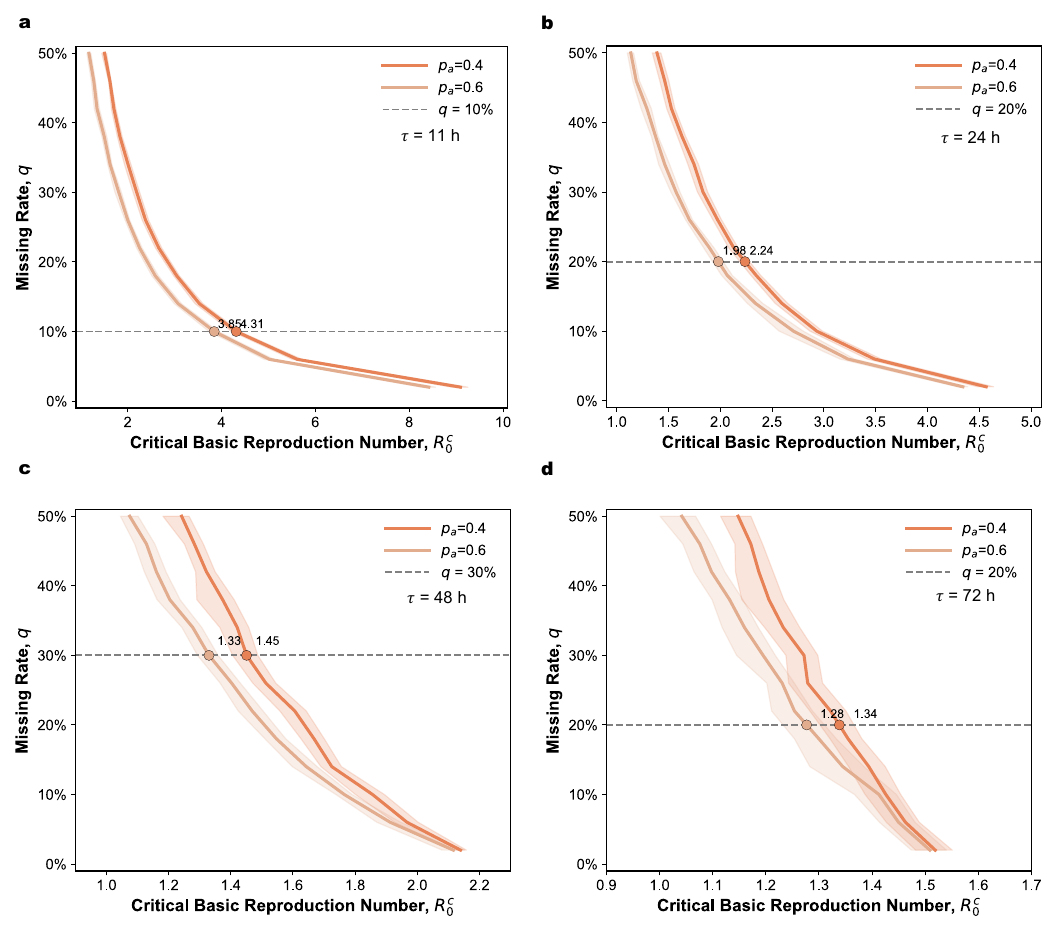}
	\caption{
        \textbf{Critical basic reproduction number $R^c_0$ in Shenzhen under different proportions of asymptomatic infections among all missed cases $p_a$.}
  Panels a–d correspond to $\tau=$ 11 h, 24 h, 48 h, and 72 h, respectively. Solid lines represent simulated mean $R^c_0$ for varying proportions of asymptomatic missing $p_a$ (0.4, 0.6); shaded areas indicate 95\% confidence intervals. 
  Horizontal dashed lines mark specific missing rates $q$. 
  Annotated circles on each curve show the corresponding $R_0^c$. 
  All simulations are under the Assumption D on contact tracing (see table~\ref{SItable_four_tracing}) and "Weak" social distancing with contact rate $\bar{k}_+=177.64$.
 }
	\label{SIfig_D_pasym}
\end{figure}

Based on the parameterization described above and considering $p=0.324$, we conduct a sensitivity analysis by setting $p_a$ to 0.4 and 0.6 to assess its impact on epidemic controllability. Results under the four assumptions on contact tracing are presented in Figures~\ref{SIfig_A_pasym}, ~\ref{SIfig_B_pasym}, ~\ref{SIfig_C_pasym}, ~\ref{SIfig_D_pasym}. The results show that the critical basic reproduction number $R_0^c$ decreases significantly as $p_a$ increases, confirming that asymptomatic cases substantially undermine contact-tracing efficiency.

\clearpage

\clearpage

\subsection{Pathogen specification, literature search and study selection for $R_0$ estimates}\label{sup:pathogen_lit_search}

\begin{figure}[ht!]
	\centering
	\includegraphics{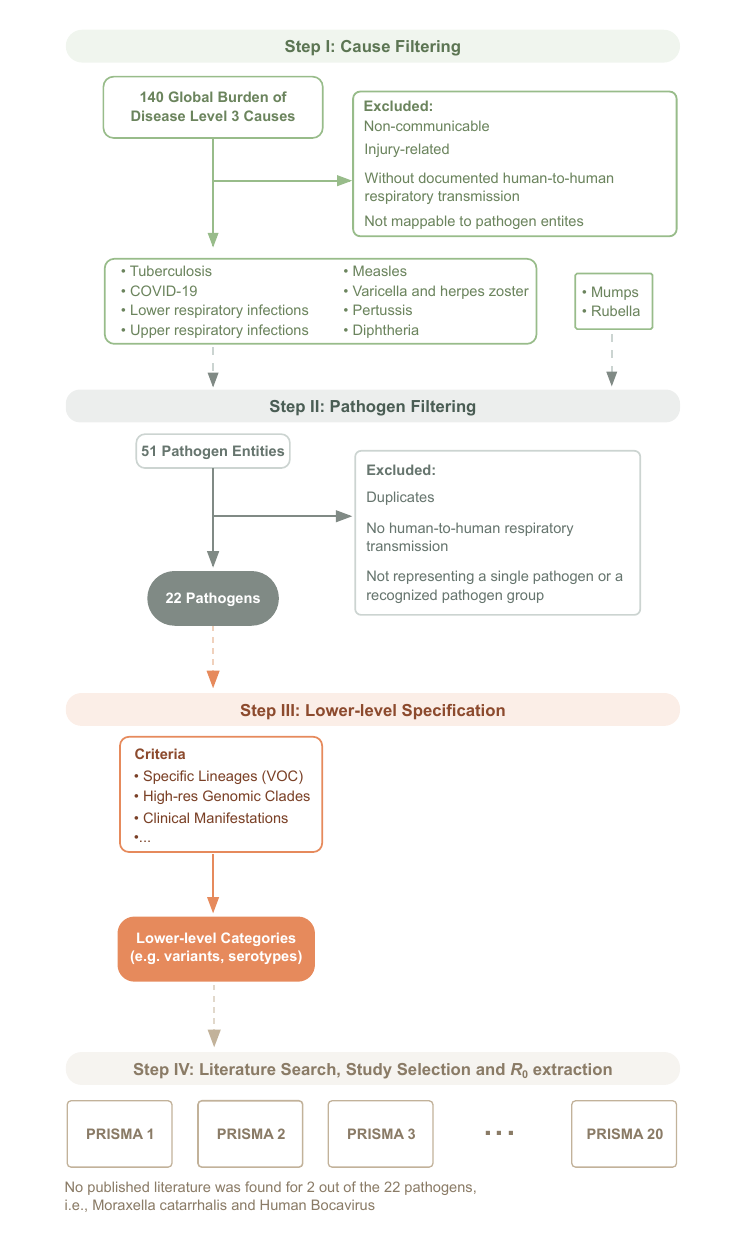}
	\caption{
  \textbf{Multi-step procedure for identifying respiratory pathogens (and their lower-level classifications, e.g., variants, serotypes, subtypes) and retrieving the corresponding $R_0$ estimates. 
 }
 }
	\label{SIfig_GBD_diagram}
\end{figure}

This entire section provides the procedures for identifying respiratory pathogens and retrieving corresponding estimates of the basic reproduction number ($R_0$), as summarized in the following four steps (also illustrated in figure~\ref{SIfig_GBD_diagram}).

\begin{enumerate}
  \item \textbf{Cause Filtering.} We started from the 140 Level 3 causes in the 2023 Global Burden of Disease (GBD) study~\cite{GBD2023Results}.
 Here a `cause' means a single disease or injury or an aggregation of diseases and injuries that causes death or disability~\cite{GBD2023Results}.
 Causes were excluded if they were non-communicable, injury-related, not mappable to pathogen entities, or linked only to pathogen entities without documented human-to-human respiratory transmission.
 We also include Mumps~\cite{cdcMumps} and Rubella~\cite{world2013rubella} as additional human-to-human respiratory infectious
disease causes.
 Finally, we identified 10 respiratory infectious disease-related causes.

  \item \textbf{Pathogen Filtering.} Based on associated pathogen entries, we systematically mapped 10 causes to 51 candidate pathogens entities through literature review~\cite{cdcTuberculosisTB,WHO_Coronavirus,sirota2025global,calderaro2022respiratory,CDC_Measles_Clinical_2026,CDC_Pertussis_About,CDC_aboutDiphtheria_2026,CDC_Shingles_2026}. 
  The final list of 22 candidate pathogen entities were obtained by applying three inclusion criteria: (1) not duplicated pathogens; (2) pathogens transmitted via human-to-human respiratory routes; (3) individual pathogens or conventionally defined, clinically recognized pathogen groups. See table~\ref{SItable_filter} for all details.

\begin{longtable}{@{}clp{5.5cm}ccc@{}}
\caption{
\textbf{Mapping respiratory infectious disease causes from the Global Burden of Disease study to their corresponding causative pathogens, followed by pathogen filtering using three inclusion criteria.} The three inclusion criteria are: R1 (not duplicated pathogens), R2 (pathogens transmitted via human-to-human respiratory routes), and R3 (individual pathogens or conventionally defined, clinically recognized pathogen groups). A cross in any inclusion criterion indicates exclusion. After filtering, 22 pathogen entities (marked in boldface) are retained.  References based on which the pathogens are identified are provided in parentheses after the causes~\cite{GBD2023Results,cdcTuberculosisTB,WHO_Coronavirus,sirota2025global,calderaro2022respiratory,CDC_Measles_Clinical_2026,CDC_Pertussis_About,CDC_aboutDiphtheria_2026,CDC_Shingles_2026}. 
}
\label{SItable_filter} \\
\hline
No. & 
\makecell[l]{Respiratory infectious\\disease causes} & 
\makecell[l]{Causative pathogen entities} & 
R1 & R2 & R3 \\
\hline
\endfirsthead
\multicolumn{6}{c}%
{{\tablename\ \thetable{} -- continued from previous page}} \\
\hline
No. & 
\makecell[l]{GBD Respiratory infectious\\disease causes} & 
\makecell[l]{Causative pathogen entities} & 
R1 & R2 & R3 \\
\hline
\endhead
\hline \multicolumn{6}{r}{{Continued on next page}} \\
\endfoot
\hline
\endlastfoot
1 & Tuberculosis (TB) \cite{cdcTuberculosisTB} & \textbf{Mycobacterium tuberculosis}   &  &  &  \\
\hline
2 & COVID-19 \cite{WHO_Coronavirus} & \textbf{SARS-CoV-2}  &  &  &  \\
\hline
3  & \makecell[l]{Lower respiratory infections (LRI) \cite{sirota2025global}} & \textbf{Streptococcus pneumoniae} &  &  &  \\
4  &   & Staphylococcus aureus  &  & \usym{2613} &  \\
5  &   & Klebsiella pneumoniae  &  & \usym{2613} &  \\
6  &   & Other  Mycobacterium species (non-TB, non-Leprosy)  &  & \usym{2613} & \usym{2613} \\
7  &   & Pseudomonas aeruginosa  &  & \usym{2613} &  \\
8  &   & \textbf{Influenza virus}  &  &  &  \\
9  &   & Escherichia coli  &  & \usym{2613} &  \\
10 &   & Other bacterial and viral pathogens  &  &  & \usym{2613} \\
11 &   & \textbf{Mycoplasma pneumoniae}  &  &  &  \\
12 &   & Other gram-negative bacteria  &  &  & \usym{2613} \\
13 &   & Aspergillus spp.  &  & \usym{2613} &  \\
14 &   & Acinetobacter baumannii  &  & \usym{2613} &  \\
15 &   & \textbf{Chlamydia pneumoniae}  &  &  &  \\
16 &   & Legionella spp.  &  & \usym{2613} &  \\
17 &   & \textbf{Haemophilus influenzae}  &  &  &  \\
18 &   & Other fungi  &  & \usym{2613} & \usym{2613} \\
19 &   & \textbf{Respiratory syncytial virus}  &  &  &  \\
20 &   & Group B streptococcus  &  & \usym{2613} &  \\
21 &   & \textbf{Group A Streptococcus}  &  &  &  \\
22 &   & Other Klebsiella species  &  & \usym{2613} & \usym{2613} \\
23 &   & Enterobacter spp.  &  & \usym{2613}  &  \\
24 &   & Other Acinetobacter species  &  & \usym{2613} & \usym{2613} \\
25 &   & Proteus spp.  &  & \usym{2613} &  \\
26 &   & Serratia spp.  &  & \usym{2613} &  \\
27 &   & Citrobacter spp.  &  & \usym{2613} &  \\
28 &   & Morganella spp.  &  & \usym{2613} &  \\
\hline
29 & \makecell[l]{Upper respiratory infections (URI) \cite{calderaro2022respiratory}} & \textbf{Human rhinovirus}  &  &  &  \\
30 &   & \textbf{Human coronavirus}  &  &  &  \\
31 &   & \textbf{Human adenovirus}  &  &  &  \\
32 &   & \textbf{Parainfluenza virus}  &  &  &  \\
33 &   & Influenza virus  & \usym{2613} &  &  \\
34 &   & Streptococcus pyogenes  & \usym{2613} &  &  \\
35 &   & Streptococcus pneumoniae  & \usym{2613} &  &  \\
36 &   & Haemophilus influenzae  & \usym{2613} &  &  \\
37 &   & Staphylococcus aureus  &  & \usym{2613} &  \\
38 &   & \textbf{Moraxella catarrhalis}  &  &  &  \\
39 &   & \textbf{Human metapneumovirus}  &  &  &  \\
40 &   & Respiratory syncytial virus  & \usym{2613} &  &  \\
41 &   & Coxsackie virus  &  & \usym{2613} &  \\
42 &   & \textbf{Human bocavirus}  &  &  &  \\
43 &   & Group A and G Streptococci  &  &  & \usym{2613} \\
44 &   & Chlamydia pneumoniae  & \usym{2613} &  &  \\
45 &   & Mycoplasma pneumoniae  & \usym{2613} &  &  \\
\hline
46 & Measles \cite{CDC_Measles_Clinical_2026}  & \textbf{Measles virus}  &  &  &  \\
\hline
47 & Pertussis \cite{CDC_Pertussis_About}  & \textbf{Bordetella pertussis}  &  &  &  \\
\hline
48 & Diphtheria \cite{CDC_aboutDiphtheria_2026} & \textbf{Corynebacterium diphtheriae}  &  &  &  \\
\hline
49 & \makecell[l]{Varicella and herpes zoster~\cite{CDC_Shingles_2026}} & \textbf{Varicella-zoster virus} &  &  &  \\
\hline
50 & Mumps~\cite{cdcMumps} & \textbf{Mumps virus} &  &  &  \\
\hline
51 & Rubella~\cite{world2013rubella} & \textbf{Rubella virus} &  &  &  \\
\hline
\end{longtable}

  \item \textbf{Lower-level Specification.} The 22 pathogens' lower-level classifications (e.g. variants, serotypes, and subtypes) were specified based on a literature search~\cite{rojas2025mycobacterium,markov2023evolution,WHO_SARS_CoV2_Variants_2023,CDC_Flu_Viruses_2025,velusamy2020expanded,fatima2025macrolide,CDC_Hib_Clinicians_2025,nuttens2024differences,CDC_StrepA_emmTyping_2024,gern2010abcs,gaunt2010epidemiology,CDC_Adenovirus_Outbreaks_2025,CDC_HPIV_Clinical_2026,sloots2006human,verduin2002moraxella,schildgen2013human,CDC_measles_2026,guiso2001fimbrial,CDC_diphtheria_2026,breuer2010proposal}. These classifications were specified to allow pathogen- and subtype-specific literature searches in Step 4.
  
  \item \textbf{Literature search, study selection and $R_0$ extraction} We implemented a detailed PubMed-based literature search strategy, including predefined keywords, screening stages (title/abstract and full-text review), and eligibility requirements, and establishing an objective rule for study selection when multiple eligible reports exist for 
  the same pathogen or lower-level category
  (see table~\ref{SItable_pathogens_variants} and Figures~\ref{SIfig_R0_TB}-\ref{SIfig_R0_rubella}). 
  In practice, the PubMed search was conducted for each pathogen and, where relevant, for each of its lower-level categories separately.
  The search records were retrieved up to March 20, 2026.
  The search terms for the title and abstract included (1) pathogen name OR variant/lineage/strain/serotype/type/subtype/subgroup/group/ specie/genotype/biotype/clade name; (2) "basic reproducti* number*" OR "basic reproducti* ratio*" OR "basic reproducti* rate*" OR "R0"; (3) 1 AND 2. 
  We excluded reviews, systematic reviews, comments, news articles, and editorials from our analyses.
  In addition, note that no published literature was found for 2 of the 22 pathogens from Step 3, which are Moraxella catarrhalis and Human bocavirus, and no reported $R_0$ estimate of Human coronavirus was found.

  To account for variation in due to biological, behavioral, and environmental factors~\cite{delamater2019complexity}, we compiled pathogen-specific estimates from the literature. For each pathogen or lower-level category, we screened the top 200 most-cited relevant articles in OpenAlex, using citation counts retrieved on March 20, 2026, and selected up to five eligible primary studies. Eligible studies were original peer-reviewed articles reporting empirical  estimates; animal-transmission studies and multi-pathogen pooled analyses were excluded. Additional studies were included for SARS-CoV-2, Mycoplasma pneumoniae, Respiratory syncytial virus, and Mumps virus.
  
  From each study, we extracted the primary overall  estimate, or, if unavailable, the maximum estimate across settings or models as a conservative outbreak-control threshold. The representative  for each pathogen or category was calculated as the arithmetic mean of study-level estimates, with the minimum and maximum defining the empirical range. Categories with fewer than two eligible studies were excluded. The final dataset includes 10 pathogen-level estimates and 6 subtype-/variant-level estimates
  (table~\ref{SItable_pathogens_variants}).
  
\end{enumerate}


This strategy establishes a transparent and reproducible framework based on a priori, well-defined criteria, thereby eliminating any potential arbitrariness, and enables consistent comparison of published \(R_0\) estimates across pathogens and their lower-level classifications.

\begin{sidewaystable}[htbp]
\centering
\footnotesize
\caption{
\textbf{Pathogens and lower-level classifications with $R_0$ estimates.} ``Overall'' indicates the pathogen-level estimate without further classification. 
Only pathogens with eligible empirical $R_0$ estimates identified under our screening criteria are listed; Human coronavirus, Human metapneumovirus, and Chlamydia pneumoniae were therefore not included. Reported values may differ from commonly cited ranges because $R_0$ is context-dependent and varies with epidemiological settings, interventions, and modelling assumptions.
}
\begin{tabular}{@{}llcccc@{}}
\hline
\makecell[l]{Pathogen/pathogen groups} &
\makecell[l]{Variant / subtype / strain / cluster \\(or equivalent lower-level classification)} &
\makecell[c]{Mean\\value} &
\makecell[c]{Range} &
\makecell[c]{Number of\\estimates} &
\makecell[c]{Reference} \\
\hline
\makecell[l]{Mycobacterium tuberculosis\\(M. tuberculosis)}  & Overall  & 4.83  & 1.13--8.34  & 5  & \cite{sanchez1997uncertainty,trauer2014construction,zhao2017analysis,brookspollock2010impact,cai2021modelling} \\
\hline
\multirow{2}{*}{\makecell[l]{SARS-CoV-2}}
   & Ancestral strain     & 3.43  & 2.20--5.80   & 6  & \cite{lai2020severe,sanche2020high,zhang2020estimation,hao2020reconstruction,jung2020real,sender2022unmitigated} \\
   & Omicron (B.1.1.529)  & 5.88  & 2.11--11.88  & 3  & \cite{liu2022effective,bi2023risk,chen2022omicron,ozkose2022fractional,khan2022mathematical} \\
\hline
\makecell[l]{Streptococcus pneumoniae\\(S. pneumoniae)}  & Overall  & 3.86  & 1.40--6.44  & 3  & \cite{hoti2009outbreaks,dekaj2024pneumococcus,le2025inference} \\
\hline
\multirow{2}{*}{\makecell[l]{Influenza virus}}
   & A (H1N1)  & 2.33  & 1.31--4.11  & 5  & \cite{fraser2009pandemic,yang2009transmissibility,balcan2009seasonal,white2009estimation,tuite2010estimated} \\
   & A (H3N2)  & 3.80  & 1.89--6.44  & 5  & \cite{longini2005containing,rvachev1985mathematical,yang2015inference,mathews2007biological,goeyvaerts2015estimating,jackson2010estimates} \\
\hline
\makecell[l]{Mycoplasma pneumoniae\\(M. pneumoniae)}  & Overall  & 4.19  & 1.60--6.78  & 2  & \cite{li2025stochastic,nielsen2025complex} \\
\hline
\multirow{2}{*}{\makecell[l]{Respiratory syncytial virus\\(RSV)}}
   & Overall  & 8.46  & 3.00--21.90  & 5  & \cite{weber2001modeling,reis2016retrospective,reis2018simulation,vanboven2020estimating,baker2022long} \\
   & RSV-A    & 5.39  & 1.03--13.80  & 3  & \cite{duvvuri2015genetic,otomaru2019transmission,suri2025analyzing} \\
\hline
\makecell[l]{Human adenovirus\\(HAdV)}  & HAdV-7  & 5.10  & 5.09--5.10  & 2  & \cite{guo2020epidemiological,guo2020artificially} \\
\hline
Measles virus  & Overall  & 19.99  & 8.25--30.80  & 5  & \cite{edmunds2000pre,metcalf2009seasonality,funk2019combining,chen2006predictive,vanboven2010estimation} \\
\hline
\makecell[l]{Bordetella pertussis\\(B. pertussis)}  & Overall  & 9.18  & 4.50--13.49  & 5  & \cite{kretzschmar2010incidence,metcalf2009seasonality,broutin2005epidemiological,mcgirr2013estimation,hsieh2014impact} \\
\hline
\makecell[l]{Corynebacterium diphtheriae\\(C. diphtheriae)}  & Overall  & 4.09  & 1.17--7.20  & 5  & \cite{metcalf2009seasonality,matsuyama2018uncertainty,fauzi2024assessing,islam2022global,kiss2024deterministic} \\
\hline
\makecell[l]{Varicella-zoster virus\\(VZV)}  & Overall  & 9.54  & 1.80--16.91  & 5  & \cite{nardone2007comparative,ogunjimi2009using,chen2006predictive,giraldo2008deterministic,santermans2015social} \\
\hline
Mumps virus  & Overall  & 10.13  & 4.28--19.30  & 6  & \cite{edmunds2000pre,glasser2016effect,kanaan2005matrix,li2017modelling,qu2016mumps,park2024modeling} \\
\hline
Rubella virus  & Overall  & 6.94  & 5.20--8.70  & 5  & \cite{edmunds2000pre,griffiths2001encouraging,glasser2016effect,kanaan2005matrix,lessler2013balancing,anderson1991infectious} \\
\hline
\end{tabular}
\label{SItable_pathogens_variants}
\end{sidewaystable}

\clearpage

\begin{figure}[ht!]
	\centering
	\includegraphics{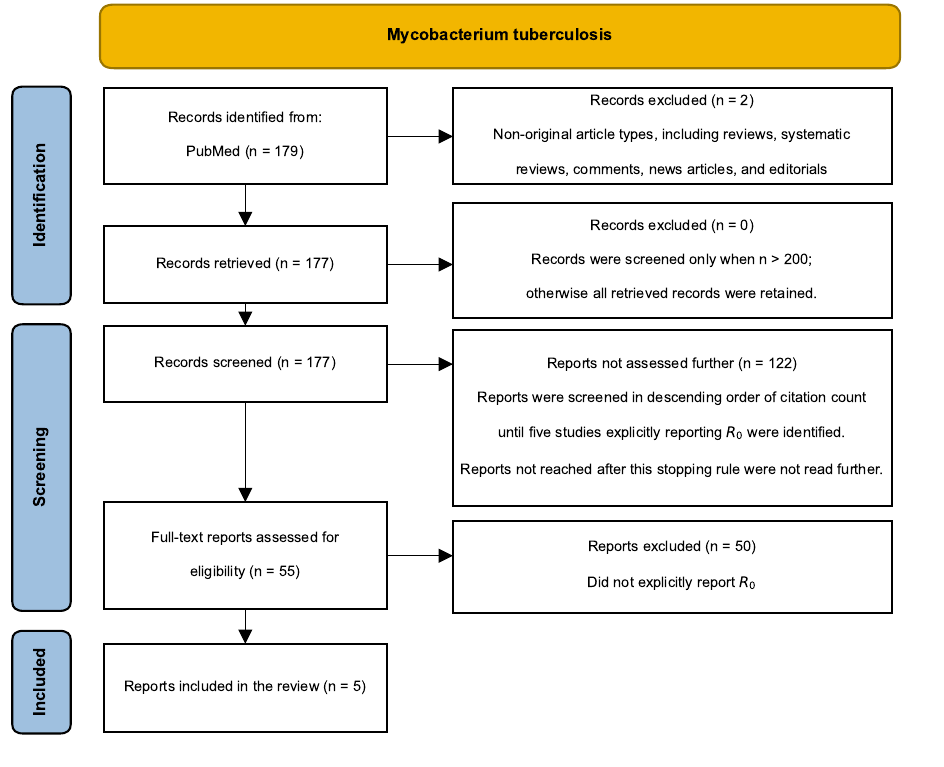}
	\caption{
  \textbf{PRISMA-style flow diagram of the literature search and study selection process for the $R_0$ estimates of Mycobacterium tuberculosis (M. tuberculosis).}
}
	\label{SIfig_R0_TB}
\end{figure}

\begin{figure}[ht!]
	\centering
	\includegraphics{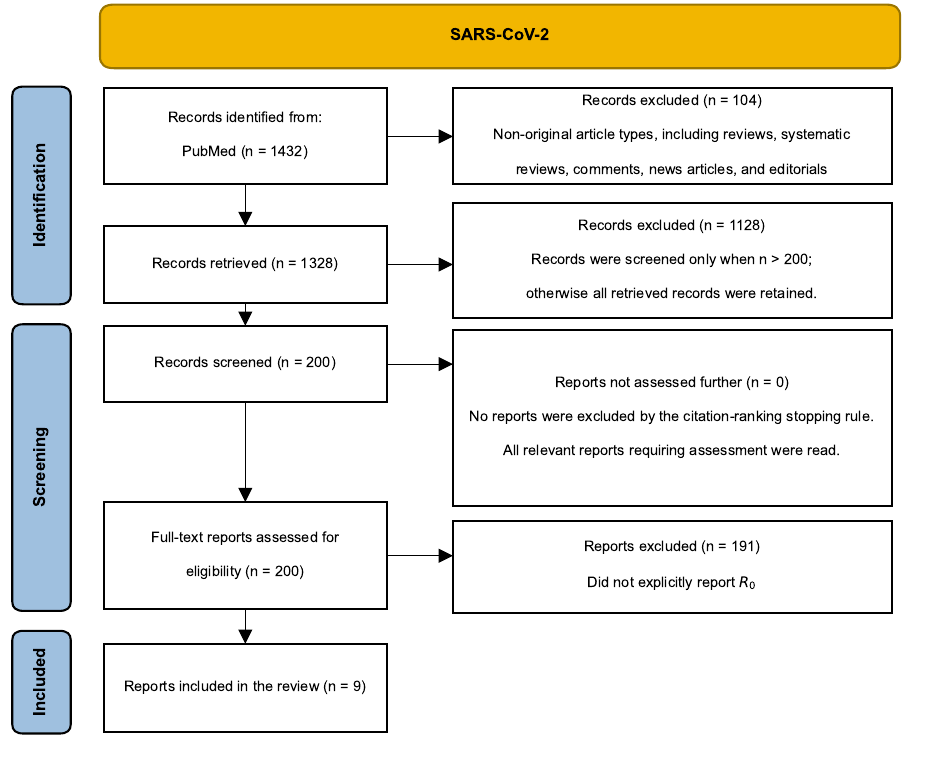}
	\caption{
  \textbf{PRISMA-style flow diagram of the literature search and study selection process for the $R_0$ estimates of SARS-CoV-2.}
}
	\label{SIfig_R0_COVID19}
\end{figure}

\begin{figure}[ht!]
	\centering
	\includegraphics{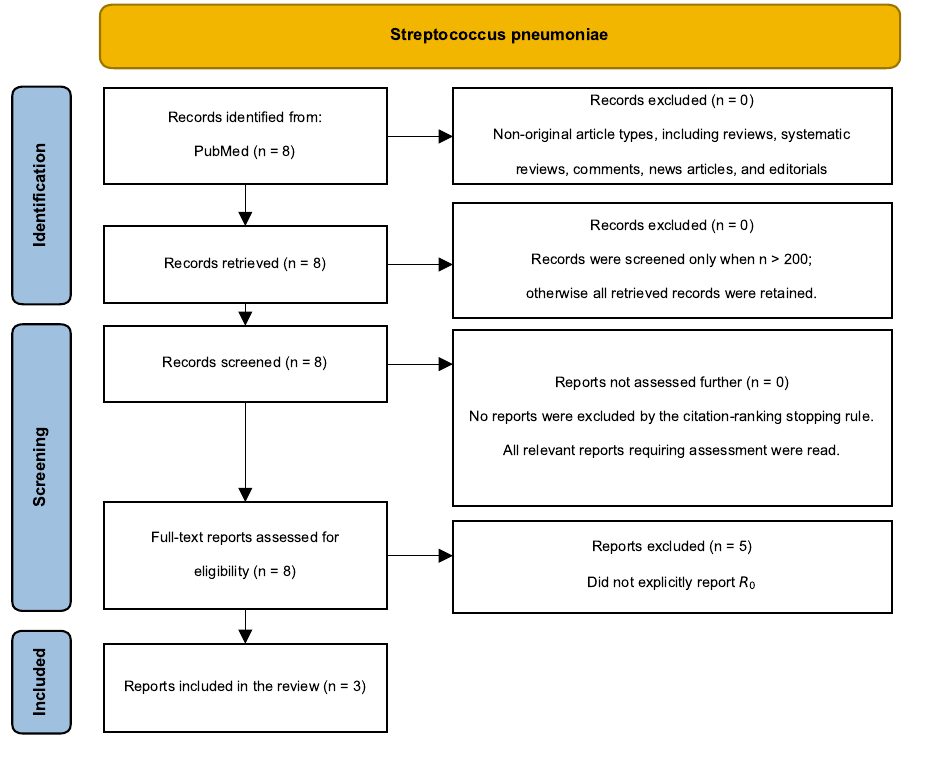}
	\caption{
  \textbf{PRISMA-style flow diagram of the literature search and study selection process for the $R_0$ estimates of Streptococcus pneumoniae (S. pneumoniae).}
}
	\label{SIfig_R0_Streptococ}
\end{figure}

\begin{figure}[ht!]
	\centering
	\includegraphics{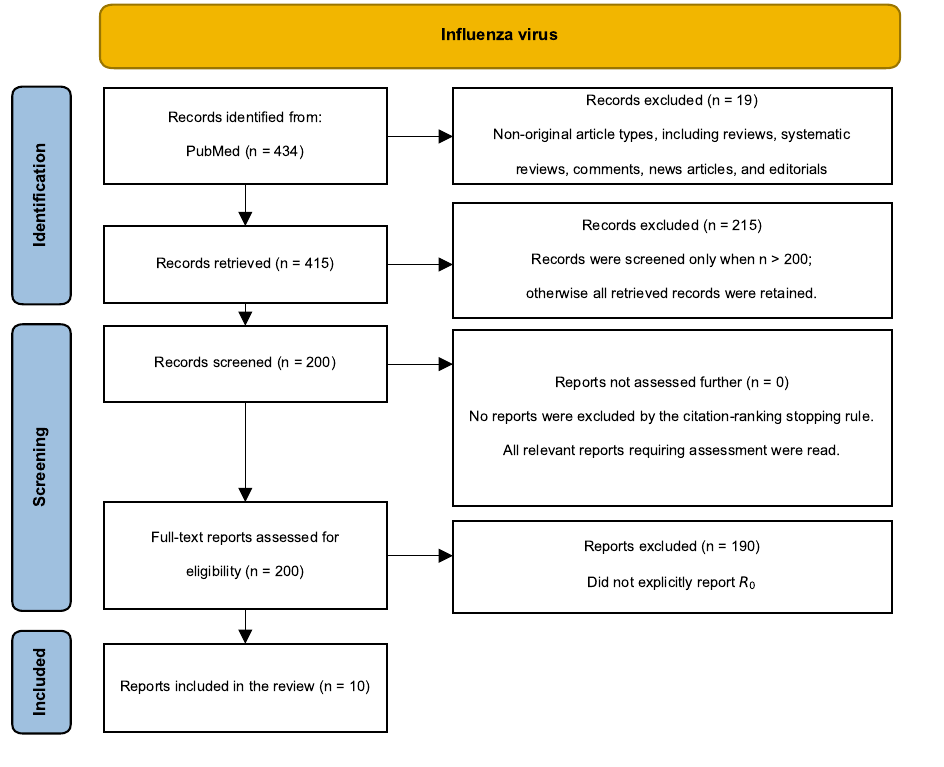}
	\caption{
  \textbf{PRISMA-style flow diagram of the literature search and study selection process for the $R_0$ estimates of Influenza virus.}
}
	\label{SIfig_R0_influenza}
\end{figure}

\begin{figure}[ht!]
	\centering
	\includegraphics{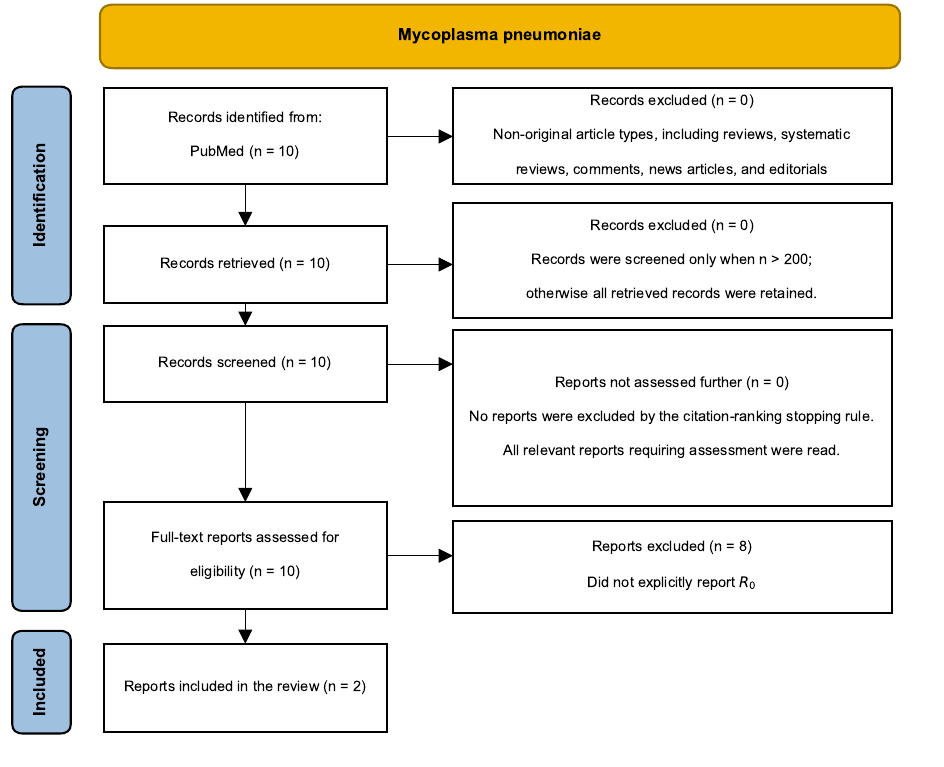}
	\caption{
  \textbf{PRISMA-style flow diagram of the literature search and study selection process for the $R_0$ estimates of Mycoplasma pneumoniae (M. pneumoniae).}
}
	\label{SIfig_R0_Mycoplasma}
\end{figure}

\begin{figure}[ht!]
	\centering
	\includegraphics{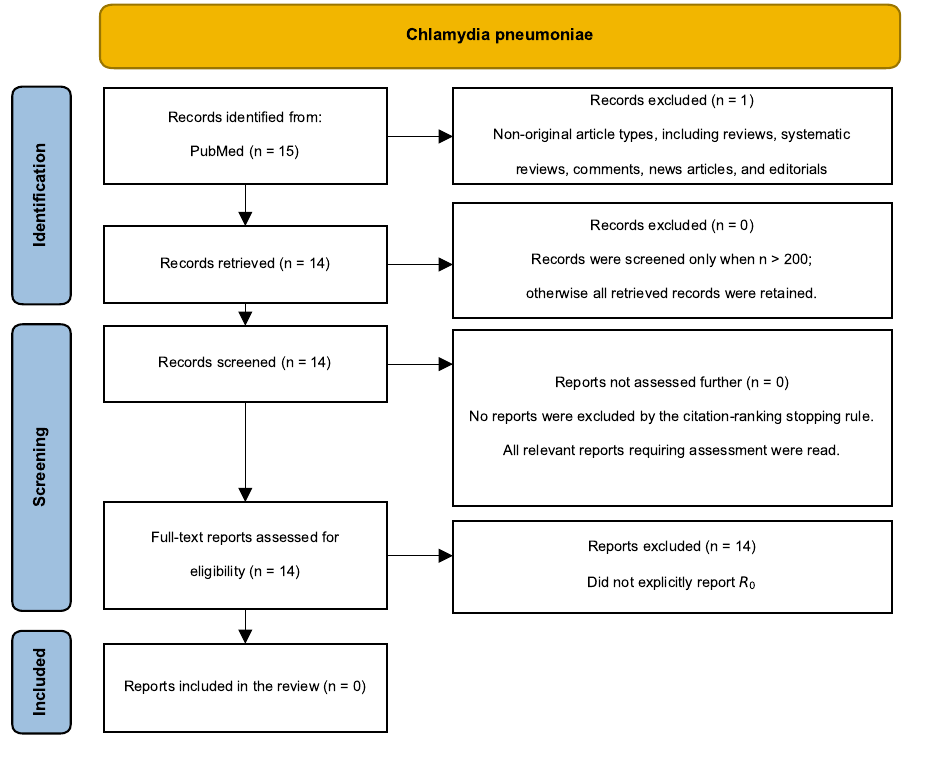}
	\caption{
  \textbf{PRISMA-style flow diagram of the literature search and study selection process for the $R_0$ estimates of Chlamydia pneumoniae.}
}
	\label{SIfig_R0_Chlamydia}
\end{figure}

\begin{figure}[ht!]
	\centering
	\includegraphics{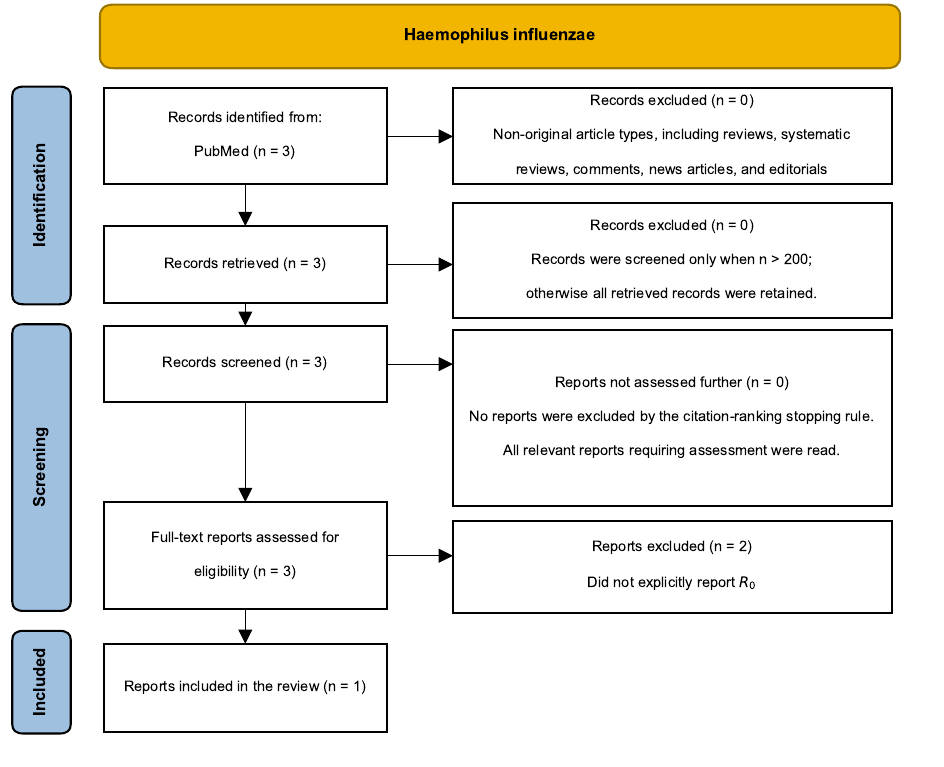}
	\caption{
  \textbf{PRISMA-style flow diagram of the literature search and study selection process for the $R_0$ estimates of Haemophilus influenzae.}
}
	\label{SIfig_R0_Haemophilus}
\end{figure}

\begin{figure}[ht!]
	\centering
	\includegraphics{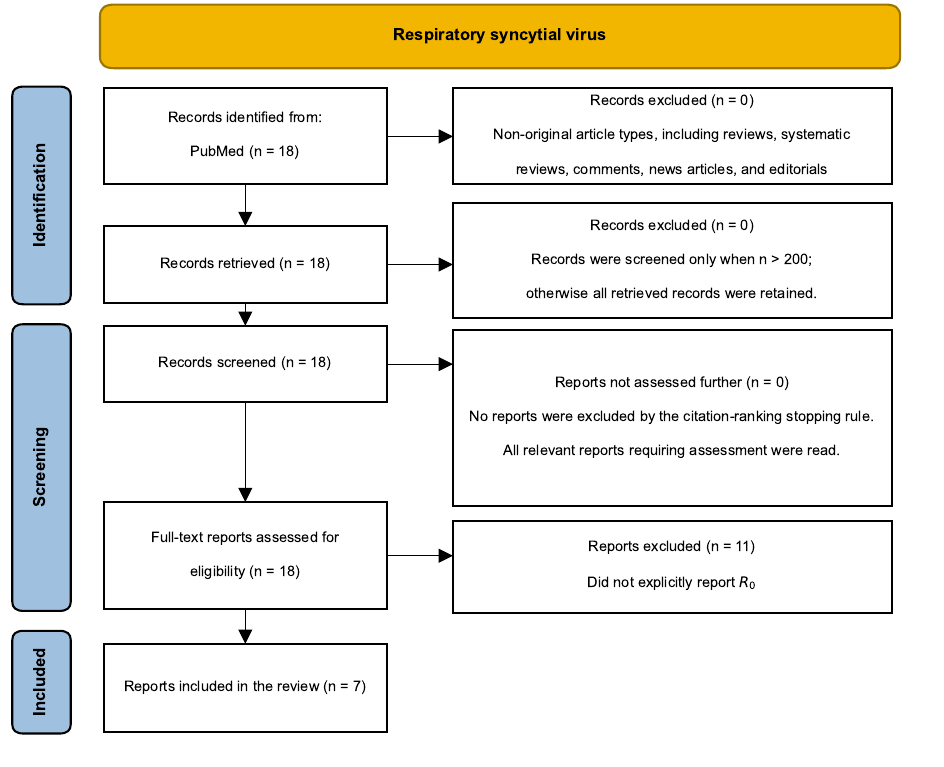}
	\caption{
  \textbf{PRISMA-style flow diagram of the literature search and study selection process for the $R_0$ estimates of Respiratory syncytial virus (RSV).}
}
	\label{SIfig_R0_RSV}
\end{figure}

\begin{figure}[ht!]
	\centering
	\includegraphics{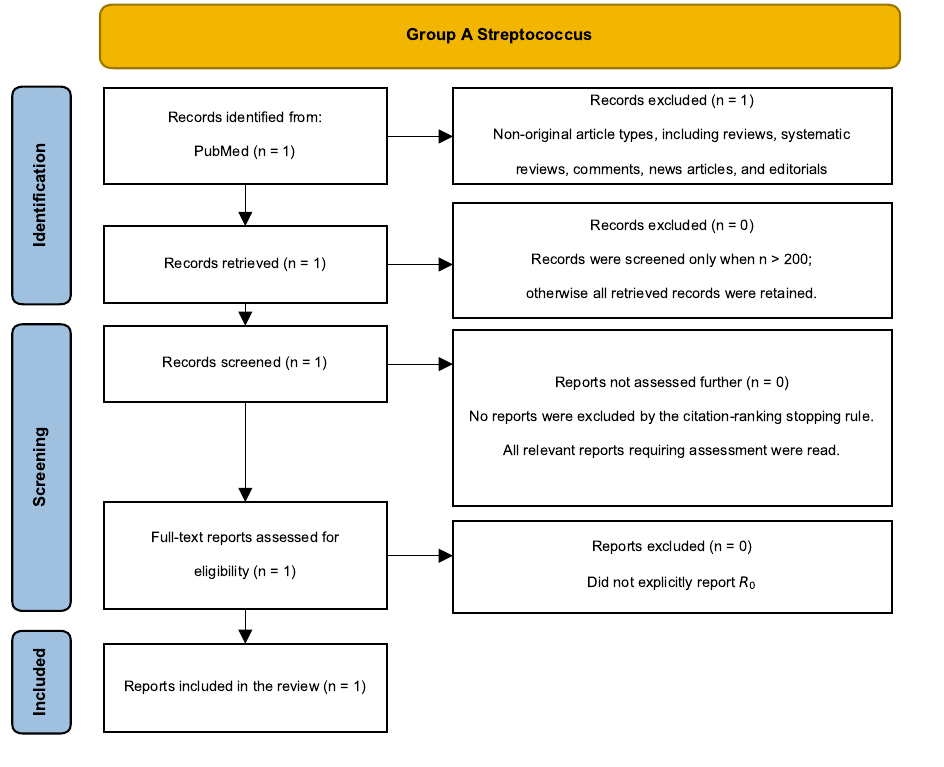}
	\caption{
  \textbf{PRISMA-style flow diagram of the literature search and study selection process for the $R_0$ estimates of Group A Streptococcus.}
}
	\label{SIfig_R0_StreptococcusGroupA}
\end{figure}

\begin{figure}[ht!]
	\centering
	\includegraphics{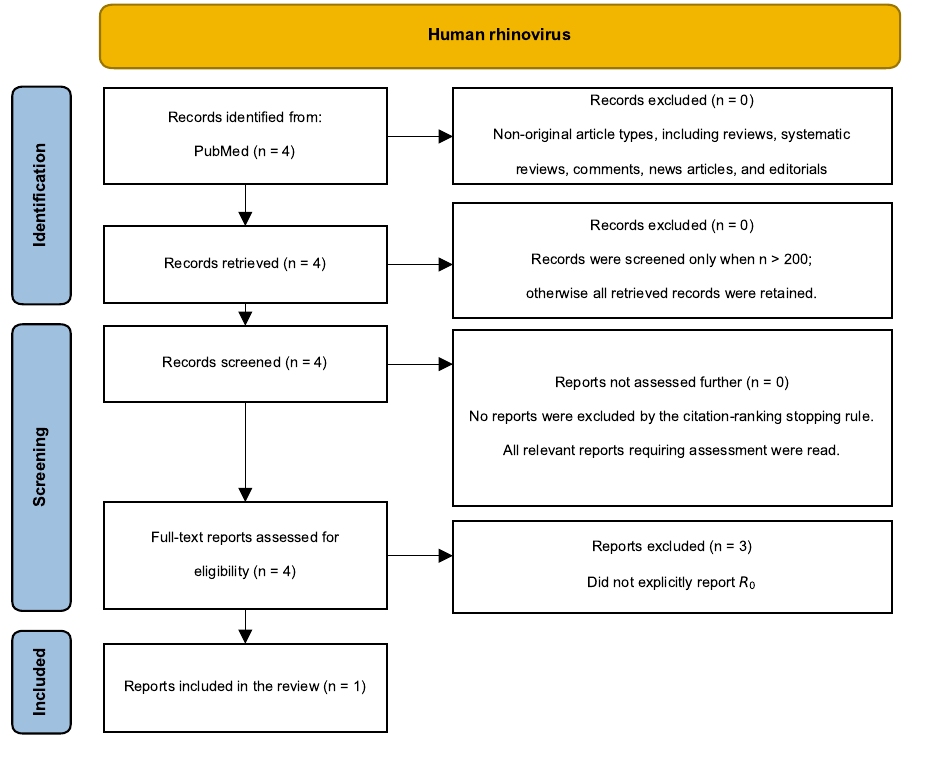}
	\caption{
  \textbf{PRISMA-style flow diagram of the literature search and study selection process for the $R_0$ estimates of Human rhinovirus.}
}
	\label{SIfig_R0_rhinoviruses}
\end{figure}

\begin{figure}[ht!]
	\centering
	\includegraphics{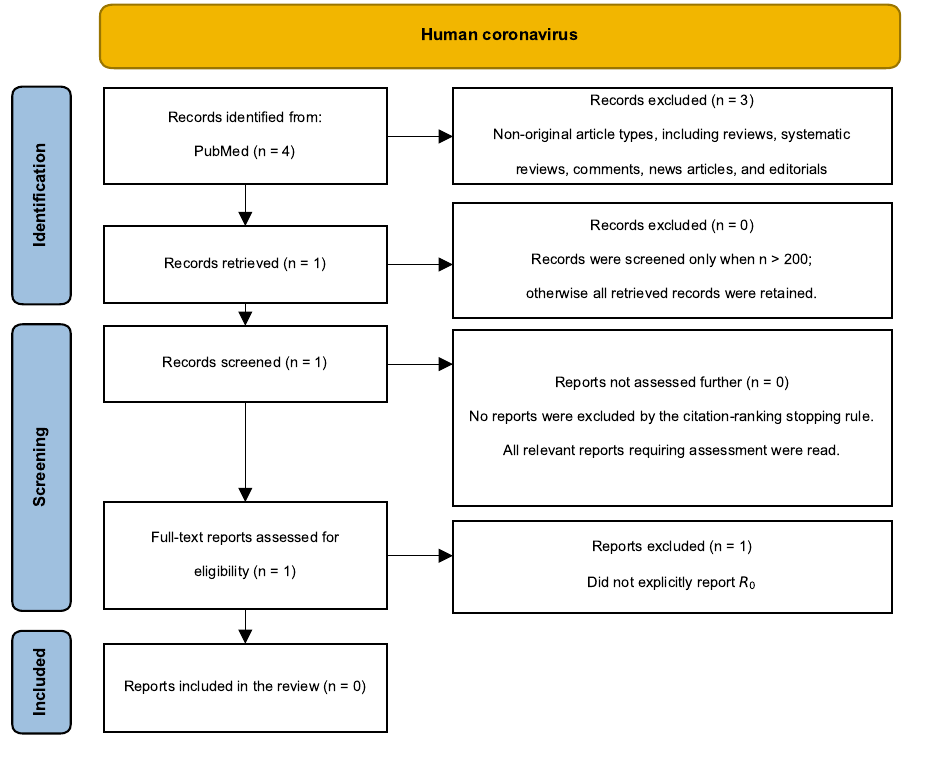}
	\caption{
  \textbf{PRISMA-style flow diagram of the literature search and study selection process for the $R_0$ estimates of Human coronavirus.}
}
	\label{SIfig_R0_HCoV}
\end{figure}

\begin{figure}[ht!]
	\centering
	\includegraphics{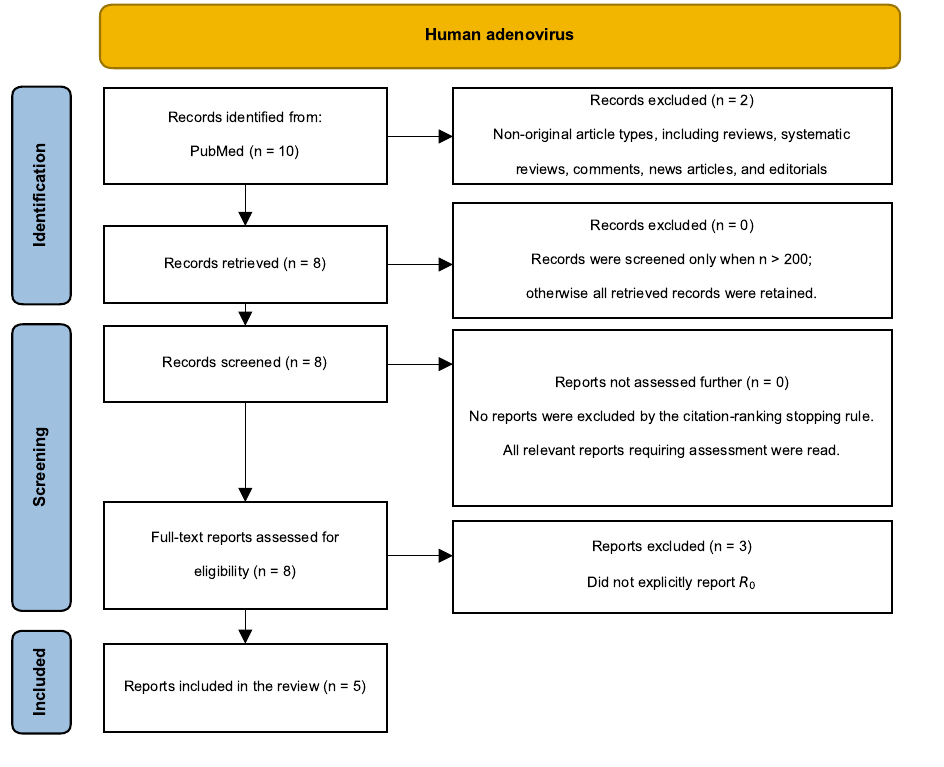}
	\caption{
  \textbf{PRISMA-style flow diagram of the literature search and study selection process for the $R_0$ estimates of Human adenovirus (HAdV).}
}
	\label{SIfig_R0_adenoviruses}
\end{figure}

\begin{figure}[ht!]
	\centering
	\includegraphics{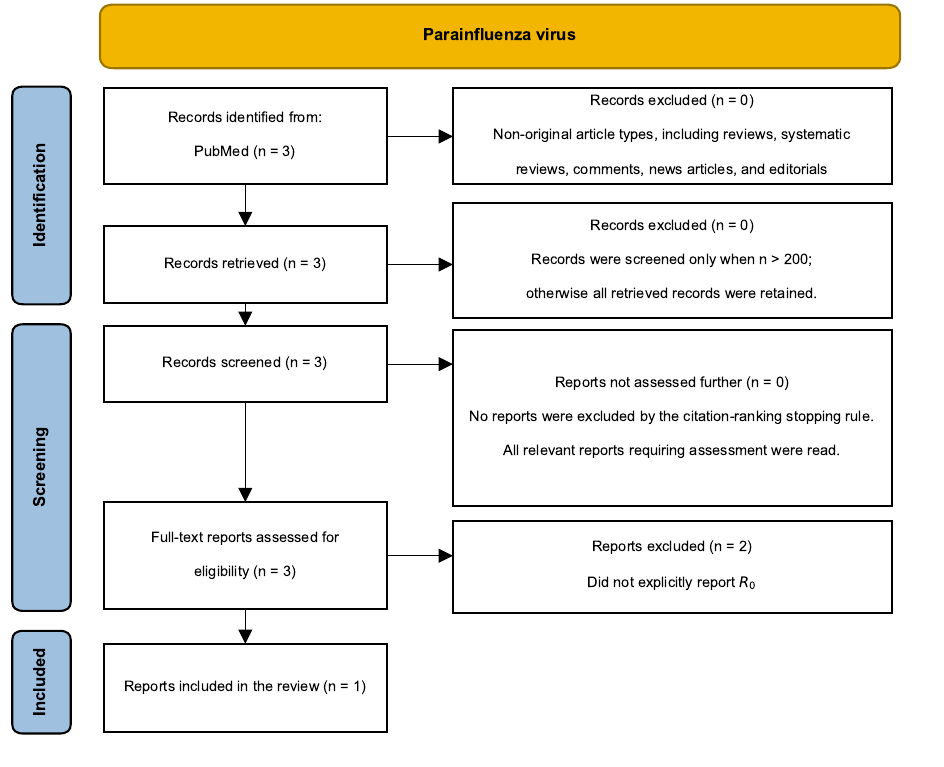}
	\caption{
  \textbf{PRISMA-style flow diagram of the literature search and study selection process for the $R_0$ estimates of Parainfluenza virus.}
}
	\label{SIfig_R0_parainfluenza}
\end{figure}

\begin{figure}[ht!]
	\centering
	\includegraphics{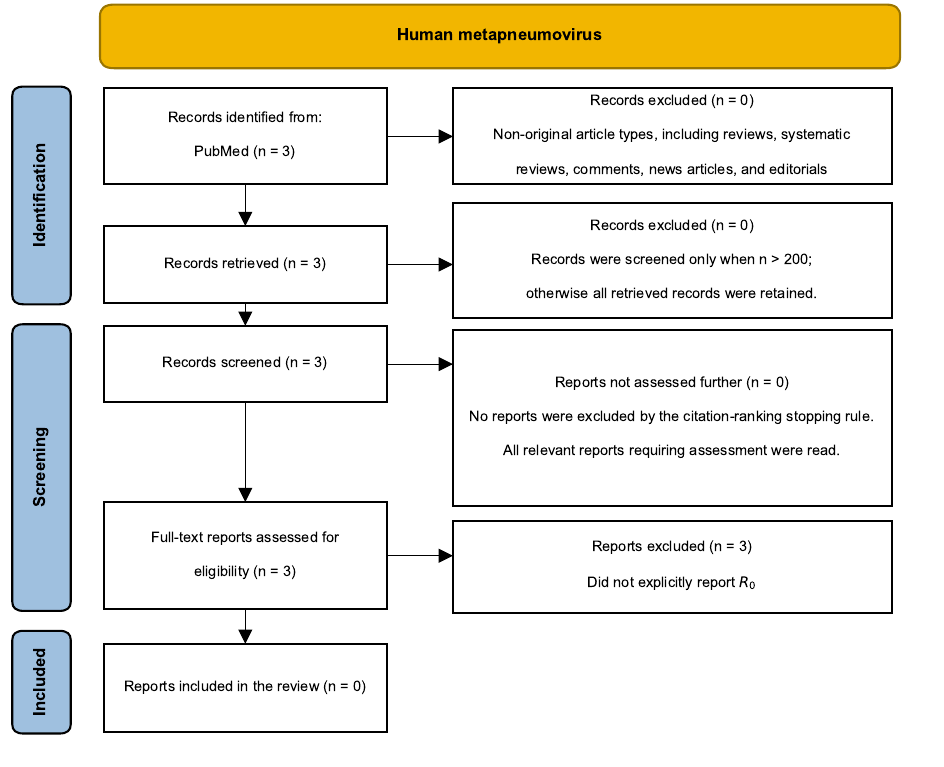}
	\caption{
  \textbf{PRISMA-style flow diagram of the literature search and study selection process for the $R_0$ estimates of Human metapneumovirus.}
}
	\label{SIfig_R0_metapneumovirus}
\end{figure}

\begin{figure}[ht!]
	\centering
	\includegraphics{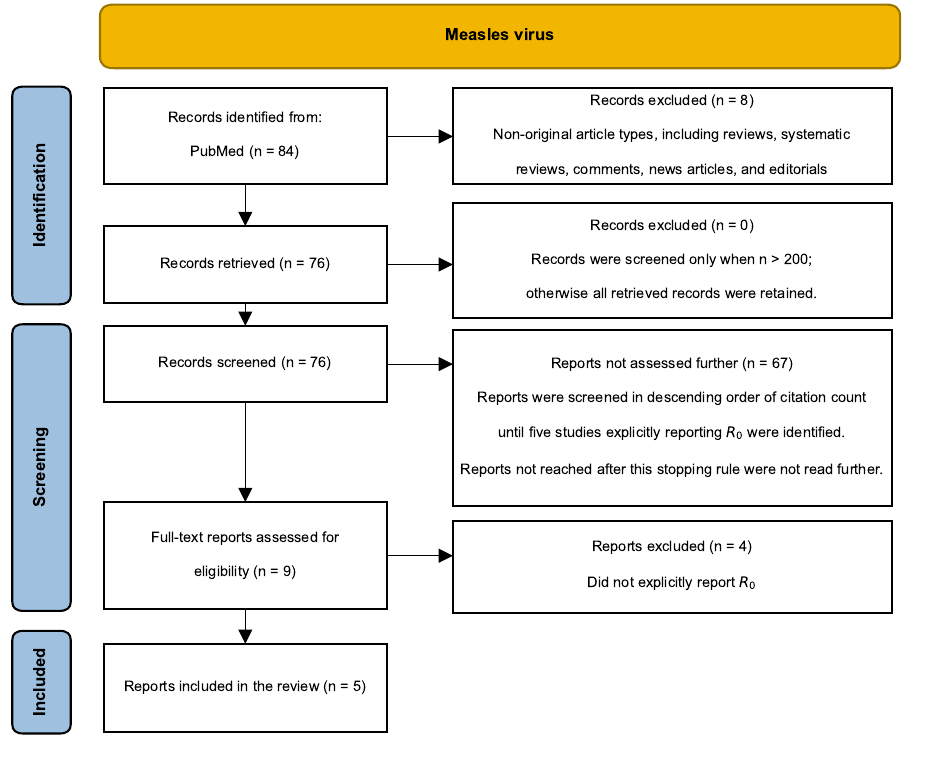}
	\caption{
  \textbf{PRISMA-style flow diagram of the literature search and study selection process for the $R_0$ estimates of Measles virus.}
}
	\label{SIfig_R0_measles}
\end{figure}

\begin{figure}[ht!]
	\centering
	\includegraphics{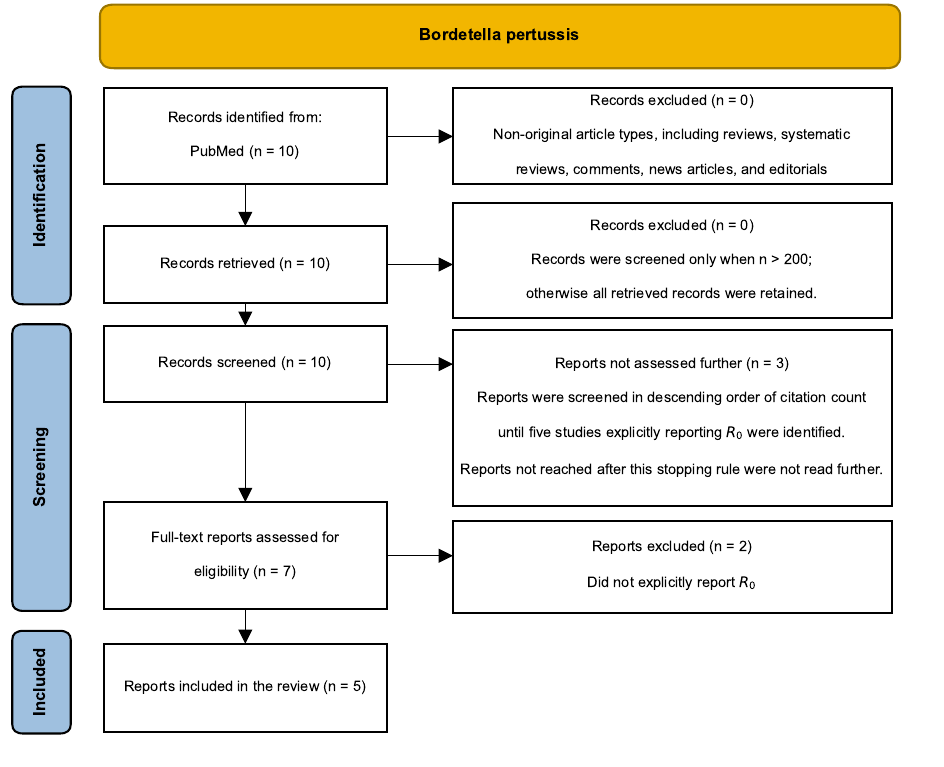}
	\caption{
  \textbf{PRISMA-style flow diagram of the literature search and study selection process for the $R_0$ estimates of Bordetella pertussis (B. pertussis).}
}
	\label{SIfig_R0_pertussis}
\end{figure}

\begin{figure}[ht!]
	\centering
	\includegraphics{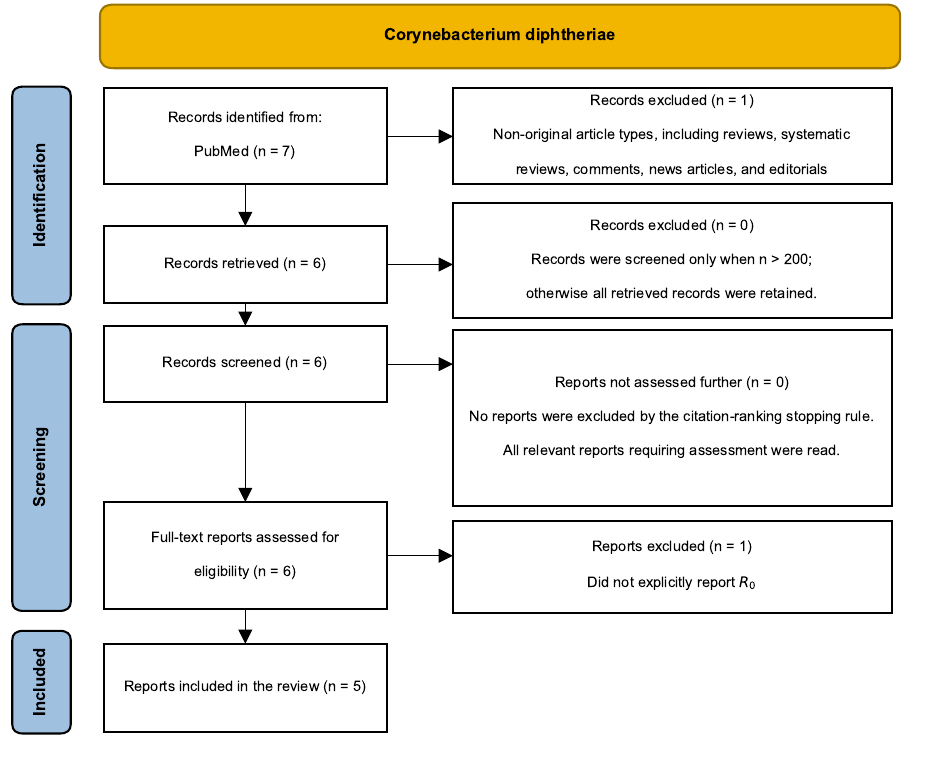}
	\caption{
  \textbf{PRISMA-style flow diagram of the literature search and study selection process for the $R_0$ estimates of Corynebacterium diphtheriae (C. diphtheriae).}
}
	\label{SIfig_R0_diphtheria}
\end{figure}

\begin{figure}[ht!]
	\centering
	\includegraphics{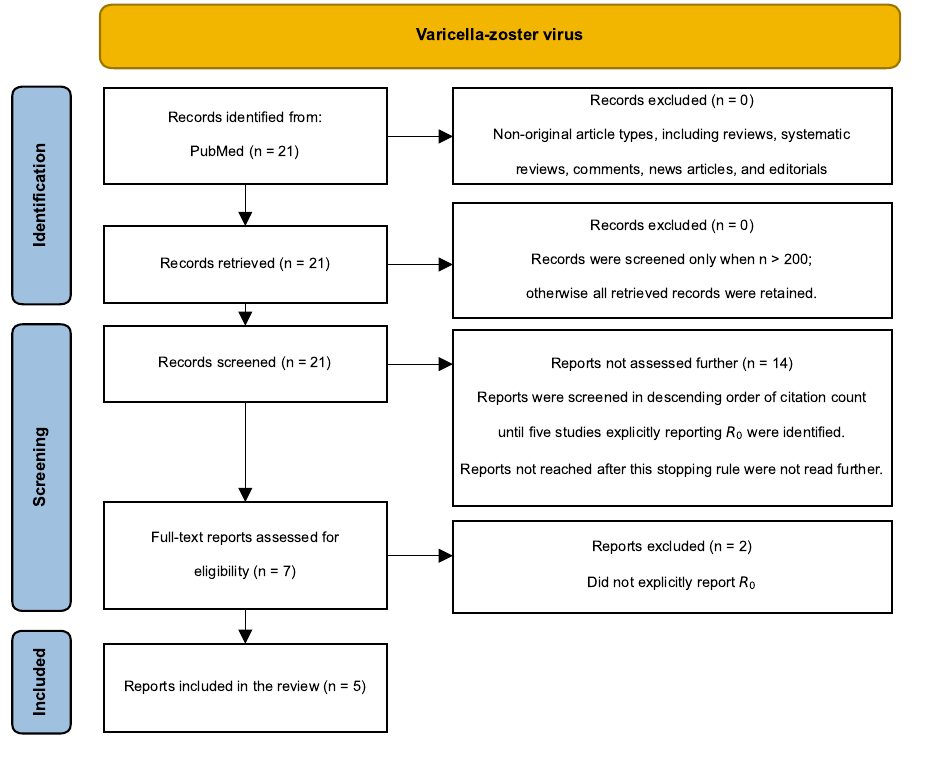}
	\caption{
  \textbf{PRISMA-style flow diagram of the literature search and study selection process for the $R_0$ estimates of Varicella-zoster virus (VZV).}
}
	\label{SIfig_R0_VZV}
\end{figure}

\begin{figure}[ht!]
	\centering
	\includegraphics{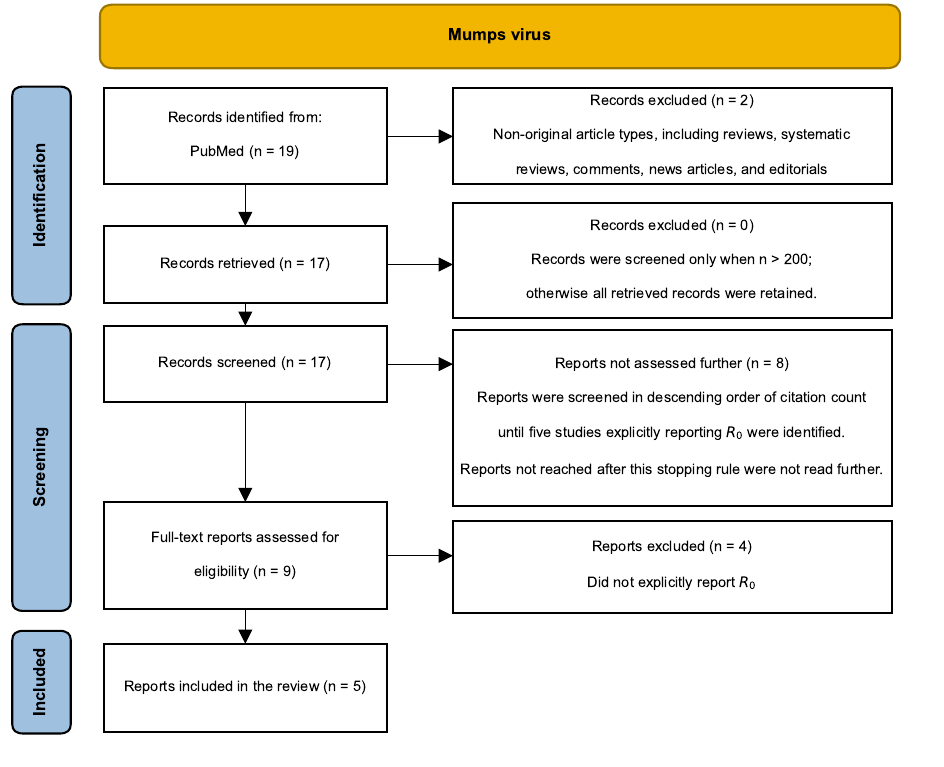}
	\caption{
  \textbf{PRISMA-style flow diagram of the literature search and study selection process for the $R_0$ estimates of Mumps virus.}
}
	\label{SIfig_R0_mumps}
\end{figure}

\begin{figure}[ht!]
	\centering
	\includegraphics{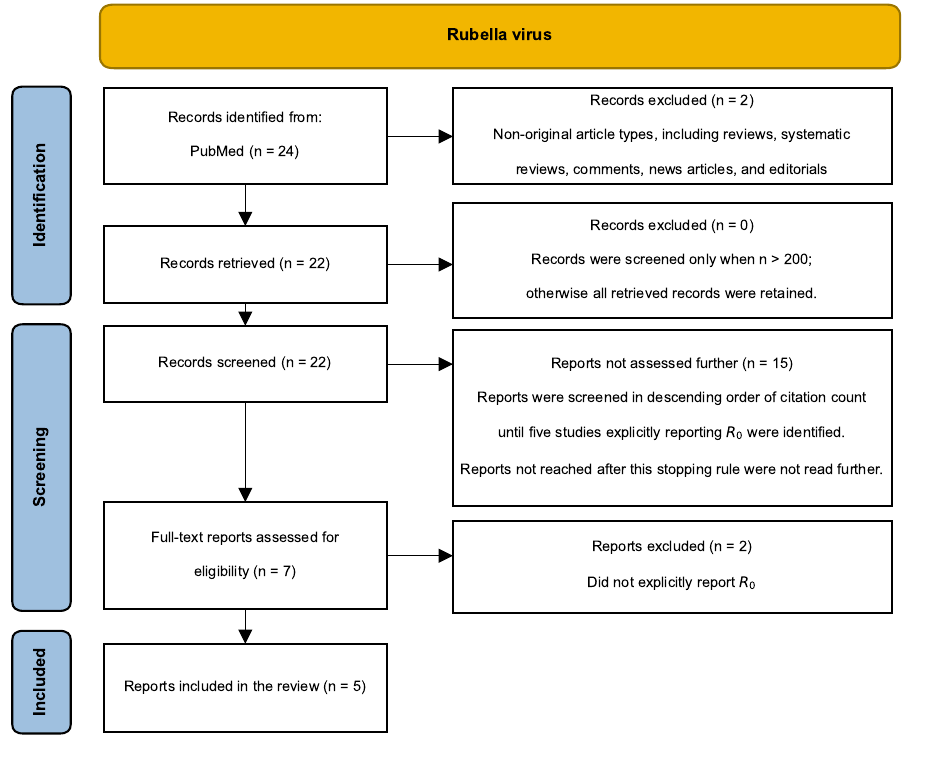}
	\caption{
  \textbf{PRISMA-style flow diagram of the literature search and study selection process for the $R_0$ estimates of Rubella virus.}
}
	\label{SIfig_R0_rubella}
\end{figure}

\FloatBarrier

\end{document}